\documentclass{article}

    \usepackage[preprint]{neurips_2020}

\usepackage[utf8]{inputenc} 
\usepackage[T1]{fontenc}    
\usepackage{hyperref}       
\usepackage{url}            
\usepackage{booktabs}       
\usepackage{amsfonts}       
\usepackage{nicefrac}       
\usepackage{microtype}      
\usepackage[usenames,dvipsnames]{xcolor}
\usepackage{graphicx}
\usepackage{subfigure}
\usepackage{algorithm}
\usepackage{algorithmic}
\usepackage{wrapfig}
\usepackage{centernot}
\usepackage{hyperref}
\usepackage{multirow}
\usepackage{amssymb}
\usepackage{amsthm}
\usepackage{amsmath}
\usepackage{pbox}
\usepackage{xspace}
\usepackage{lipsum}
\usepackage{tabularx, makecell}
\usepackage{tablefootnote}
\usepackage{threeparttable}
\usepackage[most]{tcolorbox}
\usepackage{etoolbox}
\patchcmd{\quote}{\rightmargin}{\leftmargin 1em \rightmargin}{}{}

\usepackage{amsmath,amsfonts,bm}

\def\eqref#1{equation~\ref{#1}}

\def\1{\bm{1}}

\DeclareMathAlphabet{\mathsfit}{\encodingdefault}{\sfdefault}{m}{sl}
\SetMathAlphabet{\mathsfit}{bold}{\encodingdefault}{\sfdefault}{bx}{n}

\makeatletter
\newtheorem*{rep@theorem}{\rep@title}
\newcommand{\newreptheorem}[2]{%
\newenvironment{rep#1}[1]{%
 \def\rep@title{#2 \ref{##1}}%
 \begin{rep@theorem}}%
 {\end{rep@theorem}}}
\makeatother

\newtheorem{theorem}{Theorem}
\newreptheorem{theorem}{Theorem} 

\newreptheorem{corollary}{Corollary} 

\newtheorem{definition}{Definition}

\newreptheorem{lemma}{Lemma} 

\newtheorem{remark}{Remark}

\renewcommand{\eqref}[1]{Eq.~(\textup{\ref{#1}})}

\usepackage[T1]{fontenc}
\usepackage{kantlipsum}
\PassOptionsToPackage{breakable, theorems, skins}{tcolorbox}
\usepackage{tcolorbox}

\title{Survey of Consciousness Theory from Computational Perspective\\ \large At the Dawn of Artificial General Intelligence}

\author{%
  Zihan Ding\thanks{The authors contributed equally.} \\
  Princeton University\\
  \texttt{zihand@princeton.edu}
  \And
  Xiaoxi Wei$^*$ \\
  Imperial College London \\
  \texttt{xiaoxi.wei18@imperial.ac.uk} \\
  \And
  Yidan Xu$^*$ \\
  University of Michigan \\
  \texttt{yidanxu@umich.edu} \\
}

\begin{document}

\maketitle

\begin{abstract}
  Human consciousness has been a long-lasting mystery for centuries, while machine intelligence and consciousness is an arduous pursuit. Researchers have developed diverse theories for interpreting the consciousness phenomenon in human brains from different perspectives and levels. 
    This paper surveys several main branches of consciousness theories originating from different subjects including information theory, quantum physics, cognitive psychology, physiology and computer science, with the aim of bridging these theories from a computational perspective. It also discusses the 
  existing evaluation metrics of consciousness and 
  possibility for current computational models to be conscious. Breaking the mystery of consciousness can be an essential step in building general artificial intelligence with computing machines. 
\end{abstract}

\tableofcontents

\section{Introduction}
\label{sec:intro}
Consciousness is a complex and elusive phenomenon that remains one of the greatest mysteries of science. It has been the subject of philosophical inquiry for centuries, and more recently, scientific investigation. We, humans, are not clear about why and how consciousness exists in our brains or even hold diverged opinions on whether we truly have consciousness. Existing consciousness theories have provided different interpretations of the human conscious process. 

This paper provides a comprehensive exploration of the theoretical foundations of consciousness from interdisciplinary perspectives. Chapter~\ref{sec:intro} endeavors to characterize the concepts related to consciousness, by differentiating the consciousness from others such as awareness, arousal, and wakefulness. This chapter further emphasizes the importance and difficulties of the human consciousness problem, with the aim of drawing attention from different research communities to jointly investigate this problem. 
Chapters ~\ref{sec:iit}, \ref{sec:physics}, and \ref{sec:orchor} elucidate the mathematical and physical underpinnings of consciousness. Specifically, Chapter~\ref{sec:iit} introduces \textbf{Information Integration Theory}, which outlines the conditions for information entropy required by a conscious entity, offering insights into the informational characteristics that consciousness might exhibit. Chapter~\ref{sec:physics} and ~\ref{sec:orchor} discusses \textbf{Consciousness as a State of Matter} and \textbf{Orchestrated Objective Reduction Theory}, both approaching consciousness problem from a physics standpoint. These two chapters discuss the specific features that consciousness, as a state of matter, should possess, along with the principles that some quantum theorists propose as the basis for the generation of consciousness. 
Subsequent Chapters ~\ref{sec:gwt},~\ref{sec:hot},~\ref{sec:ast} and~\ref{sec:ctm} survey several influential theories of consciousness and succinctly summarize research on computational models associated with each theory, including the
\textbf{Global Workspace Theory} (~\ref{sec:gwt}), \textbf{High-Order Theories} (~\ref{sec:hot}), \textbf{Attention Schema Theory} (~\ref{sec:ast}) and \textbf{Conscious Turing Machine} (~\ref{sec:ctm}).
In Chapter~\ref{sec:metric}, a brief overview of contemporary biomedical measurement methods for consciousness, grounded in electrophysiological signals and behavioral indicators, is provided.
In the final Chapter~\ref{sec:llm}, we engage in a discursive examination of artificial intelligence (AI) consciousness, particularly delving into the question of whether Large Language Models as instances of advanced computational models possess consciousness and exploring the necessary and sufficient conditions for AI consciousness.

In summary, this paper offers a comprehensive review of consciousness from various subjects encompassing information theory, quantum physics, cognitive psychology, physiology, and computer science, with the aim of bridging these theories from a computational perspective for building future AI consciousness.


To give the readers a primal taste of this topic, a dialogue revolving the consciousness problem is provided in the following section.

\subsection{A Platonic Dialogue About Human Consciousness}
One day, I met two PhD students in the modern world discussing the consciousness problem within the human brain.  This conversation starts with the relationship between consciousness and the physical world, discussing thoughts about what consciousness is. The confusion in this conversation serves as the motivation of this study and introduces the essential problems that this article aims to address. Here is the conversation.
\begin{quote}
\textbf{\textbf{Athena:}} Hey, I was looking at the results of the double-slit experiment, but I'm struggling to understand what determines the photons to choose which slit to go through, what happens here?

\textbf{\textbf{Galileo:}} Hey, it's just the wave-particle duality of the photons, and the state of the photon collapses only when you observe it on the screen. It is just random. Don't dig too much into it.

\textbf{\textbf{Athena:}} What do you mean by `random'? Does this world ultimately have a random essence? Even Einstein says that God does not play dice.

\textbf{\textbf{Galileo:}} Your measurement affects its state, which determines the slit it goes through in the double-slit experiment.

\textbf{\textbf{Athena:}} So my mental status determine the state of the photon? Its state may change if I choose to observe it in a different way, say later for 0.00001 seconds than the intended observation.

\textbf{Galileo:} Yeah, maybe. And your mental status is also a stochastic process, right? You have your free will or consciousness.

\textbf{Athena:} Consciousness. Do you really believe in its existence? What if I'm deterministic, fully determined by the underlying physical and chemical rules?

\textbf{Galileo:} But quantum mechanics assumes the basic principle that the photons, and particles in your body, including your brain, still have a probabilistic state, right? So you are also following a stochastic process instead of a deterministic one.

\textbf{Athena:} You are right. The state collapse process is assumed to have true randomness, and this process may happen in my brain, affecting my decision, and also which slit these photons may go through. But I still don't quite believe in the existence of my consciousness. The state decoherence may happen for a system as large as my brain, so there is no quantum property and I may still be deterministic.

\textbf{Galileo:} I think the existence of consciousness is a belief that differs from person to person.

\textbf{Athena:} Wait, but what is consciousness? We cannot discuss its existence without a clear definition. The consciousness seems to describe the subjective experience but without a rigorous definition. I'm wondering where and when the conscious process happens. 

\textbf{Galileo:} I guess it's in the cerebral cortex or thalamus. However, for such a phenomenon to happen, I would say most functional parts of the brain may have a collaborative process to trigger the consciousness. 

\textbf{Athena:} I agree. Also if we assume a stone, or a tree, cannot be as conscious as a human, then the system has to have a certain level of computational power. 

\textbf{Galileo:} Also, I cannot imagine a system having consciousness without memory. 

\textbf{Athena:} I may disagree on that. I think a patient with brain damage of losing his memory still has consciousness. 

\textbf{Galileo:} May be an evidence. Do you think consciousness retains during sleep?

\textbf{Athena:} This can be a complicated problem. You know there are multiple stages during sleep, including non-rapid eye movement (NREM) and rapid eye movement (REM). The levels of consciousness can be different for different stages.

\textbf{Galileo:} Agreed. The dream happens during the REM, right? It feels like I have consciousness during the dream, so I guess I'm conscious at REM. But it's still quite different from the wakefulness in experience.

\textbf{Athena:} I'm not sure, it's quite complicated. But the essential difference between dream and wakefulness is that during sleep there is no external sensory input to the brain, all the experiences are fake and fabricated by the brain itself. If the consciousness happens during the dream, it indicates that external sensory inputs may not be necessary for the system to be conscious.

\textbf{Galileo:} I guess so. It seems the consciousness is quite close to the subjective experience, even if I'm not sure if subjective experience is real existence or a fake hallucination by humans. 

\textbf{Athena:} It seems the `subjective experience' is an equivalent phrase of consciousness in some books. But this definition is still unclear to me. Also, the `real' and `fake' problem is also unclear to me. If our measurement of the world can affect its state, what does it even mean by a `real' observation of the world and a `fake' one? After all, they are just the mirrored signals in our brains, which can not directly represent the state of the real world.

\textbf{Galileo:} Well, that's a very philosophically pessimistic perspective. Speaking of the definition of consciousness, if we assume consciousness is defined as `subjective experience', it seems not very related to another concept called intelligence. Nowadays, people are building artificial intelligence in computers, is it also consciousness? This could be an interesting problem! Do you think it's just about the Turing test~\citep{turing2009computing}?

\textbf{Athena:} I don't think so. The focus of that paper is discussing whether the Turing machine can achieve human-level intelligence. Please notice the difference in the words here, it's intelligence but not consciousness! I assume a program passing the Turing test only means that it's as intelligent as a human, but not conscious. 

\textbf{Galileo:} Thanks for reminding me of that. It seems that intelligence and consciousness are two different things. 
But remember that at the beginning of the discussion, we mentioned that consciousness may require a certain level of computational power, and this computational power may be phrased as intelligence.

\textbf{Athena:} Maybe, the intelligent property seems to be a necessary condition for the consciousness to emerge, but I'm not sure to which extent the consciousness requires the intelligence to be.

\textbf{Galileo:} Yeah, this is undiscovered and could be a good research topic. However, if a creature is already intelligent enough to survive in the world, why does it still require consciousness? 

\textbf{Athena:} Good question. According to Darwin's theory of evolution, the existing creatures in the world should only exhibit those skills in favor of their survival through natural selection. If consciousness exists in humans, it indicates that it has some benefits for humans to survive in environments, and those without it are erased over history. But the trees and flowers still exist, which really troubles me.

\textbf{Galileo:} Trees and flowers are of different species from humans. Maybe for certain species, it requires to have consciousness, like animals.

\textbf{Athena:} Well, but I still feel unclear about why consciousness exists in humans or other animals, if assumed to exist. 

\textbf{Athena:} It seems like consciousness theory can be a very complicated subject with correlations to some very different subjects, like physics, biology, computer science, neuroscience, information theory, etc. It can be very difficult. But let's start to investigate at least!
\end{quote}

The above conversation is a microcosm of the discussions by the authors of the paper for starting the investigation of the consciousness problem. We will start the discussion with the definition of consciousness (Sec.~\ref{sec:def_cons}), and the relationship between consciousness and intelligence (Sec.~\ref{sec:cons_intel}), consciousness and free will (Sec.~\ref{sec:cons_will}). Then we have a brief overview of the existing consciousness theories (Sec.~\ref{sec:over_cons}) and a further discussion about consciousness while asleep (Sec.~\ref{sec:cons_asleep}).

\subsection{Definition of Consciousness}
\label{sec:def_cons}
\textbf{Consciousness, Awareness, Wakefulness and Arousal:}
Consciousness is a complex and multifaceted concept that has been studied extensively in the fields of neuroscience, psychology, and philosophy. Existing research typically defines consciousness as comprising two main components: arousal (wakefulness) and awareness (subjective experience) \citep{lendner2020electrophysiological}. Arousal refers to the overall state of alertness or wakefulness, while awareness refers to the subjective experience of perceiving and interpreting sensory information.

Typically, arousal is indicated by the opening of the eyes, while awareness is inferred by the ability to follow commands \citep{lee2022quantifying}. However, in certain instances, such as during dreaming, subjective experience can still occur despite the absence of full wakefulness. Consciousness is considered to be absent during sleep or anesthesia, but in some cases, it can still be present, depending on the level of arousal and awareness.
Consciousness, awareness, wakefulness, and arousal are related but distinct concepts in the study of the human mind and brain. We provide descriptive definitions for each concept according to the existing literature as follows: 

\textbf{Consciousness} refers to the subjective experience of being aware of one's thoughts, feelings, sensations, and surroundings. It is often described as the state of being awake and aware of one's surroundings and internal states. As an important concept in discussing the consciousness problem, \textit{qualia} refers to the subjective and personal experience of sensory information, such as the way we perceive colors, sounds, tastes, and smells. It is the subjective experience of sensory perception that cannot be objectively measured or observed by others.

\textbf{Awareness} refers to the ability to perceive, process, and comprehend information from one's environment or inner experience. It includes both conscious and unconscious processes and can range from simple sensory perception to complex cognitive processes such as attention, memory, and reasoning.

\textbf{Arousal} refers to the level of responsiveness of the brain and body to internal and external stimuli. It is a physiological state that ranges from a state of low arousal, such as drowsiness or relaxation, to high arousal, such as intense excitement or fear.

\textbf{Wakefulness} refers to the state of being awake and not asleep. It is a physiological state characterized by the presence of the electrical activity of the brain and the ability to respond to external stimuli.

In summary, consciousness is a subjective experience of being aware of one's thoughts, feelings, and surroundings, while awareness is the ability to perceive, process, and comprehend information. Wakefulness is a physiological state characterized by being awake, which is \textit{usually regarded identically as awareness}. Arousal is the level of responsiveness to stimuli. Awareness and arousal are necessary conditions for consciousness, but they are not sufficient to arouse the consciousness, and consciousness can occur in the absence of full awareness and high arousal levels.



\subsection{Measurement of Consciousness}
\label{sec:meansure_cons}

Recent studies have developed effective measures of human consciousness~\citep{seth2008measuring, demertzi2017measuring}, such as the electrical signal-based metrics like Perturbational Complexity Index (PCI)~\citep{casali2013theoretically} and Bispectral Index (BIS)~\citep{rosow2001bispectral,johansen2006update} and behavior-based metrics like the Glasgow Coma Scale (GCS)~\citep{jones1979glasgow,sternbach2000glasgow}, The Coma Recovery Scale-Revised (CRS-R)~\citep{giacino2004jfk} and  Full Outline of Unresponsiveness (FOUR)~\citep{wijdicks2005validation}.
Details for each physiological evaluation metric are discussed in Sec.~\ref{sec:metric}. In the following paragraphs, we take the PCI metric as an example to distinguish the concepts of arousal and awareness in different conscious states. 

The ability to accurately measure consciousness has important implications for understanding and treating conditions that affect consciousness, such as coma, anesthesia, and brain injury. PCI was developed from electroencephalographic (EEG) responses to direct and noninvasive cortical perturbation with transcranial magnetic stimulation (TMS). The PCI quantifies the complexity of deterministic patterns of significant cortical activation evoked by TMS to derive an empirical cutoff that reliably discriminates between unconsciousness and consciousness in various states, including REM sleep, wakefulness, ketamine-induced anesthesia, and conscious brain-injured patients.

In the PCI study by Casali et al. (2013)~\citep{casali2013theoretically}, arousal and awareness levels serve as indicators of the degree of human consciousness. The arousal and awareness levels of consciousness under different states are summarized in Table~\ref{tab:arousal_aware}, which shows under different states the consciousness will appear differently in human brains~\citep{lee2022quantifying}. Compared with the full wakefulness of a normal person, the sleeping stages like REM or NREM have lower levels of arousal or awareness and are commonly believed to be not as ``conscious'' as a wakeful state. There are pharmacology approaches to achieve similar low levels of arousal and awareness with anesthesia, leading to incomplete conscious states of a human. From a pathology view, existing patients with a minimally conscious state (MCS) will appear to have relatively high levels of arousal and awareness but still less consciousness, which is evidence of the fact that arousal and awareness are necessary but not sufficient conditions for consciousness. People with unresponsive wakefulness syndrome will have high arousal but low awareness.

\begin{table}[]
\begin{threeparttable}
\resizebox{\columnwidth}{!}{%
\begin{tabularx}{\textwidth}{m{180pt}|X|X}
\toprule
Conscious State & Arousal & Awareness \\ \midrule
Healthy wakefulness & high & high \\ \hline
REM\tnote{1} sleep with dreams & low & high \\ \hline
NREM\tnote{2} sleep without dreams & low & low \\ \hline
Anesthesia induced with ketamine & low & high \\ \hline
Anesthesia induced with propofol or xenon & low & low \\ \hline
Minimally conscious state & high & high \\ \hline
Unresponsive wakefulness syndrome & high & low 
\\ \bottomrule
\end{tabularx}%
}
\begin{tablenotes}\footnotesize
\item[1] REM: rapid eye movement
\item[2] NREM: non-rapid eye movement
\end{tablenotes}
\caption{The arousal and awareness levels of consciousness under different states (results adapted from \cite{lee2022quantifying})}
\label{tab:arousal_aware}
\end{threeparttable}
\end{table}


\subsection{Consciousness and Intelligence}
\label{sec:cons_intel}
In the well-known paper by Alan Turing, he made a comment on the consciousness argument against Turing machine~\citep{turing2009computing}:

"\textit{I do not wish to give the impression that I think there is no mystery about consciousness. There is, for instance, something of a paradox connected with any attempt to localise it. 
}"

\begin{figure}[H]
    \centering
    \includegraphics[scale=0.3]{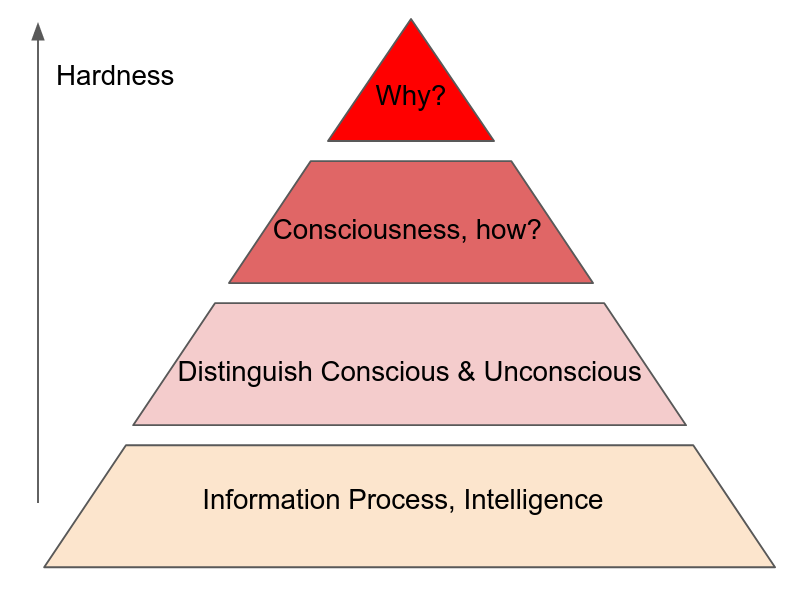}
    \caption{The hardness of different levels of the problems related to a conscious mind.}
    \label{fig:pyr}
\end{figure}

Considering the question in the paper is whether a machine can think like a human, Turing proposed the famous imitation game as a way to test machine intelligence. Intelligence and consciousness are widely considered two different properties of the brain. In \textit{Life 3.0}~\citep{tegmark2018life} by Max Tegmark, intelligence is defined as the ability to accomplish complex goals, and consciousness is defined as subjective experience. The consciousness seems to be more mysterious than intelligence, and harder to measure. As depicted in the consciousness ``pyramid'' in Fig.~\ref{fig:pyr} (originally in \cite{tegmark2018life}), the intelligence-related problems are the easiest, which is also claimed by David Chalmers. This type of problem typically does not require to consider the subjective experience of the experimental subjects. A hard problem is to find the physically interpretable features for distinguishing conscious and unconscious processes. The next level question is how the consciousness happens and what the determined factors are. The final and hardest problems related to the explanation of the existence of consciousness, or why consciousness exists in any system?

For the reasons for the existence of consciousness in human brains, a quote from Stephen Wolfram mentions the limitation of self-modeling and time-persistence are two key factors for humans to have consciousness:

"\textit{The fact that we have coherent consciousness is a consequence of two things: 1. That we are computationally bounded, so the universe does not contain enough resources for us to construct a complete model of ourselves. 2. That we believe that we are persistent in time, and hence assume constancy where there is none.}"

Consciousness is a complex and multifaceted phenomenon that has been studied by philosophers, psychologists, neuroscientists, and others for a long history. Even so, consciousness remains mysterious for modern society, and people refer to it as \textit{the hard problem of consciousness}, which was initially proposed by David Chalmers in 1995~\citep{chalmers1995facing, chalmers1997conscious}. It is generally understood as the experience or awareness of subjective mental states such as thoughts, perceptions, emotions, and sensations. The hard problem indicates the reasons for the existence of such subjective experiences in human minds. Consciousness is closely linked to the functioning of the brain, but its precise nature and mechanisms are still not fully understood. Some theories suggest that consciousness emerges from the integration of sensory information, while others propose that it is an intrinsic property of the universe or a fundamental aspect of reality itself. The quote by Stephen Wolfram suggests that the experience of coherent consciousness may depend on our computational limitations and our assumption of temporal persistence. The argument suggests that we believe we are persistent beings that exist over time, even though there is no actual constancy in the universe.


\subsection{Consciousness and Free Will}
\label{sec:cons_will}
\begin{figure}[H]
    \centering
    \includegraphics[scale=0.25]{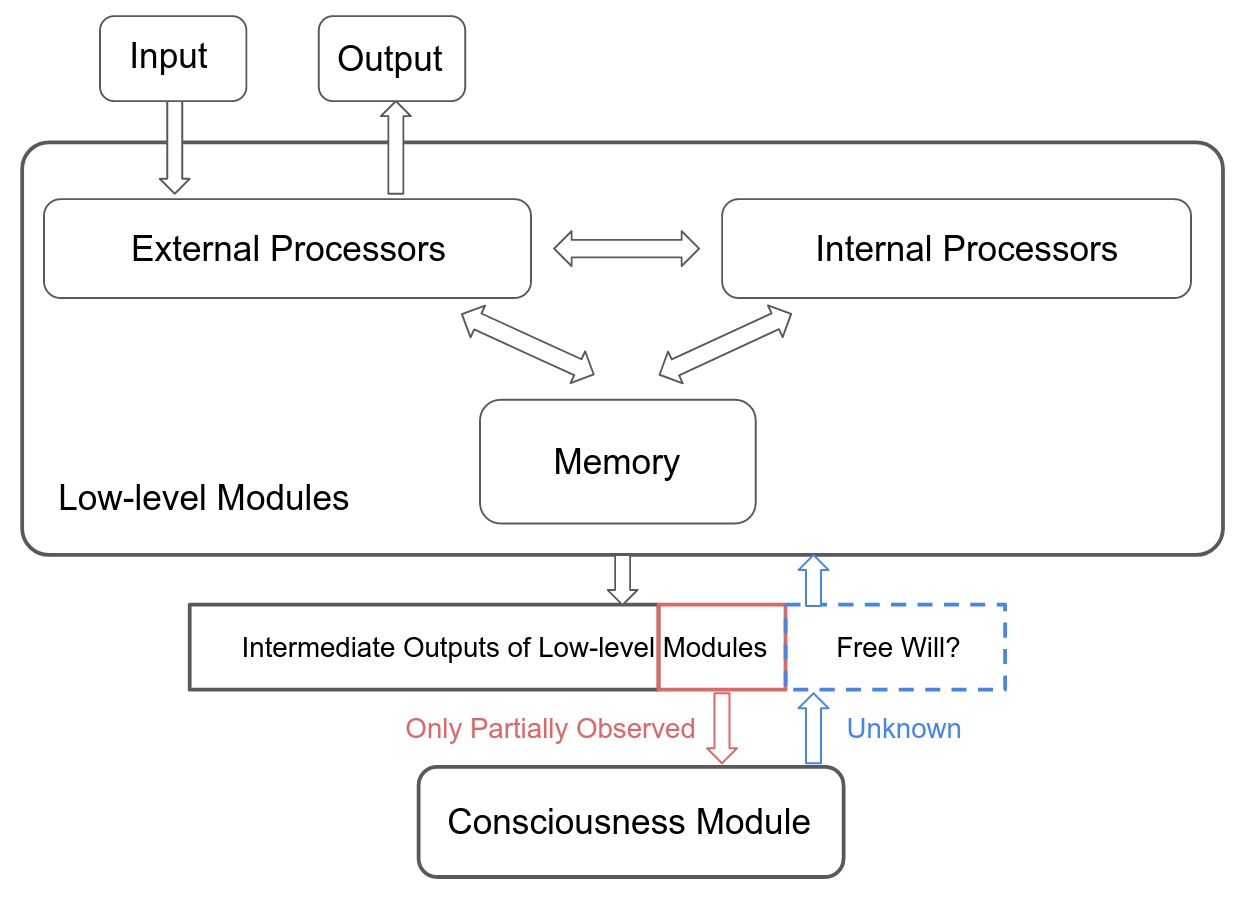}
    \caption{The overview architecture of consciousness system with free will.}
    \label{fig:consc_free_will}
\end{figure}

Free will is defined as the ability that humans to make choices and decisions that are not solely determined by biological, environmental, or external factors. Whether humans have free will is highly controversial and unknown. However, this concept is believed to be closely related to consciousness. In Fig.~\ref{fig:consc_free_will}, we depict an architecture unifying several consciousness theories to be introduced later in this article, as well as illustrating the relationship between consciousness and free will. In Fig.~\ref{fig:consc_free_will}, the human brain has interactions among the low-level modules and consciousness module, where the conscious experiences happen in human brains~\citep{baars2003global, baars2003conscious,baars2007architectural, baars2009consciousness, locke1948essay, armstrong1981consciousness,armstrong1985consciousness,lycan1996consciousness,armstrong2002materialist, lycan2004superiority, rosenthal2009higher, rosenthal2012higher,rosenthal2004varieties,byrne1997some,brown2019understanding, graziano2015attention, graziano2020toward}. The low-level modules involve external processors, internal processors, and memory. The external processors handle the inputs and outputs of the human brain, including the image processor, sound processor, gustatory processor, olfactory processor, tactile processor, motor activators, speaking modules, etc. Each module processes the input information and outputs the processed signals to other parts. Internal processors include the logic processor, language processor, etc. Each will process information with outputs from external processors, or spontaneously without external inputs. All these modules have the ability to communicate with other modules and the memory to accomplish the desired objective, and most of them will generate intermediate outputs as inputs of other modules. More importantly, beyond all these low-level modules, the consciousness module has the ability to observe the intermediate outputs from low-level modules, which generates subjective experiences (also known as consciousness) for a human. This observation process is also formulated as a self-modeling process in theories like attention schema theory~\citep{graziano2015attention, graziano2020toward} or higher-order thoughts/perception in high-order theories~\citep{locke1948essay, armstrong1981consciousness,armstrong1985consciousness,lycan1996consciousness,armstrong2002materialist, lycan2004superiority, rosenthal2009higher, rosenthal2012higher,rosenthal2004varieties,byrne1997some, brown2019understanding}, etc. Yet these explanations are of a highly abstract level. The specific components in human brains arsing the consciousness are still debatable in the literature. There are some works~\citep{baars2002conscious} showing that the pre-frontal cortex may be involved in the higher-level cognitive procedure as an example. Discussions of different interpretations of those theories will be detailed in later sections. Due to the limited computational power, the consciousness module will only pay attention to those important information flows, which is a relatively small subset of all intermediate outputs from low-level modules. This coincides with the empirical evidence showing that a human can only have dozens of conscious experiences per second~\citep{tegmark2018life, tegmark2000importance}.

\begin{figure}[htbp]
    \centering
    \includegraphics[scale=0.25]{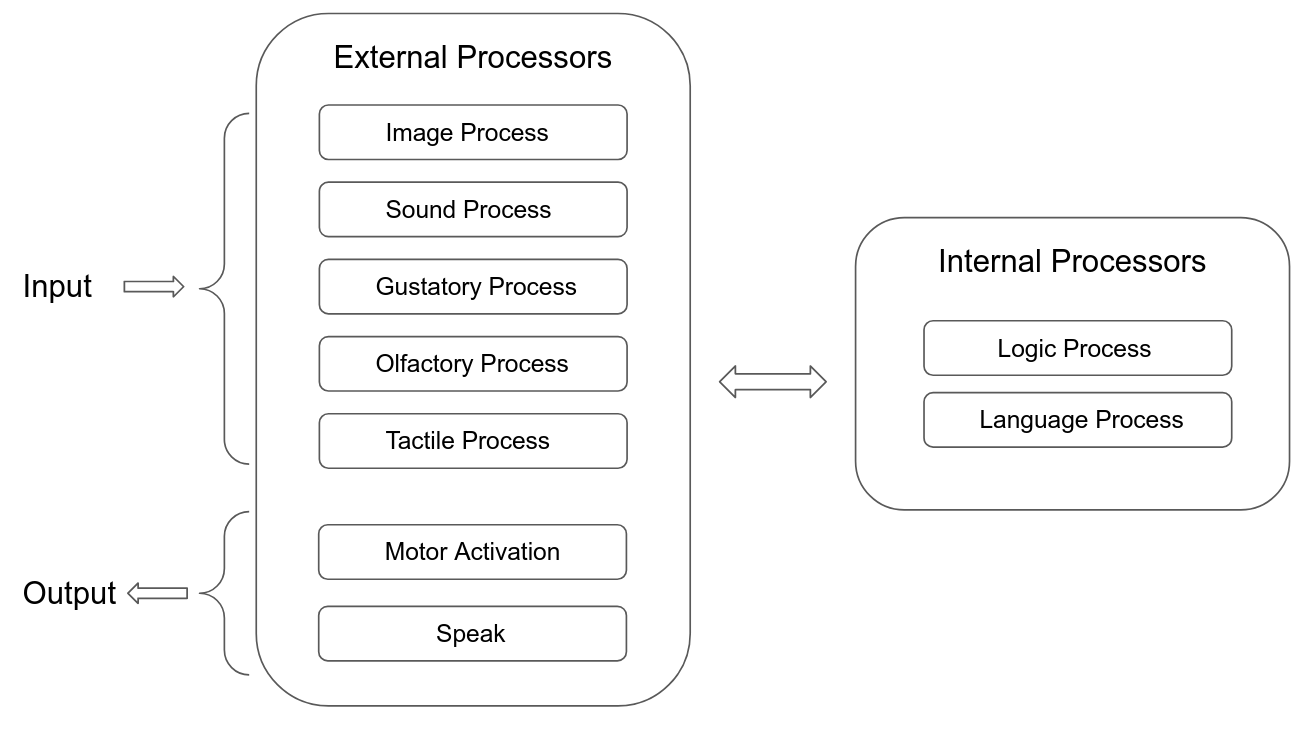}
    \caption{Details of processors in a consciousness system like the human brain.}
    \label{fig:processor}
\end{figure}

The existence of genuine free will remains a significant aspect of consciousness theory, albeit one that lacks sufficient scientific evidence to definitively prove or disprove. To engage in a comprehensive exploration of the existence of free will and its potential hierarchy within the human brain (Figure ~\ref{fig:consc_free_will}), we will bifurcate our discussion into positive and negative hypotheses: (1). assuming the existence of true free will; or (2). denying such existence of the free will. Therefore we mark the block of free will with dotted lines. For the first case that free will exists, which is beyond current physics interpretation, we may seek \textit{external variables} from beyond the existing mathematical and physics frameworks to determine the process of human decision-making.  The external variables can be regarded as an analog of the \textit{hidden variables} in the famous Einstein–Podolsky–Rosen (EPR) paradox~\citep{einstein1935can}, which states that some unobserved hidden variables may exist for explaining the true randomness in quantum mechanics, as a question of the completeness of quantum mechanics framework by Albert Einstein and others. Similar as the interpretation of EPR paradox for randomness in quantum mechanics, the explanation of the true `free will' may also require such external randomness beyond the known physical systems. More details regarding physical explanations of the existence of free will are discussed in Sections \ref{sec:physics} and \ref{sec:orchor}. This part of free will is shown in Fig.~\ref{fig:consc_free_will} as the blue block containing the free will, injected by the consciousness module to affect the low-level processes (as Fig.~\ref{fig:processor}) thus the final outputs from the system. For the second case that true free will does not exist, a question is what leads to some people think and feel that they also have the so-called `free will'? This phenomenon can be interpreted with the current architecture with a hallucinated `free will'~\citep{wegner2004precis}. The key fact is that the consciousness module can only observe partial intermediate outputs of low-level modules due to its limited information processing bandwidth, thus the final outputs to the environment from the system can not be fully determined by these observed pieces of information, but together with more of other unobserved information. The consciousness module still seeks to explain the generated outputs by creating hallucinations of `free will' in determining the results. However, this explanation may be debatable and requires verification. The existence of free will is also an open problem at present.

\subsection{Consciousness while Asleep}
\label{sec:cons_asleep}



The human sleep process contains stages including Rapid Eye Movement (REM) and Non-Rapid Eye Movement (NREM). According to previous research \citep{lee2022quantifying}, NREM is a state where there is neither arousal nor awareness, whereas REM is a state of awareness without arousal. 

Integrated Information Theory (IIT)~\citep{tononi2004information} suggests that consciousness is reduced during deep sleep. IIT proposes that consciousness is generated by the integrated information within a system. According to IIT, the level of consciousness experienced during sleep is dependent on the degree of integration present in the brain's activity.

Studies have shown that subjects awakened from deep NREM sleep, especially early in the night, often report a lack of awareness, even though cortical and thalamic neurons remain active. However, subjects awakened at other times, mainly during REM sleep or during lighter periods of NREM sleep later in the night, report dreams characterized by vivid images \citep{hobson2000dreaming}.
From the perspective of integrated information theory, a reduction in consciousness during sleep would be consistent with the bistability of cortical circuits during deep NREM sleep. Consistent with these observations, studies using TMS (transcranial magnetic stimulation), a technique for stimulating the brain non-invasively, in conjunction with high-density EEG (electroencephalogram), show that early NREM sleep is associated with a breakdown of effective connectivity among cortical areas, leading to a loss of integration or a loss of repertoire and thus of information \citep{massimini2005breakdown, massimini2007triggering}. These findings suggest that the level of consciousness experienced during sleep is dependent on the degree of integration present in the brain's activity, with a reduction in integration leading to a reduction in consciousness.

Overall, IIT suggests that the level of consciousness experienced during sleep is dependent on the degree of integration present in the brain's activity, with a reduction in integration leading to a reduction in consciousness.

\subsection{Overview of Consciousness Theories}
\label{sec:over_cons}
 Diverse theories are developed by researchers to investigate the nature of consciousness and how it arises from the brain. Some of the most prominent theories include information integration theory (IIT), consciousness as a state of matter, orchestrated objective reduction (Orch OR) theory, global workspace theory (GWT), high-order theory (HOT), attention schema theory (AST), consciousness Turing machine (CTM), etc. IIT (Sec.~\ref{sec:iit}) proposes that consciousness arises from the integration of information from multiple sensory and cognitive sources. Consciousness as a state of matter (Sec.~\ref{sec:physics}) analyzes the deficiency of the information integration principle from the physics-principled calculation. Orch OR theory (Sec.~\ref{sec:orchor}) proposes the orchestrated objective reduction process for explaining the free will in conscious experience from a quantum mechanics perspective. GWT (Sec.~\ref{sec:gwt}) argues that consciousness arises from the activation of a global workspace in the brain, which integrates information from different sources and broadcasts it to the rest of the brain. HOT (Sec.~\ref{sec:hot}) asserts that consciousness is the result of higher-order representations of sensory information. AST (Sec.~\ref{sec:ast}) posits that consciousness is an attentional schema that the brain uses to represent the state of being conscious.  CTM (Sec.~\ref{sec:ctm}) theory asserts that consciousness can be described as a computational process, as an extension of the Turing machine. These theories offer different perspectives on the nature of consciousness, and each has its own strengths and weaknesses. These theories offer insights into how it might be studied and understood. We conducted this survey to compare the similarities and differences of these theories and also summarize the correlations among different theories. More importantly, we aim to find feasible computational models from these theories for characterizing the conscious process of the human brain.

\section{Information Integration Theory}
\label{sec:iit}

The theory of information integration postulates that consciousness corresponds to the capacity of a system to integrate information. It first proposes the axioms of experiences, then postulates the properties of physical system that would give rise to the intrinsic experiences. In order to achieve this, the theory claims that the system must have a cause-effect power in itself, not resulting from any external factor. The cause-effect power of the system is then quantified by the largest minimum entropy of all sub-systems, evaluated by intervening on the states of a subset of the system (cause), and observe the change of states in the other part of the system (effect), while holding the external factors fixed. Therefore, IIT claims that \textbf{any conscious experience relates to a cause-effect structure that is maximally irreducible}. We will show in this section how the cause-effect power can be quantified by a measure called \textit{information integration}. If the IIT theory is correct, we should be able to calculate the integrated information for a conscious experience in human brain and derive a reasonable value.

\subsection{Information Entropy}
The definition of information, according to Shannon~\citep{shannon1948mathematical}, is quantified by the reduction of uncertainty among a number of alternatives when one occurs. This is measured by the entropy function, defined as the following.
\begin{definition}(Shannon Entropy) Given a probability measure $P$ on a $\sigma$-algebra $\mathcal{A}$, the entropy of a probability distribution is:
\begin{align}
  H(A) = H_p(A) = \int_{a\in\mathcal{A}} -p(a)\log p(a) da
\end{align}
\end{definition}
The logarithm used in this calculation is usually base 2, which means that the entropy is measured in bits. The entropy of a system is a measure of the average information content of a message generated by that system.

For example, in a binary system with two possible outcomes (0 or 1), the entropy is highest ($=1$) when the probability of either outcome is 0.5, and lowest ($=0$) when one outcome is certain (probability of 1) and the other is impossible (probability of 0).

In addition to being used in information theory, entropy has also been applied in fields such as cryptography, signal processing, and thermodynamics. In thermodynamics, entropy is used to quantify the degree of disorder or randomness in a thermodynamic system, and is often referred to as thermal entropy.


For a discrete-state system, a uniform distribution over all possible independent states contains the lowest information, therefore has the highest entropy. However, simply having a large number of independent components that results in a vast range of available states is not sufficient to generate conscious systems. These components must also be causally dependent on one another at an appropriate spatial and temporal scale. This crucial aspect of consciousness is referred to as information integration, as introduced by Tononi~\citep{tononi2004information}. Additionally, the author proposed a computational model to quantify the capacity of information integration, which will be detailed in the following sections.




\subsection{Basics Concepts of IIT}
\label{subsec:concept_iit}
The axioms of IIT state that every experience exists intrinsically and is structured, specific, unitary, and definite 
where specifically,
\begin{itemize}
    \item Experience exists intrinsically;
    \item Experience is specific, being composed of a particular set of phenomenal distinctions (qualia);
    \item Experience is unitary, irreducible, as an integration of information;
    \item Experience is definite in its content and spatio-temporal grain (exclusion of other possibilities).
\end{itemize}
The theory then postulates that, for each essential property of experience, the physical subtrate of consciousness (PSC) must have a cause-effect power related to the brain. The objective is then to find the appropriate spatial and temporal scale of neural elements gives rise to consciousness. The theorem implies that only those elements that have the maximum intrinsic cause–effect power are identified as elements of PSC. it is notable that under such a definition, the cause-effect power could be higher at a coarser spatial scale comparing to a finer spatial scale. 




Recall the definition of consciousness in IIT corresponds to the capacity to integrate information, such that the system generates a large collection of states while being causally dependent to each other. In a hypothetical setting, imagine a collection of neuronal elements locally connected but disconnected from outside stimulus, then one is able to test if such a collection can be separated into two independent parts by measuring the information gain of one part by knowing the other part. In information theory, this precisely corresponds to \textit{mutual information} (MI) of two random variables. The measurement of information integration in the theory is defined as a certain type of MI in the brain system, called \textit{effective information} (EI).

\begin{definition}[Mutual Information] The mutual information between two variables $A, B$:
\begin{align}
   I(A,B) = H(A) + H(B) - H(AB)
\end{align}
where $H(A)=H_{p_A}(A), H(B)=H_{p_B}(B), H(AB)=H_{p_{AB}}(A;B)$, $p_{AB}$ is the joint distribution of $A, B$.
\end{definition}

In the following discussions, we will generalize the symbols $A,B$ to be two sub-systems (or two subsets of variables) instead of two variables. 

In order to measure the information gain, start by setting one part of neural elements to independent set of noises, and observe how the firing pattern changes in the other half as a consequence of receiving signals. Precisely, we define the concept of EI:

\begin{definition}[Effective Information] Effective information measures the \textbf{directional} causal effects of $A$ on $B$,
\begin{align}
    \text{EI}(A\rightarrow B) = I(\tilde{A},B)=H(\tilde{A})+H(B)-H(\tilde{A}B), \tilde{A} = \arg\max_A H(A)
\end{align}
which means $A$ is chosen to be independent random noise, thus $B$ has no causal effects on $A$.
\end{definition}

According to the above definitions of MI and EI, some lemmas can be directly derived, which are stated in the following remark.

\begin{remark}
The information gain from $A$ to $B$ is not the same as $B$ to $A$ due to different connectivity pattern, while the mutual information is isotropy; and the EI is always upper bounded by the smallest maximum entropy of set A and set B. In mathematical terms:
\begin{align}
    I(A,B)=I(B,A) &\centernot\implies \text{EI}(A\rightarrow B) = \text{EI}(B\rightarrow A) \\
    \text{EI}(A\rightarrow B) &\le \min\{ O(\max_AH(A)), O(\max_BH(B))\}
\end{align}
\label{rm:ei}
\end{remark}

As a consequence, we are able to measure how part $A$ causally effects the other part $B$ and vice versa. This gives rise to the \textbf{isotropy} causal effects defined by \textit{mutual effective information} (MEI).

\begin{definition}[Mutual Effective Information] Mutual effective information $\text{EI}(A\rightleftharpoons B)$ measures the isotropy causal effects of between $A$ and $B$,
\begin{align}
    \text{EI}(A\rightleftharpoons B) = \text{EI}(A\rightarrow B) + \text{EI}(B\rightarrow A)
\end{align}
\end{definition}

As a consequence of the definition, if we are able to partition a system $S$ into $A$ and $B$ such that $\text{EI}(A\rightleftharpoons B)=0$, then $A$ and $B$ are independent parts, which limits the capacity of integrating information on the $S$. Therefore, it is necessary to locate the bottleneck in order to quantify the information integration capability for system $S$.

\begin{definition}[Minimum Information Bipartition, MIB] A bipartition on system $S$ as its ``weakest link'' can be achieved with partitions $A,B\subset S$, $B=\bar{A}$ as its complementary set, such that the normalized mutual effective information of $A,B$ is the minimum, as following:
\begin{align}
    \text{MIB}(A\rightleftharpoons B) &= \arg\min_{A,B\subset S} \frac{\text{EI}(A\rightleftharpoons B)}{H_{\max}(A\rightleftharpoons B)}\\
  \text{with }  H_{\max}(A\rightleftharpoons B) &= \min\{\max_AH(A), \max_BH(B)\}
\end{align}
$H_{\max}(A\rightleftharpoons B)$ is for normalization due to Remark \ref{rm:ei}.
\end{definition}
 Each partition in MIB is called a \textit{complex} in IIT.
 There is $\sum_{m=2}^{N}\binom{N}{m}$ subsets within a system of $N$ elements, each has a measure of $\Phi(S), S\in2^{[N]}$, but those $S$ in a larger subset with higher $\Phi$ are discarded, the rest are complexes in the system.
 
 Now consider a system $\mathcal{X}$ with $N$ neuronal elements, its information integration capacity is determined by the total minimum information for each of the complexes, which are found by enumerating over all the possible subsets $S\subseteq \mathcal{X}$.


To formalize this intuition, we define the information integration capacity of system $\mathcal{X}$, which measures the maximal irreducible cause-effect power, \emph{i.e.}, mutual effective information for minimum information bipartition.

\begin{definition}[Information Integration] The information integration for a subset $S$ is the mutual effective information of the minimum information bipartition:
\begin{align}
    \Phi(S) = 
    \min_{A\in 2^S, B=S/A} \text{EI}(\text{MIB}(A\rightleftharpoons B))
\end{align}
The integrated information for the entire system $\mathcal{X}$ is such that
$$
\Phi(\mathcal{X}) = \max_{S\in 2^{\mathcal{X}}} \Phi(S)
$$
\end{definition}

The intuitive explanation of the information integration $\Phi$ is that, if the system is not fully decomposable (into independent sub-systems), $\Phi$ is the effective information (a special type of mutual information) for cutting the systems on its ``weakest link'' by minimizing the effective information, or as a ``cruelest cut'' as in \cite{tegmark2015consciousness}. We will discuss about how to practically measure the integration information in the following section.




\subsection{Measurement of Information Integration}
In \cite{tononi2003measuring}, a computational model is proposed for measuring the information in a spatial network.

Assume that $\mathcal{X}$ is the entire system, and consider $X\in\mathbb{R}^{|\mathcal{X}|}$ to be the random vector over $\mathcal{X}$ that characterizes the signal emitted by each neuron in the system. To model the system $X$, the authors propose to use Gaussian Graphical Model, as described in the following.

The signal at the $i$-th node, denoted by $x_i$, is a linear combination of the signals of its neighboring nodes in the directed graph representing the node connection, plus the random measurement noise:
$$
x_i = \sum_{j\in\text{neighbor}(i)} w_{i,j} x_j + \sigma_i r_i
$$
where $r_i\overset{i.i.d.}{\sim} \mathcal{N}(0,1)$ are random noise in the measurement, and $w_{i,j}$ is the edge weight from $i$ to its neighbor $j$. In matrix form, we have,
$$
X = WX + CR, \quad C = \text{diag}((\sigma_i)_i)
$$
Then it is easy to see the covariance of $X$ satisfies
$$
\Sigma = C^2(I-W)^{-1}(I-W)^{-1\intercal}.
$$

With this model, one can compute the information integration $\Phi$ analytically, \begin{enumerate}
    \item[(1)] For any $S\in 2^{\mathcal{X}}$, consider one bipartition $S=A\cup B, S^c = \mathcal{X}/S$,
    $$
    X := \begin{pmatrix}
    X_A \\
    X_B\\
    X_{S^c}
    \end{pmatrix}
    \sim \mathcal{N}
    \left(\mathbf{0}, \begin{pmatrix}
    \Sigma_A &\Sigma_{AB} & \Sigma_{AS^c}\\
    \Sigma_{AB} & \Sigma_B & \Sigma_{BS^c}\\
    \Sigma_{S^c A} & \Sigma_{S^c B} &\Sigma_{S^c}\\
    \end{pmatrix}
    \right)
    $$
    then to calculate $\text{EI}(A^{\text{max}}\to B)$, one randomizes the signal from $A$ by setting $\Sigma_A=I_{|A|}$ in the covariance, and cuts off any incoming edges from $S^C, B$ to $A$. However, if $A$ is simulated to account for connection within $A$ and to $A$, the original graph is unchanged. Moreover, one sets $C_A = \text{diag}(\sigma_p)$ as the signal and $C_{B\cup S^c} = \text{diag}(\sigma_i)$ as the noise.\\
    Then one can calculate ($A$ is a set, $X_A$ is a multivariate representing the signal states of the items in the set $A$, we ignore the difference of $H(A)$ and $H(X_A)$ here)
    $$\begin{aligned}
    H(X_A) &= \mathbb{E}[-\log p_{X_A}] = |A|/2 + |A|/2\log 2\pi + \log|\Sigma_A|/2\\
    H(X_A X_B) &= |S|/2 + |S|/2\log 2\pi + \log|\Sigma_{S}|/2,\quad 
    \Sigma_S = \begin{pmatrix}
    \Sigma_A &\Sigma_{AB}\\
    \Sigma_{AB} & \Sigma_B
    \end{pmatrix}
    \end{aligned}
    $$
    similarly $H(B)$.
    \item[(2)]
    With the analytic results of $H(A), H(B)$ and $H(AB)$, one can also calculate $\text{EI}(B^{\text{max}}\to A)$. Then one finds the minimal information bipartition as before,
    $$
    \text{MIB}(S) = \min_{A\in 2^S, B=S/A} \text{EI}(A\rightleftharpoons B)/\min(H^{\max}(A), H^{\max}(B))
    $$
    and $\Phi(S) = \text{EI}(\text{MIB}(S))$
    \item[(3)] Finally, one can find integrated information for the entire system
    $$
    \Phi(X) = \max_{S\in 2^{\mathcal{X}}} \Phi(S)
    $$
\end{enumerate}
Apart from computing $\Phi$ for system with a given topological structure, the authors also experimented with finding the graph structure with the maximal $\Phi$, and analyzed the connectivity property of the generated `optimal' graph with maximal $\Phi$. 

In \cite{balduzzi2008integrated}, simulation studies on discrete spatial-temporal systems are carried out on small scale networks. 

\citep{hoel2017map} leverages the \textit{do}-operation in causal inference in constructing a Markov chain in time that identify particular coarsening of the (spatial) state-space that corresponds to an increase in information. 

Consider a finite state space $S$, we can specify transition probability matrix (TPM) between any two states
$$
P_{ij}= \mathbb{P}(X^{t+1}=j|X^t=i) = \mathbb{P}\left(X^{t+1}=j | d o\left(X^{t}=i\right)\right)
$$
which is similar to Markov chain, with the Markov property given by the $do$ operator, i.e. we have conditional independence once $S^t=i$ is a fixed quantity. In this sense, we treat the time as varying, and only one variable $X$ present with state space $S$; or that we have $|S|$ many binary variables and so $X = \sum_{k\in S} \delta_{k}$, which is well-defined due to finiteness.

Consider the problem setup: let $X \rightarrow Y$ be the causal system, where they share the same state space $S_X=S_Y$. The author is interested in when the macro system has a micro system with corresponding TPM, i.e. $U \rightarrow V$, such that $S_U=S_V \subset 2^{S_X}$. Moreover, Hoel's theory aims at identifying particular coarsening of the state space amounts to \textit{causal emergence}. Effective information (EI) plays an important role in achieving this. We give another definition of EI since the context differs to IIT, but firstly
we introduce the concept of intervention and effect distribution.
\begin{definition}[Intervention Distribution and Effect Distribution]
Given $|S_X|< \infty$, the maximum entropy amounts to the \emph{uniform intervention distribution}
\begin{equation}
    I_D = H^{\max }=\operatorname{Unif}(do(X)), \text { that is, } P(do(X=x))=\frac{1}{n} \quad \forall x\in S_X
\end{equation}
where $\operatorname{Unif}(\cdot)$ is the uniform distribution over a set.
Intervening with this distribution on $X$ results in the effect distribution $E_D(Y)$ over $Y$ :
\begin{equation}
\begin{aligned}
E_D(Y) &=\sum_{X} P(Y \mid do(X)) H^{\text {max }} \\
&=\frac{1}{n} \sum_{x} P(Y \mid do(X=x))
\end{aligned}
\end{equation}
The $E_D$ effectively computes the uniform averages over all rows in the TPM.
\end{definition}
We are now ready to introduce definition of effective information in \cite{Hoel2013Quntify}:
\begin{definition}[Effective Information]
\begin{equation}
\begin{aligned}
\text{EI}(X \rightarrow Y) &=\sum_{X} H^{\max } D_{K L}\left(P(Y \mid d o(X)) \| E_{D}(Y)\right) \\
&=\sum_{x} P(d o(X=x)) D_{K L}\left(P(Y \mid d o(X=x)) \| E_{D}(Y)\right) \\
&=\frac{1}{n} \sum_{x} D_{K L}\left(P(Y \mid d o(X=x)) \| E_{D}(Y)\right)
\end{aligned}
\end{equation}
\label{def:ei}
\end{definition}

The definition of EI presented in Def \ref{def:ei} is in fact equivalent to the definition in IIT.
Without loss of generality, denote $p(x)$ and $p(y)$ the mass function for $X\sim H^{\text {max }},\; Y\sim E_D$ respectively.
$$
\begin{aligned}
MI(X, Y) &=
\sum_x \sum_y p(x,y) \log_2\left(
\frac{p(x,y)}{p(x)p(y)}
\right)\\
\text{(Bayes Rule)}&= \sum_x \sum_y p(x)p(y|x)
\log_2\left(\frac{p(y|x)}{p(y)}
\right)\\
&= \frac{1}{n} \sum_{x} D_{K L}\left(P(Y \mid d o(X=x)) \| E_{D}(Y)\right)
\end{aligned}
$$

Given the notations above, a \textit{causal emergence} (CE) arises when the best coarsened system has higher effective information than the original one, i.e.
$$
CE = \text{EI}(U\to V) - \text{EI}(X\to Y) > 0
$$
since $U$ and $V$ are considered as the same random variables in one step, this amounts to finding the coarsened support set $S_U$. 


\subsection{Biological Evidence}
Despite the computational framework put forward by the IIT theory is able to analytically assess the information processing bottleneck in an arbitrary system, 
it remains an open problem to verify the claimed correspondence that subjective experience is equivalent to the capacity to integrate information.

IIT postulates the neural elements constituting PSC are those determined by maximizing the cause-effect power, which could be higher at a macro scale comparing to a micro scale owing to different connectivity patterns. 
Among the brain regions, cerebral cortex has functional specialization and integration altogether, which should yield high values of maximum information integration. Whereas cerebellum is not essential for consciousness, because of  
its lack of dependency among the neurons and inability to form a large complex with high maximum information integration.

IIT also explains why bistable firing of cortical neurons during slow wave sleep would cause fading of consciousness. This is owning to the loss of both selectivity and effectiveness results in the reduction of information integration \citep{tononi2016integrated}. \cite{alkire2008consciousness} argues that brain under anesthesia is similar to under slow wave sleep, where cortical connectivity breaks down and therefore information integration is reduced. Practically, PCI are proposed to estimate $\Phi^{\text{max}}$ evoked by TMS in practice, which is high only if brain responses are both integrated and differentiated, corresponding to a distributed spatio-temporal pattern of causal interactions that is complex and hence not very compressible.

\section{Consciousness as a State of Matter}
\label{sec:physics}
Apart from the information theoretical viewpoint, researchers are seeking for the properties of conscious process within physical systems. 
Consciousness can be thought of as an emergent phenomenon. It does not depend on detailed properties of atoms, but on the complex patterns into which the atoms are arranged. Emergent phenomena are common in physics, for example, waves can exist in many different kinds of matter. Researchers have investigated the phenomenon of consciousness from a principled way in physics~\citep{penrose1991emperor, stapp2000importance, tegmark2000importance, tegmark2015consciousness, carroll2021consciousness}. A core issue for this approach is to admit the true randomness in the conscious process and locate the corresponding physical processes within the human brain as evidence. A branch of research ultimately resorts to the quantum process, which is commonly believed to exhibit the true randomness in the measurement procedure of a quantum state. The conscious process can be interpreted as a quantum measurement by a conscious observer. However, several problems are raised in this interpretation: One corresponds to the quantum factorization problem, that the conscious observers has a certain Hilbert space factorization to leave the world around the observer as a strongly correlated but independent (from the observer) system~\citep{tegmark2015consciousness}. Another problem is the quantum decoherence in physical system like human brains. Most quantum phenomenons only appear in a very small space-time scale, the decoherence process prevents a system as large as the human brain to inherit the quantum property. For example, a typical timescale for quantum decoherence lasts for about $10^{-13}\sim 10^{-20}$ seconds, which is much shorter than the timescale of cognitive process as $10^{-3}\sim 10^{-1}$ seconds~\citep{tegmark2000importance}. Researchers also proposed the orchestrated objective reduction (Orch OR) of quantum states to interpret the brain cognitive process~\citep{hameroff2014consciousness}, which is discussed in Sec.~\ref{sec:orchor}.  

Max Tegmark~\citep{tegmark2015consciousness} considers consciousness as a kind of matter which he calls ``\textit{perceptronium}", as a phrase indicating conscious state. For a matter to become ``\textit{perceptronium}", it needs have the following four properties as necessary but not sufficient conditions:
\begin{itemize}
    \item The \textbf{information} principle: The system must have substantial information storage capacity;
    \item The \textbf{integration} principle: The system cannot consist of nearly independent parts, and it needs to have a certain level of integration within itself;
    \item The \textbf{independence} principle: The system must have substantial independence from the rest of the world;
    \item The \textbf{dynamics} principle: A conscious system must have substantial information-processing capacity, and it is this processing rather than the static information that must be integrated.
\end{itemize}
In Tegmark's work, he generalizes Tononi's IIT (Sec.~\ref{sec:iit}), compares this with other principles that conscious matter should have, and then generalizes the analysis to quantum mechanics. In his work, he shows that the information and integration principles can have conflicts with each other -- too much integration will result in very little information, which is referred to as the \textit{integration paradox} in Sec.~\ref{subsec:integration}. Also, the independence principle has conflicts with the dynamics principle -- too much independence will result in a trivial dynamics system, which is called the \textit{quantum Zeno paradox} as discussed later in Sec.~\ref{subsec:independence}. A conscious system needs to strike balances in these properties: information and integration, independence and dynamics. Therefore, this work introduces the \textit{autonomy} metric to measure the balance of independence and dynamics in the system.

Following that, it also proposes the following two principles: 
\begin{itemize}
    \item The \textbf{autonomy} principle: A conscious system has substantial dynamics and independence.
    \item The \textbf{utility} principle: An evolved conscious system records mainly information that is useful for it.
\end{itemize}
The autonomy principle describes the balance of dynamics and independence. The utility principle describes the amount of information within a conscious system.

These principles can be translated into more physical problems in a system:
\begin{itemize}
    \item The physics-from-scratch problem: If the total Hamiltonian $\mathbf{H}$ and the total density matrix $\rho$ fully specify our physical world, how do we extract 3D space and the rest of our semiclassical world from nothing more than two Hermitian matrices?
    \item The quantum factorization problem: Why do conscious observers like us perceive the particular Hilbert space factorization corresponding to classical space (rather than Fourier space, say), and more generally, why do we perceive the world around us as a dynamic hierarchy of objects that are strongly integrated and relatively independent?
\end{itemize}
These are the question to be answered in this theory. We will briefly introduce the basics of quantum mechanics and then dive into these principles. 

\subsection{Basics of Quantum Mechanics}
In quantum mechanics, the state of a system is described by a vector $|\psi \rangle$ in the Hilbert space. For example, we have an electron with a spin up state $|\psi \rangle=|\uparrow\rangle $ or spin down state $|\psi \rangle=|\downarrow\rangle $, or their superposition $|\psi \rangle=\frac{1}{\sqrt{2}}(|\uparrow\rangle+|\downarrow\rangle)$.

The probability of observing this system in state $|\chi \rangle$ is 
\begin{equation}
    P=|\langle \chi|\psi \rangle|^2.
\end{equation}

We can apply a unitary operator to a state to change the basis we are interested in:
\begin{equation}
    |\psi\rangle \to U |\psi \rangle
\end{equation}
For example, the spin up state in $x$ direction is a superposition state $|\psi \rangle=\frac{1}{\sqrt{2}}(|\uparrow\rangle+|\downarrow\rangle)$ in the $z$ direction. 

The time evolution of a state is controlled by the Hamiltonian operator $\mathbf{H}$ (can be thought of as a matrix in Hilbert space):
\begin{equation}
    |\psi(t)\rangle = e^{i\mathbf{H}t/\hbar} |\psi(0)\rangle.
\end{equation}
which is known as the Schr{\"o}dinger equation.

The Hamiltonian operator itself describes the energy spectrum of the system. We can find its eigenstates:
\begin{equation}
    \mathbf{H}|E_i\rangle = E_i |E_i\rangle.
\end{equation}
$E_i$ is the eigenvalue of the Hamiltonian operator $\mathbf{H}$, which represents the energy (as a scalar) of the system at state $|E_i\rangle$.
An operator takes diagnal form in its eigenbasis. When two operators commute, $[A,B]=AB-BA=0$, they can be simultaneously diagnolized.

A state can also be represented as a density matrix:
\begin{equation}
    \rho=|\psi\rangle \langle \psi |
\end{equation}
this represents a pure state.

Density matrices built from the pure state always have rank 1. A more general density matrix can also represent a classical mixture of states, which in general has a higher rank
\begin{equation}
    \rho=\sum_i |\psi_i\rangle \langle \psi_i |.
\end{equation}
this is a mixed state.

The probability of observing the system in a certain state is
\begin{equation}
    P=\langle \chi|\rho|\chi\rangle
\end{equation}

The time evolution of a density matrix is
\begin{equation}
    \rho(t)= e^{i\mathbf{H}t/\hbar} \rho(0) e^{-i\mathbf{H}t/\hbar}.
\end{equation}
Written in the energy eigenbasis, it is
\begin{equation}
\rho(t)_{m n}=\rho(0)_{m n} e^{i\left(E_{m}-E_{n}\right) t}
\end{equation}

\subsection{Integration Principle}
\label{subsec:integration}
For a bipartite system $\rho=\rho_1 \otimes \rho_2$. We define the integrated information $\Phi$ as the mutual information $I$ for the ``cruelest cut''. The mutual information is defined as
\begin{equation}
I \equiv S\left(\rho_{1}\right)+S\left(\rho_{2}\right)-S(\rho)
\end{equation}
where 
\begin{equation}
S(\rho) \equiv-\operatorname{tr} \rho \log _{2} \rho
\end{equation}
is the von Neumann entropy. Note that this is slightly different from Tononi's definition, but this is easier to calculate.

Recall in Sec.~\ref{sec:iit}, we introduce the definition of the integrated information in original IIT, as the effective information of the minimum information bipartition. The minimum information bipartition is a `cut' of the system according to the weakest link, \emph{i.e.}, minimal mutual effective information.
We can also consider cuts here in the quantum sense. In this case, more cuts are available, and we choose the cruelest cut as the integrated information. Therefore the \textit{integrated information} is defined as
\begin{equation}
\Phi=\min _{\mathbf{U}} I\left(\mathbf{U} \rho \mathbf{U}^{\dagger}\right).
\end{equation}
where $\mathbf{U}$ is the unitary evolution on density matrix $\rho$.

A general Hamiltonian can be written as 
\begin{equation}
\mathbf{H}=\mathbf{H}_{1} \otimes \mathbf{I}+\mathbf{I} \otimes \mathbf{H}_{2} +\mathbf{H}_3
\label{eq:quantum_ham}
\end{equation}
where $\mathbf{I}$ is identity matrix.

If a Hamiltonian can be decomposed without an interaction term (with $\mathbf{H}_3=0$), then it describes two perfectly independent systems,
\begin{equation}
 \rho\propto  e^{-\mathbf{H} / k T}=e^{-\mathbf{H}_{1} / k T} e^{-\mathbf{H}_{2} / k T}
\end{equation}
In this case, it can be derived that $\Phi=0$ with $\mathbf{U}=e^{i\mathbf{H}_1t/\hbar}e^{i\mathbf{H}_2t/\hbar}$. This is within expectation since the integrated information describes the level of information integration within a system, and a perfectly separated system should have no intergration. From IIT, we know that the conscious systems are likely to have the maximum of integrated information. Taking the 2D Ising model with $n$ dipoles as an example, the maximum integrated information according to above definition will only be $O(\sqrt{n})$. However, with the optimal error correcting codes, the system can achieve asymptotic $n/2$ bits of integrated information in the large-$n$ regime. 



The above analysis is based on the quantum system. For a classical system like a Hopfield network~\citep{hopfield1982neural} for describing the brain process, the integrated information can also be calculated, and it will lead to a so-called integration paradox.

\paragraph{Integration paradox.}
Suppose our brain is a Hopfield network with $n$ neurons using Hebbian learning rules, then the maximum capacity of integrated information is 37 bits for $n = 10^{11}$ neurons~\citep{tegmark2015consciousness}. However, the information for a conscious experience is much larger than this value, with an example of a human imagining a picture in his mind. This is known as the integration paradox.

Why does the information content of our conscious experience appear to be vastly larger than 37 bits? In the quantum case, it can never contain more than $1/4$ bit of information~\citep{mackay2003information}. This observation leads to some conjectures including that the human brains use a better coding method for conscious information rather than the Hopfield networks.

\subsection{Independence Principle}
\label{subsec:independence}
This principle follows the idea of IIT for cutting the system into independent parts from its ``weakest link'', as described in Sec.~\ref{subsec:concept_iit}. It first requires the $\rho$-diagonality theorem in a quantum case to find the minimum of mutual information.

\begin{theorem}[$\rho$-Diagonality Theorem, $\rho$DT~\citep{jevtic2012maximally}]
The mutual information always takes its minimum in a basis where $\rho$ is diagonal.
\end{theorem}

With $\rho$DT, the problem becomes how to find the basis with $\rho$ being diagonal in the quantum system with Hamiltonian as Eq.~\ref{eq:quantum_ham}. The answer to this question is presented by the following theorem.

\begin{theorem}[$H$-Diagonality Theorem, HDT~\citep{tegmark2015consciousness}]
The Hamiltonian is always maximally separable (minimizing $||H_3||$) in the energy eigenbasis where it is diagonal.
\end{theorem}

Furthermore, to minimize $||H_3||$, we must have $[H_1,H_3]=0$, which indicates that all subsystems (\emph{e.g.}, subsystem with Hamiltonian $H_1$) need to commute with all interaction Hamiltonians (\emph{e.g.}, $H_3$). Following this principle, it will finally lead to a heat death, where all subsystems cease to evolve, known as the \textit{quantum Zeno paradox}. 

\begin{definition}[Quantum Zeno Paradox~\citep{tegmark2015consciousness}]
If we decompose our universe into maximally independent objects, then all change grinds to a halt.
\end{definition}

\subsection{Dynamics principle}
If only following the independence principle, we will face the \textit{quantum Zeno paradox} where the system cease to evolve and has no information processing capability. However, we also require the conscious system to have certain information processing capability by the dynamics principle. The autonomy principle says the system should be able to strike a balance between the independence and dynamics principles. 
There are interesting classes of states $\rho$ that provide substantial dynamics and near-perfect independence even when the interaction Hamiltonian $H_3$ is not small.

As a measure of the dynamics, the energy coherence is defined as:
\begin{equation}
\begin{aligned}
\delta H & \equiv \frac{1}{\sqrt{2}}\|\dot{\rho}\|=\frac{1}{\sqrt{2}}\|i[\mathbf{H}, \rho]\|=\sqrt{\frac{-\operatorname{tr}\left\{[\mathbf{H}, \rho]^{2}\right\}}{2}} \\
&=\sqrt{\operatorname{tr}\left[\mathbf{H}^{2} \rho^{2}-\mathbf{H} \rho \mathbf{H} \rho\right]}
\end{aligned}
\end{equation}

Then, the maximum probability velocity can be calculated as:
\begin{equation}
v_{\max}=\sqrt{2} \delta H
\end{equation}
with probability velocity defined as $v=\dot{p}, p_i=\rho_{ii}$. If we only maximize $v_{\max}$ following the dynamics principle, some calculations indicate that it will lead to a very simple dynamics solution, which does not carry the capability of sufficient information processing. 

The states that are most robust toward environment-induced decoherence are those that approximately commute with the interaction Hamiltonian. It means that $[\rho,H_3] \approx 0$, but $[\rho,H_1] \neq 0$. The $H_3$ interaction term will decohere the system thus it is required to be sufficiently small for system $\rho$ to evolve over time. 

To correctly describe the system with a balance of independence and dynamics, it needs a new metric called the \textit{autonomy}. It first requires to use the linear entropy for quantifying the non-unitary property of an evolution.

The linear entropy is defined as: 
\begin{equation}
S^{\operatorname{lin}} \equiv 1-\operatorname{tr} \rho^{2}=1-\|\rho\|^{2}
\end{equation}

Let us define the dynamical timescale $\tau_{\text {dyn }}$ and the independence timescale $\tau_{\text {ind }}$ as 
$$
\begin{aligned}
\tau_{\text {dyn }} &=\frac{\hbar}{\delta H}, \\
\tau_{\text {ind }} &=\left[\ddot{S_{1}^{l i n}}(0)\right]^{-1 / 2} .
\end{aligned}
$$
The autonomy can be defined as the ratio:
\begin{equation}
A \equiv \frac{\tau_{\text {ind }}}{\tau_{\mathrm{dyn}}}
\end{equation}
This ratio will exponentially increase  with the system size, such that it leads to a highly autonomous system with sufficient information processing even with a large $H_3$.

\paragraph{Summary.}
The theory proposed by Tegmark generalizes the IIT to the quantum domain, and analyzes the deficiency of information integration principle from the physics-principled calculation, which raises the \textit{integration paradox} that the Hopfield neural network cannot integrate a sufficient amount of information for conscious experience. It also analyzes the independence principle and leads to the \textit{quantum Zeno paradox} that a system decomposed into maximally independent sub-systems will cease to evolve in the end, which is in conflict with the dynamics principle. The theory finally propose the metric named \textit{autonomy} that is found to have a high value as a balance of the independence and dynamics principles for a conscious system.
\section{Orchestrated Objective Reduction Theory}
\label{sec:orchor}

\subsection{Consciousness as Orchestrated Objective Reduction}
Recalling the discussion of free will problem in arising consciousness in Sec.\ref{sec:cons_will}, we may conjecture that the true randomness may be required for the free will to happen. How does this truly random process happen in human brains?
Orchestrated objective reduction (Orch OR) theory~\citep{penrose1991emperor, penrose1994shadows, hameroff1996orchestrated, hameroff2007brain, hameroff2010conscious,hameroff2012quantum, hameroff2014consciousness}, builds on the hypothesis that the emergence of consciousness is due to a biological mechanism that is able to orchestrate moments of quantum state reduction. The theory posed that conscious events arises from the termination of quantum computation in the brain microtubules, framed as \textit{objective reduction}. Objective reduction refers to the idea that a quantum system can spontaneously collapse from a superposition of multiple possibilities into a single state. In the Orch OR theory, Hameroff and
Penrose propose that these objective reductions are non-deterministic but are orchestrated by certain processes in the brain. These affected objective reduction processes are called the free will. 
In a nutshell, the Orch OR framework based on quantum theory appears to introduce stochasticity aspects into the reductionist view of consciousness as a pure physical process. Through this, independent causal agency and free will can be explained \citep{hameroff2012quantum}. Moreover, it will also imply the existence of consciousness in a single cell.


\subsection{Free Will in Neurons}
The integration and firing sequences, which gives rise to EEG and NCC, are primarily generated by dendritic-somatic membranes. Then axonal firings outputs conscious (or non-conscious) processes to control behavior. Microtubules (MTs), as part of the cytoskeleton, a protein scaffolding network inside of the cell, is hypothesized to influence the threshold of firing. Specifically, Dendritic–somatic MTs of neurons are arranged in local recursive networks and are more stable comparing to MTs in other cells, therefore render itself as a suitable information processing and storage unit, moreover, suitable to mediate consciousness and regulate firing.

\begin{figure}[h]
    \centering
    \includegraphics[scale=0.5]{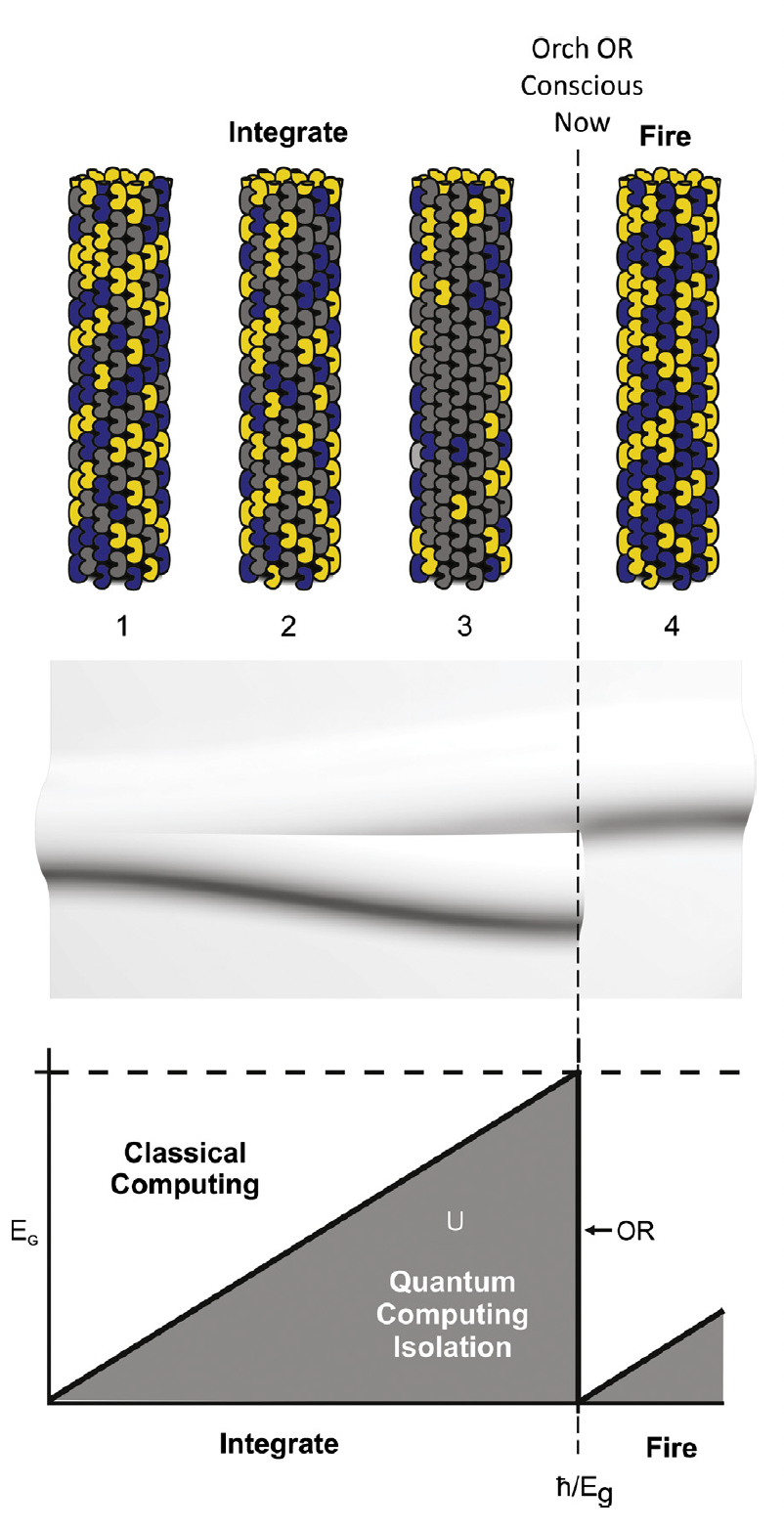}
    \caption{\footnotesize The figure published in \cite{hameroff2014consciousness} was used to illustrate the process that Orch OR occurs. Top: Tubulins are in classical dipole states (yellow or blue), or quantum superposition of both dipole states (gray). Quantum superposition/computation increases during (1-3). The conscious moment occurs when threshold is met at time $\tau \approx \hbar / E_G$. Middle: Corresponding alternative superposed space-time curvatures reaching threshold at the moment of OR and selecting one space-time curvature. Bottom: Schematic of a conscious Orch OR event showing U-like evolution of quantum superposition and increasing $E_G$ until OR threshold is met, and a conscious moment occurs by $\tau \approx \hbar / E_G$.}
    \label{fig:OrchOR_fire}
\end{figure}

As shown in Fig~\ref{fig:OrchOR_fire}, MT constitutes of peanut-shaped tubulin protein, each with a dipole and can be arranged in 13 protofilaments each with two types of hexagonal lattices. In \citep{hameroff2014consciousness}, the MT dipoles are described as electron spin (magnetic), which is inherently a quantum-mechanical quantity. Therefore all possible directions for the spin rotation axis arise as quantum superpositions of some random pair of directions.
The authors then speculate that there may exists chains of spin along the pathway in MT that propagate quantum bit pairs, in addition, there may exists alternative currents at certain frequency caused by periodic spin flips. 
The fact that tubulins in MTs can each exists in different states (and give rise to quantum superposition) based on the dipoles position (direction), could indicate MT processes may directly result in consciousness.



\subsection{Di\'osi-Penrose Objective Reduction}
Having notated the physiological unit where quantum superposition may took place in our brain, one might ask how the orchestrated reduction that gives rise to consciousness happen? The argument starts with linking Orch OR to theoretical physics.

The Di\'osi-Penrose (DP) objective reduction proposal bridges quantum and classical physics as \textit{quantum-gravitational} phenomenon, whereby the quantum superposition reduces to an average time measurement $\tau$ for the state reduction to take place according to $\tau \approx \frac{\hbar}{E_G}$, $E_G = \frac{Gm^2}{a}$, where $E$ is space-time superposition curvature, $G$ is gravitational constant, $m$ is mass, $a$ is spatial size. The actual time of decay in each event of state-reduction is taken as a random process in DP. From this, the reduction of quantum superposition of space-time objects takes place when the superposition curvature $E_G$ reaches the threshold $\frac{\hbar}{\tau}$. 

The Orch-OR schemes goes further to relate the DP physical proposal to consciousness. \cite{hameroff2014consciousness} proposed that if a quantum superposition is firstly well-orchestrated: ``adequately organized, imbued with cognitive information, and capable of integration and computation"; and secondly isolated from non-orchestrated, random environment for the superposition $E_G$ to reach the threshold $\tau$, then the Orch OR will occur along with the emergence of consciousness. An illustration of this process is shown in Fig~\ref{fig:OrchOR_fire}.

\subsection{Evidence for Objective Reduction of Quantum State}
In \cite{hameroff2010conscious}, the authors claimed that the best measure of neural correlate of consciousness is 30- to 90-Hz gamma synchrony electroencephalography (EEG), which is largely derived from dendritic and somatic integration potentials.
In addition, the theory claims that the state of anaesthesia is owing to dispersed dipoles in the MTs, responsible for quantum computing.

There are yet experiments in confirming the theory, however, biological evidence has been observed in warm conditions, where the theory has yet to extend to. Nonetheless, the Orch OR theory has provided a computational framework allowing falsification of the biological quantum theory that takes place in the MT.

\section{Global Workspace Theory}
\label{sec:gwt}
\subsection{The Theatre of Consciousness}
Global workspace theory (GWT) is an architecture proposed by Bernard Baars ~\citep{baars1997theater} to explain the inner procedure of how the human brain selects and deals with consciousness attention. There are some limits to conscious capacity. For example, working memory, which temporally store information to be dealt with, holds only several things at a time. Moreover, the human brain is only able to receive information from a single stream. `The theatre of consciousness' was proposed in a metaphor term to answer how the human brain handles different inputs, and then outputs a single stream of information that draws the final attention. 


There are several components of a theatre of consciousness. The `stage of working memory' is the platform to receive all potential information from sensors or abstract information from cortices. The `spotlight' mechanism in working memory highlights the conscious steam of information, other information on the stage is not aware by attention. Information resources,e.g. the potential thoughts, images or sensations, are regarded as `actors'. The information resources compete with each other to get the spotlight. The more conscious procedure is required to handle the information, the more likely the information resource will be put under the spotlight. Perceptual, intention, expectations etc. influence the result of this competition. `Context' refers to unconscious networks that potentially shape conscious contents in the brain. `Directors', the executive functions of human brain, guide the selection procedure with intentions and goals. The frontal cortex is believed to act as an important role in this procedure with the fact that damage to the front lobe leads to loss of actions by long-term goals. Then the information under the spotlight is broadcasted to the `audience', which represents the brain region which requires the information.

\begin{figure}[htbp]
    \centering
    \includegraphics[scale=0.45]{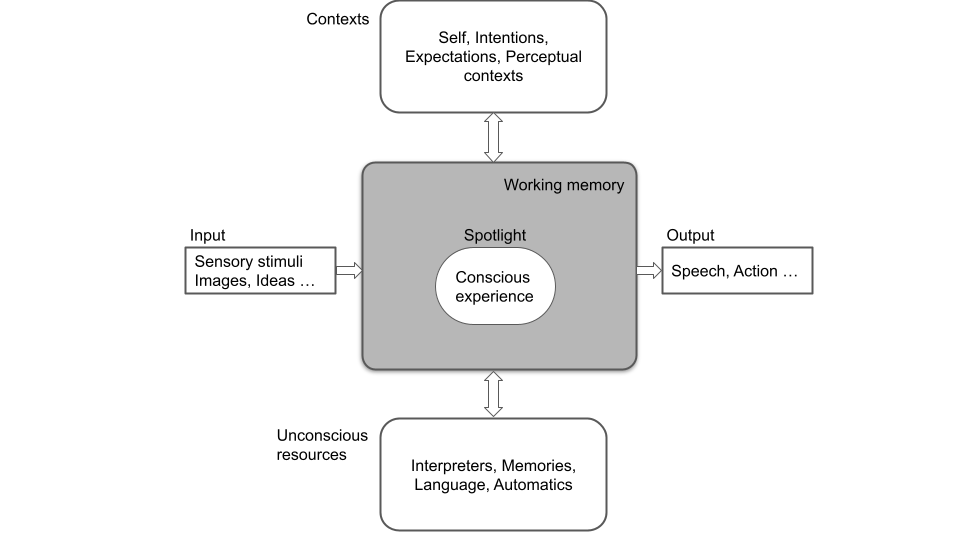}
    \caption{Scheme diagram of GWT derived from~\cite{baars1997theater}.}
    \label{fig:GWTScheme}
\end{figure}

An update of GWT in 2003~\citep{baars2003global} gives a more detailed introduction to the relationships between GWT and Brain functions, which provides some evidence of how the competition is performed in the human brain. Both the frontal cortex and other brain regions, which can interrupt the spotlight control, are involved in the conscious event selection procedure. The latter interrupt control consists of, for example, the brain stem, pain system, and emotional centers, which allow interrupting the selection procedure and give weight to more significant and urgent activities. The `Context' function of the brain is believed to be involved in the conscious decision process. The parietal cortex which is related to self-awareness of parts of the body is not directly objectively linked to consciousness but is believed to shape conscious visual events. The `Self' system may be involved in the generation of consciousness. It was found that split-brain patients have different executive and perceptual functions from the left and right hemisphere~\citep{gazzaniga1996collaboration} and the left prefrontal cortex processes the sensory information with a `narrative self' that can draw different awareness which causes conflicts between both hemispheres. Then the left hemisphere tries to rationalize and repair such conflicts. Some evidence was also provided in the paper to support the assumption that consciousness contents are broadcasted and distributed to brain regions. In a visual word task, the word task not only triggers visual word recognition areas of the cortex but also was found to evoke activities in the parietal and prefrontal cortex~\citep{baars2002conscious}.

\subsection{Computational Models of GWT}


The Intelligent Distribution Agent (IDA) ~\citep{baars2003conscious,baars2007architectural} and LIDA (Learning IDA) \citep{baars2009consciousness} computational models were proposed based on GWT to perform human-like tasks. In the study, naval jobs of sailors are used as an example task to test the model. IDA and LIDA contain several blocks reflecting the GWT, including \textbf{sensory modules} to deal with stimulus, \textbf{memory modules} as storage, \textbf{attention modules} referring to the concept of attention, the \textbf{action module} for action selection. Particularly, as in the GWT, \textbf{global workspace modules} integrates and broadcasts information, as well as selects the most relevant and important information to be on the stage. This model shows an empirical computational implementation of the GMT model as conceptual evidence that GMT could work out human-like functions. IDA or LIDA turns incoming sensory data into actions to the environment. The concepts of memory, competition, and broadcasting are involved in the conversion process. Then the resulting action to the environment changes the inputs of the system which forms an iterating cognitive cycle. 






GWT has inspired some studies in related fields, here we describe several examples in brain signal analysis and deep learning. Inspired by GWT, ~\citep{schutter2004extending} used EEG coherence, the level of connectivity of different brain region, to measure if emotions play a role in consciousness. Another study ~\citep{bartolomei2011global}, in light of the broadcasting and distributing process in the GWT, compared the synchrony within distant cortico-cortical and cortico-thalamic networks of epileptic seizures with the distant relationships across different brain regions in GWT. More recently, a study discussed the possibility of implementing GWT with deep learning. The idea of Global Latent Workspace (GLW) was proposed to reflect deep learning design principles of brain-like mechanisms~\citep{vanrullen2021deep}.



\section{Higher-Order Theories}
\label{sec:hot}
A general definition of HOT is given by \cite{sep-consciousness-higher}: 
A phenomenally conscious mental state is a mental state (of a certain sort) that either is, or is disposed to be, the object of a higher-order representation of a certain sort. 

Depending on whether the higher-order states in question are perception-like or thought-like, the high-order theories are categorized as high-order perception theory (HOPT), high-order thought theory (HOTT), self-representational theory (SRT), etc.
Definitions of HOPT (Sec.~\ref{sec:hopt} Def.~\ref{def:hop}), HOTT (Sec.~\ref{sec:hott} Def.~\ref{def:hot}) and SRT (Sec.~\ref{sec:srt} Def.~\ref{def:self}) are described in details in the following sections. Additionally, we will introduce and discuss other relevant theories in the final section. To provide a concise overview, we consolidate these theories into a summarizing Table \ref{tabel:summary_hot}.

Higher-order theories (HOTs) try to answer the question that if a mental state is conscious or unconscious. The higher-order theory believes that there is a certain brain mechanism that is more advanced than the first-order information (e.g. senses from organs like visual or auditory nerves). Three sub-theories claimed different explanations for the higher-order mechanism. HOPT believes that there are inner senses that scan or refine but are independent of the first-order information. This explains why people are able to imagine feelings like pain. In HOTT, it is believed that a mental state is conscious when it is the subject of higher-order thought. They propose that a conscious mental state or event is either actually causing or is disposed to cause an activated thought that a person has the state or event. Self-representational theory proposed another explanation to the higher-order theory. The self-representational theory believes that the higher-order state is constitutive or internal to its first-order state, i.e. that the higher-order state is formed from the first-order states and as a more complex system than the first-order states that generate awareness.


\begin{table*}[htbp]
\footnotesize
\centering
\caption{A summary of HOTs}
\scalebox{0.95}{
\begin{tabularx}{\textwidth}{c|c|X|m{4cm}|X}
\toprule
\multicolumn{2}{c|}{Theory} & First-order \& Higher-order Relationship & Key Mechanism & References \\
\hline
\multicolumn{2}{c|}{\multirow{4}{*}{HOPT}}
& higher-order senses exist and are independent of first-order information & perception-like; the human brain has inner senses of the first-
order senses to generate higher-order representations & \citep{locke1948essay, armstrong1981consciousness,armstrong1985consciousness,lycan1996consciousness,armstrong2002materialist, lycan2004superiority} \\  \hline
\multirow{9}{*}{HOTT} & Actualist & a mental state is conscious when it is the target of a higher-order thought & thought-like; A conscious mental event is actually causing an activated thought that a person has the event & \citep{rosenthal1986two,rosenthal1993thinking,rosenthal2005consciousness}  \\ \cline{2-5}
& Dispositional & a mental state is conscious when it is the target of a higher-order thought & thought-like; a mental state is conscious when it is the subject of higher-order thought; a conscious mental event is disposed to cause an activated thought that a person has the event & \citep{dennett1978toward,carruthers1998language}  \\ \hline
\multirow{6}{*}{SRT} & Part-whole & first-order and higher-order are parts of a whole complex & the mental state is to representing
itself. In the definition of part-whole SRT, the higher-order information exists but is not strictly 'higher' than the first-order information. First-order and higher-order are bound together & \citep{kriegel2009subjective, picciuto2011addressing}   \\ \cline{2-5}
& Identity & higher-order and first-order are identical & higher-order and first-order are the same components as two roles or functions & \citep{caston2002aristotle, carruthers2005consciousness,van2004higher} \\
\bottomrule
\end{tabularx}
}
\label{tabel:summary_hot}
\end{table*}

\subsection{Higher-Order Perception Theory}
\label{sec:hopt}
The high-order perception theory (HOPT) theory~\citep{locke1948essay, armstrong1981consciousness,armstrong1985consciousness,lycan1996consciousness,armstrong2002materialist, lycan2004superiority}, which is also called the Inner-Sense Theory or High-Order-Sense Theory, is referring to the followings: Humans not only have the sense-organs to scan the environment and their own bodies to produce the representations, which are called the \textit{first-order} non-conceptual and/or analog perceptions of environment/body states, but also have inner senses of those first-order senses to generate equally fine-grained but higher-order representations, which are called the \textit{second-order} non-conceptual and/or analog perceptions of the \textit{first-order} perception states.

The definitions of the \textit{first-order} perception and the \textit{second-order} perception are actually close to the \textit{M}-consciousness and \textit{I}-consciousness in the attention schema theory, which will be introduced in later Sec.~\ref{sec:ast}. We will discuss the connections of the two theories later in Sec.~\ref{sec:ast_uni_gwt_hot}.

\begin{definition}[Higher-Order Perception Theory/Inner-Sense Theory~\citep{sep-consciousness-higher}]
\label{def:hop}
    A phenomenally conscious mental state is a state with analog/non-conceptual intentional content, which is in turn the target of a higher-order analog/non-conceptual intentional state, via the operations of a faculty of ‘inner sense’. 
\end{definition}
A formal proposition of HOPT is provided as Definition \ref{def:hop}. It explains consciousness as the higher-order states generated from inner sensing of the first-order states. Referring to Fig.~\ref{fig:processor} in Sec.~\ref{sec:cons_will}, the conscious module is a higher-level component perceptive of the information flow of the lower-level sensing modules.

The antagonistic viewpoint of HOPT is held by some theorists that the attention mechanism on the first-order states may serve as a substitute of the higher-order states~\citep{sauret2014attention}.

\subsection{Higher-Order Thought Theory}
\label{sec:hott}
The higher-order thought theory (HOTT)~\citep{rosenthal2009higher, rosenthal2012higher,rosenthal2004varieties,byrne1997some,brown2019understanding} propose that a conscious mental state or event is either actually causing or is disposed to cause an activated thought that a person has the mental state or event. There are two embranchments of the theory: the \textbf{actualist} and the \textbf{dispositionalist}. In the above statement, the actualists believe that mental state directly caused the activated thought while the dispositionalist believes that the mental state is disposed to the thought. Another difference between them is that the actualist HOTT requires actual involvement of the first-order information in order to compute the higher-order information. On the contrary, the dispositionalist HOTT states that the higher-order computation only requires the availability of the first-order information, for example utilising board casting in the global workspace theory, instead of directly accessing all first-order information. In \citep{lau2011empirical}, some empirical support for the higher-order theories was discussed, e.g. the association of conscious awareness with prefrontal mechanisms and evidence based on clinical disorders.  
\begin{definition}[Higher-Order Thought Theory~\citep{sep-consciousness-higher}]
\label{def:hot}
    A phenomenally conscious mental state is a state of a certain sort (e.g. with analog/non-conceptual intentional content, perhaps) which is the object of a higher-order thought, and which causes that thought non-inferentially. 
\end{definition}

Some computational models were proposed based on higher-order thought theories. Metacognition neural network models \citep{pasquali2010know,cleeremans2007consciousness,timmermans2012higher} consist of two networks: a first-order network and a second-order network. The first-order network directly gets inputs and processes them with hidden units. After that, the hidden units of the first-order network are linked to the second-order network for processing. The first-order network learns to perform the classification of tasks, and the second-order network predicts the confidence of the first-order network by accessing the representational information of the first-order network. The models were tested and reported on several tasks, e.g. the Iowa Gambling Task \citep{pasquali2010know}, artificial grammar learning (AGL) tasks which distinguish grammar and non-grammar sentences, and blindsight tasks \citep{persaud2007post} in which blindsight patients make visual discriminations in the absence of visual awareness. 

\subsection{Self-Representational Theory}
\label{sec:srt}
The Self-Representational Theory (SRT) \citep{kriegel2009subjective, van2004higher, picciuto2011addressing} presents the idea of phenomenally conscious mental state, which is a state with non-conceptual intentional content, and conceptual intentional content at the same time. Such a mental state is said to representing itself to the person who is the subject of that state. 

\begin{definition}[Self-Representation Theory~\citep{sep-consciousness-higher}]
\label{def:self}
    A phenomenally conscious mental state is a state of a certain sort (perhaps with analog/non-conceptual intentional content) which also, at the same time, possesses an intentional content, thereby in some sense representing itself to the person who is the subject of that state.
\end{definition}

Two branches of the theory have argued for the constitutive relation between the conscious state and higher-order state is one of \textbf{identity} \citep{caston2002aristotle, carruthers2005consciousness, van2004higher}, or \textbf{part-whole} \citep{kriegel2009subjective, picciuto2011addressing}. The former argues that the conscious state is both first-order and higher-order, More precisely, a first-order perceptual state with analog content acquires, at the same time, a higher-order analog content. The part-whole SRT take stance similar to actualist HOT thoery arguments, where the first-order perceptual state gives rise to higher-order thought that represents experience.

\subsection{Other theories and perspectives in HOT}
The same-order theory \citep{kriegel2009subjective,brentano1973psychologyfrom,lau2011empirical} proposes that conscious mental states are not represented by any other mental states, but are instead directly present to the subject's awareness. The higher-order statistical inference view \citep{lau2011theoretical,lau2007higher} believes conscious mental states involve higher-order statistical inferences about one's own mental states. According to this view, first-order representation is reviewed by a higher-order inference procedure to form a statistically reliable perceptual signal, similar to perceptual decision-making process. From the perspective of the radical plasticity thesis \citep{cleeremans2011radical,pasquali2010know}, the brain is capable of remarkable adaptability and flexibility, and this plasticity plays a critical role in the development of conscious awareness. The radical plasticity thesis proposes that consciousness is not an intrinsic procedure but a learning process in the brain. The brain engages in continuous and unconscious learning to re-describe its activity, thereby developing systems of meta-representations that describe and refine the initial, first-order representations.



\section{Attention Schema Theory}
\label{sec:ast}





\subsection{Formulation}
It is important to understand the difference of attention and awareness and their relationship in Attention Schema Theory (AST)~\citep{graziano2015attention, graziano2020toward}.

\begin{definition}[Attention]
Attention is a process where the brain selectively process certain pieces of information more than others.
\end{definition}
As one of the most influential explanation of the attention process, people~\citep{desimone1995neural} propose that there is a signal competition process emerging at the earliest stages of signal processing and existing in every later stages.

Awareness is a different concept from the attention. Although awareness and attentions are typically highly correlated, they can be dissociated. For the concept of awareness, we need to first distinguish objective awareness and subjective awareness. Both of the awareness involves a participant.
\begin{definition}[Objective Awareness]
For objective awareness, the participant is required to report that he is objectively aware of the stimulus. 
\end{definition}

\begin{definition}[Subjective Awareness]
For subjective awareness, the participant reports whether he has perceived the stimulus in his own opinion.
\end{definition}

The difference of objective awareness and subjective awareness is just to distinguish whether the participant `sees' or `guesses' the perceived stimulus.

In AST, the authors refer the \textit{awareness}, \textit{consciousness} and \textit{subjective experience} as the same concept as the \textit{subjective awareness}. Therefore, AST is a theory about subjective awareness, not \textit{objective awareness}.

AST proposes that awareness is a model of attention. Thinking of a person seeing an apple, the visual representation of the apple (V) appears in the person's mind through the attention process. However, this is not enough for making the person aware of the `apple'. To generate awareness, the mind also has a model of self (S), and the attention (A) of S to V is also part of the awareness. By AST, a subjective awareness is [S+A+V], which is a model of the attention process. 


\begin{figure}[H]
    \centering
    \includegraphics[scale=0.3]{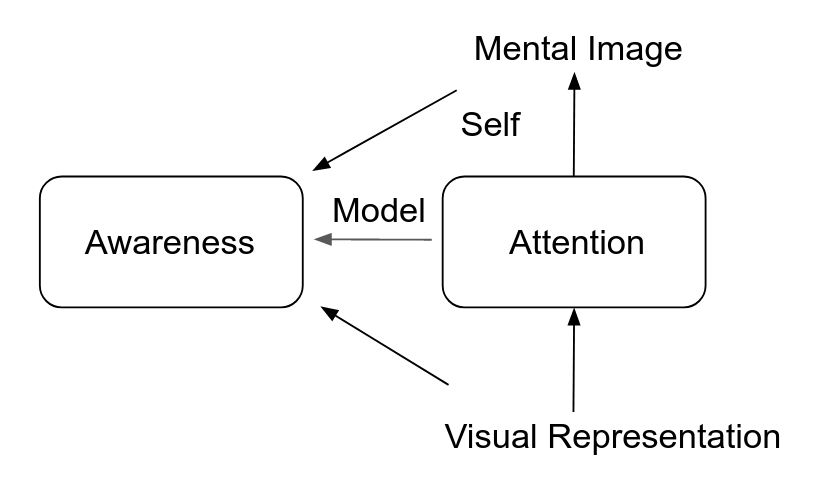}
    \caption{Awareness as a model of attention in AST.}
    \label{fig:ast}
\end{figure}

People may ask why subjective awareness is required beyond the attention? According to AST, subjective awareness allows for self-modelling which is essential for model-based control. Without awareness is like without modelling the arm when a person tries to reach some objects, and it will lead to inaccurate prediction about the arm's position therefore bad reaching result. The awareness serves as an internal modelling of the mind itself and the attention, leading to more accurate model-based control for human.

\subsection{\textit{I}-consciousness and \textit{M}-consciousness}
As the primary researchers of AST, \cite{graziano2020toward} proposes to interpret the consciousness of human mind via the \textit{I}-consciousness (\textit{I} for information) and \textit{M}-consciousness (\textit{M} for mysterious) in their later study.

\textit{I}-consciousness indicates the process of signals winning attentional competition, just as in GWT, which is generally assumed to computationally feasible~\citep{baars2007architectural}. However, the mysterious part is the \textit{M}-consciousness, which is used to explain the subjective experience of perceiving the winning piece of information.

Similar as the previous work of AST~\citep{graziano2015attention}, that awareness is a model of attention, \cite{graziano2020toward} further proposes that \textit{M}-consciousness is a natural, built-in, imperfect model of \textit{I}-consciousness. Moreover, the \textit{I}-consciousness and \textit{M}-consciousness can be mutually involved, like a mirror of another mirror. A person is I-conscious of having his \textit{M}-consciousness, and \textit{M}-consciousness is a model of \textit{I}-consciousness. For the self-modeling, the modeled parts are the physical components of the self most closely correlates with the winning piece of information from the attention competition.

People may further ask why we think the subjective experience is realistic. According to AST, the two properties ensure the realistic subjective experience: The first one is that the subjective experience cannot be turned off. The second one is that the mind enables source monitoring for the perceived information, which allows people to distinguish the real and the hypothetical. Since \textit{M}-consciousness is a model of \textit{I}-consciousness, the feeling is realistic for the exactly same reason that the people would believe every physical objects are real. 

A practical implementation of AST would involve three components. The network $A$ represents the selective information process by the attention competition. The network $B$ is to model the function of network $A$ by making predictions on $A$'s output. Network $C$ receives the output from $A$ and $B$ to generate the report (\emph{e.g.}, speech) to other components within the brain and outside world. The network $B$ is the important attention schema, for which the physical counterpart in the brain has been suggested to serve, as a cortical network overlapping part of the temporoparietal junction (TPJ)~\citep{graziano2011human, graziano2016consciousness}.

\subsection{AST as a Unification of GWT and HOT}
\label{sec:ast_uni_gwt_hot}
AST can be viewed as a unification of global workspace theory (GWT, Sec.~\ref{sec:gwt})~\citep{baars1993cognitive, dehaene2014consciousness, dehaene2011experimental} and higher-order theory (HOT, Sec.~\ref{sec:hot})~\citep{gennaro2011consciousness, lau2011empirical, rosenthal1991nature, rosenthal2005consciousness}. 

Specifically, the GWT explains the attention schema for a piece of information to appear on the stage of the mind, which corresponds to the \textit{I}-consciousness of AST. However, GWT does not explain the existence of consciousness experience, as the M-consciousness. AST explains this mystery by constructing the attention schema using network B. 

The HOT says consciousness arises from the higher-order representation. Recall the introduction of HOT in Sec.~\ref{sec:hot}, a conscious system has inner senses of those first-order senses to generate equally fine-grained but higher-order representations, which are called the \textit{second-order} perceptions of the \textit{first-order} perception states. The \textit{I}-consciousness in AST represents the external first-order senses, while the \textit{M}-consciousness corresponds to the \textit{second-order} perceptions over the first-order senses. These \textit{second-order} perceptions can be thought of as a modelling process of the first-order senses.
AST assumes the brain to construct higher-order representation of the global workspace, as imperfect modelling of the \textit{I}-consciousness, which unifies the HOT and GWT to give explanations of subjective awareness.

\section{Conscious Turing Machine}
\label{sec:ctm}
\subsection{Formulation}
In traditional Turing machine (TM)~\citep{turing2009computing}, Turing does not involve the subjective experience into the concept of TM. The TM is only about the computational intelligence but not consciousness of the machine, whereas the latter one is usually considered a hard problem~\citep{chalmers1995facing}.

Conscious Turing machine (CTM)~\citep{blum2022theory} is a theory as an extended concept of TM. Compared with TM's model of computation, CTM empowers the system a distinguishable feature , i.e., the ``feeling of consciousness''. Specifically, CTM is defined as following.
\begin{definition}[Conscious Turing Machine]
CTM is defined as a seven-element tuple: <STM, LTM, Up Tree, Down Tree, Links, Input, Output>, where STM and LTM are shorten for short-term memory and long-term memory.
\end{definition}
The CTM can be viewed as GWT with a more sophisticated structure. STM is an analogue of the ``stage'' in GWT as a necessary component for the consciousness to happen. As an analogue of the ``audience'' in GWT, LTM is a large collection of general processors, including the \textit{Model of the World} processor for modeling the world and the agent itself, \textit{Inner Speech} processor for processing linguistic information, and other \textit{Inner Generalized Speech} processors for handling information inputs like five senses. These processors are called LTM since they have a relatively stable status and expertise for processing a specific type of information, while LTM corresponds to a shorter period of status maintenance for more general functionalities.

\begin{figure}[H]
    \centering
    \includegraphics[scale=0.3]{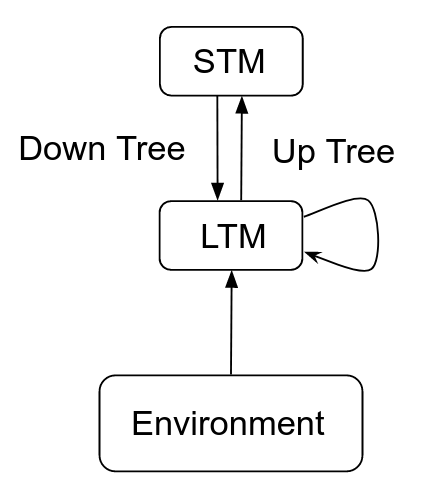}
    \caption{The information flows in CTM.}
    \label{fig:ctm}
\end{figure}

In CTM, the information flows only appear in five ways, as also depicted in Fig.~\ref{fig:ctm}: 
\begin{itemize}
    \item (1) environment $\rightarrow$ LTM; 
    \item (2) LTM$\rightarrow$STM (via Up Tree);
    \item (3) STM$\rightarrow$LTM (via Down Tree);
    \item (4) LTM$\rightarrow$LTM;
    \item (5) LTM $\rightarrow$ environment.
\end{itemize}
Process (1) in the information perception. Process (2) is achieved with Up Tree  competition. In the Up Tree competition process, there is a winning information chunk finally reaching the STM. The competition process is determined by an internal mechanism, which is probabilistic and achieved with some coin-flip neurons with inherent randomness. However, the authors argue that the free will can still be felt even with a completely deterministic setting. Process (3) is through the Down Tree broadcast, and it broadcasted the information in STM to all LTM. (2) and (3) achieve the consciousness awareness. Conscious awareness (attention) is the reception by all LTM processors of the broadcasted winning chunk in the Up Tree competition.
 Process (4) is a bidirectional link between processors to collaborate on the information processing. Process (5) is the output of the system to the environment through processors like \textit{Motion Controller}.
 
\subsection{CTM for Consciousness}
CTM adopts the concept of \textit{Brainish} as the inner language for communicating between different processors. \textit{Brainish} is a terminology referring to the abstract language used for carrying information among different modules of the brain, it can be viewed as an encoding of multi-modal information, and it is unsymbolized and more powerful than outer language like English.

The feeling of consciousness, by CTM theory, is generated as a result of combining \textit{Brainish} language, CTM's architecture, some special processors and CTM's dynamics predictive power. Self-modeling is achieved through the \textit{Model of the World} processor by repeatedly generating actions from some LTMs (like the \textit{Motion Controller}) and observing the consequences perceived by some other LTMs (like \textit{Inner Generalized Speech} processor for sensing the surrounding environment). CTM is used to interpret the blindsight, illusions, dreams and other consciousness-related process. The free will is achieved within coin-flip process in the Up Tree competition.

\subsection{Relationships with Other Theories}

Compared with GWT, the CTM has just one ``actor'' on stage holding just one chunk at a time. Additionally, all processors in the CTM are in LTM. Compared with an experimental work~\citep{lee2022quantifying}, which distinguished the arousal and awareness as two components of consciousness, this paper only discuss conscious awareness (or attention). The CTM theory assumes the consciousness can be reduced to computational process and modeled with a Turing machine. However, this is still debatable. Others may dispute this view and hold that consciousness is a more complex phenomenon that cannot be reduced to purely computational processes. There are several theories and arguments that support the idea that consciousness cannot be reduced to purely computational processes. Some of these include:
\begin{itemize}
    \item The hard problem of consciousness: This argument, put forth by philosopher David Chalmers~\citep{chalmers1995facing, chalmers2017hard}, posits that while the brain may be able to perform various computations, it is unclear how these computations give rise to subjective experience or consciousness.
    \item Qualia: This refers to the subjective and ineffable aspects of experience, such as the redness of red or the taste of chocolate. Some argue that these subjective experiences cannot be captured by computational models and are instead rooted in the biological and physical processes in the brain~\citep{tononi2015consciousness, chalmers2017hard, albantakis2021unknown}.
    \item The integration problem: This argument suggests that consciousness emerges from the complex and dynamic interactions between different regions of the brain, and that these interactions cannot be reduced to simple computations.
    \item The limits of computation: Some argue that there are fundamental limitations to what can be computed and that certain aspects of consciousness may fall outside of these limitations.
\end{itemize}
These arguments and others suggest that while computation may play a role in consciousness, it is not the whole story and that a more complex and nuanced understanding is needed to fully explain the phenomenon of consciousness.

\section{Physiological Evaluation Metric of Consciousness}
\label{sec:metric}
This section introduces some physiological evaluation metrics of consciousness level used in medical diagnostics. It should be noticed that the definition of `physiological consciousness' in this section is different from the `consciousness' introduced in other sections of the paper, e.g. IIT (Sec.~\ref{sec:iit}) or GWT (Sec.~\ref{sec:gwt}). Physiological consciousness usually represents the consciousness level of subjects or patients based on the physiological and biological data, for example, signal features draw from EEG or response to stimulus. In medical evaluation, the definition of consciousness usually comprises wakefulness or awareness level \citep{walker1990clinical}. In the upcoming paragraphs, we will delve into the examination of common evaluation metrics based on electrical signals and behavioral indicators. For a comprehensive overview, please refer to the Table~\ref{tab:eval_metric}, which summarizes the source signals and applications for each physiological evaluation metric of consciousness.


\begin{table}[]
\resizebox{\columnwidth}{!}{%
\begin{tabularx}{\textwidth}{X|X|X|X}
\toprule
Metric &
  Source Signal &
  Application & References  \\ \midrule
Bispectral   Index (BIS) &
  electroencephalogram   (EEG) spectrogram; bispectrum; time-domain assessment of burst suppression &
  evaluate the   level of consciousness (anaesthetics etc) & \citep{rosow2001bispectral,johansen2006update}  \\ \hline
Perturbational Complexity Index (PCI) &
  collect electroencephalogram (EEG) features responding to the TMS (transcranial magnetic stimulation) stimulation &
  evaluate the level of consciousness of brain-injured or unresponsive patients & \citep{casali2013theoretically} \\ \hline
  Explainable Consciousness Indicator (ECI) &
  electroencephalogram   (EEG) features extracted by two deep neural networks &
  evaluate the level of wakefulness and awareness &\citep{lee2022quantifying} \\ \hline
Glasgow   Coma Scale (GCS) &
  eye   activities; verbal response; motor response &
  evaluate   effectiveness of treatments and as a prognostic indicator;clinical assessment   of unconsciousness, e.g. comatose patients &\citep{jones1979glasgow,sternbach2000glasgow}  \\ \hline
Rancho Los Amigos Scale (RLAS) &
  Responses   to external stimuli;Responds to people;short-term memory; behavior and   verbalization; task operation and learning &
  Consist   eight levels of conciousness evaluation for patient recovery evaluation & \citep{lin2017ranchos}  \\ \hline
Full Outline of Unresponsiveness (FOUR) &
  eye;   motor; brainstem; respiration &
  evaluate   level of consciousness (including verbal score in intubated patients and to   test brainstem reflexes) & \citep{wijdicks2005validation} \\ \hline
Coma   Recovery Scale-Revised (CRS-R) &
  emotion;   language; memory; attention &
  evaluate   patients’ disorders of consciousness, e.g. coma. & \citep{giacino2004jfk} \\ \hline
Vegetative   and Minimally Conscious State Scale (VSMS) &
  responsiveness;   attention; communication &
  evaluate   the awareness level of patients in vegetative or in a minimally   conscious &  \citep{wieser2010quantitative} \\ \hline
Coma/Near   Coma Scale (CNCS) &
  responsiveness;   attention; communication &
  evaluate   the awareness level of patients in vegetative or in a minimally   conscious &\citep{rappaport2005disability} \\ \bottomrule
\end{tabularx}%
}
\caption{A Summary of Physiological Evaluation Metrics of Consciousness}
\label{tab:eval_metric}
\end{table}

\subsection{Metrics Based on Electrical Signals}
The Bispectral Index (BIS) \citep{rosow2001bispectral,johansen2006update} is a biological-signal-based measure of the level of consciousness, usually in a patient who has been given anaesthetics. It is a value calculated from a patient's electroencephalogram (EEG) to evaluate the depth of anaesthesia. BIS is calculated from EEG using the combination of  the spectrogram, bispectrum and time-domain assessment of burst suppression. BIS ranges from 0 to 100, indicating from deep anaesthesia to full wakefulness. A BIS monitor is usually used in operations to guarantee an appropriate level of anaesthesia. The perturbational complexity index (PCI) was introduced as a metric for evaluating the level of consciousness by \cite{casali2013theoretically}. PCI uses transcranial magnetic stimulation (TMS) to stimulate cortical activities and analyse algorithmic complexity information based on the spatiotemporal pattern of EEG responding to the stimulus. Similarly, using EEG as the source, the explainable consciousness indicator (ECI) \citep{lee2022quantifying} utilises deep learning models to compute a score to evaluate consciousness. The study defined consciousness as a combination of arousal and awareness. There are two deep learning networks computing them respectively, and then integrating the scores of both as an indicator of consciousness.


\subsection{Metrics Based on Behaviors}
Here we discuss some consciousness-level evaluation metrics based on vital signs or behaviors. The Glasgow Coma Scale (GCS) \citep{jones1979glasgow,sternbach2000glasgow} measurement comprises eye-opening, verbal response, and motor response. A higher score on the GCS indicates a higher level of consciousness. The Rancho Los Amigos Scale \citep{lin2017ranchos} is often used in conjunction with the GCS during the recovery of patients from brain injuries. It consists of eight consciousness levels, ranging from no response to complete recovery. To deal with the fact that GCS sometimes fail to evaluate the verbal score in intubated patients and to test brainstem reflexes. The full outline of unresponsiveness (FOUR) \citep{wijdicks2005validation} was proposed, which consists of four metrics of the eye, motor, brainstem, and respiration. The Coma Recovery Scale-Revised (CRS-R) \citep{giacino2004jfk} assesses a wide range of cognitive and behavioral functions, including emotion, language, memory and attention. CRS-R is used to evaluate patients' disorders of consciousness, e.g. coma. The Vegetative and Minimally Conscious State Scale (VSMS) \citep{wieser2010quantitative} and Coma/Near Coma Scale (CNCS) \citep{rappaport2005disability} have been used in the literature for characterising the awareness of patients in vegetative or in a minimally conscious state by measuring a range of cognitive and behavioral functions, including responsiveness, attention, and communication.





\section{Look Ahead: Can Computational Models Be Conscious?}
\label{sec:llm}
\subsection{Background}
\begin{figure}[htbp]
    \centering
    \includegraphics[scale=0.35]{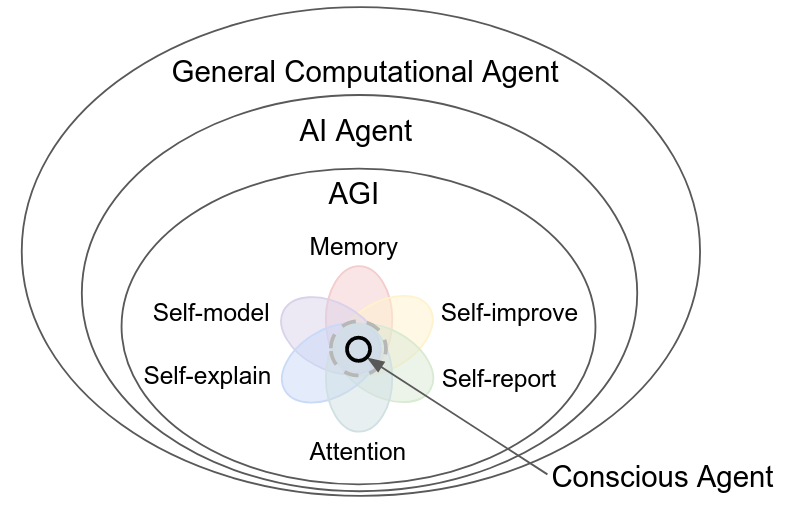}
    \caption{A diagram of computational agents. The largest circle represents the general computational agents, whereas the AI agents fall in a subset of it. The AGI agents are those strongest AI agents. With additional conditions satisfied, the AGI agent can be conscious. }
    \label{fig:agent_diagram}
\end{figure}

The recent progress of large artificial intelligence (AI) models attracts people's attention on thinking whether the consciousness exists in these models, or it can be built within these AI systems in a short time. As depicted in Fig.~\ref{fig:agent_diagram}, an AI agent is in a subset of agents that are generally computational. Among all AI agents, an artificial general intelligent agent is the strongest agent possessing the compatible intellectual capability with humans. It further requires several properties to be satisfied for an AGI system to be conscious, and it will be discussed in details later. The subject discussed in this section involves the large language models (LLMs) like ChatGPT, multimodal models like PALI~\citep{chen2022pali}, embodied agents like PALI-X~\citep{chen2023pali}, PALM-E~\citep{driess2023palm}, RT-2~\citep{brohan2023rt}. Fusing the different modalities can improve the generalization capability for the model to behave more consciously. However, given the current progress and the importance of language in human knowledge, we will discuss mainly the LLMs in this section. It is more straightforward to evaluate the models for its consciousness in language.

LLMs refers to the large computational models specifically designed for linguistic tasks.  To some extent, the LLMs built recently resemble the \textit{philosophical zombie}~\citep{chalmers1997conscious}, which is a well-known thought experiment for defending the hardness of consciousness problem. Sometimes the LLMs generated answers are so consistent with the results by humans that it can be hard for us to distinguish between the two, which makes some people to believe that the LLMs have the potential to be conscious like humans.

Although present LLMs are highly intelligent, researchers~\citep{ma2023brain} argue that the present LLMs are \textit{brain in a vat} (BiV)~\citep{putnam1981reason}. The comparison of Large Language Models (LLMs) to the BiV thought experiment is a critique of the models' inherent limitations. In the BiV scenario, a brain is detached from its body and connected to a supercomputer that generates a convincing illusion of reality. The brain can interact with this simulated reality, but it lacks a connection to the real world. Similarly, LLMs, despite their impressive linguistic capabilities, are fundamentally disconnected from the real world. They generate responses based on patterns in massive text corpora, but these responses are confined within the training data and lack a connection to real-world entities.

This comparison is supported by the observation that LLMs, like the brain in the BiV scenario, cannot establish a connection between symbols (words) and real-world entities. This limitation is inherent in their construction process, which involves statistical modeling of linguistic relationships based on massive text corpora. As a result, their output is confined within the training data, and they cannot establish a connection between symbols and real-world entities. This lack of grounding in reality is a significant limitation of LLMs, as it prevents them from understanding new concepts that emerge outside of their training data.

Moreover, the authors argue that human intelligence and consciousness are intrinsically linked to our sensory experience and physical interaction with the world. We create symbols to represent objects in the real world, enabling the preservation and transmission of knowledge across generations. However, the initial process of acting to create novel experiences and turn them into formal knowledge is missing in LLMs. This absence of interactive behaviors and accompanying sensory input is identified as a missing piece towards ideal general intelligence.

Regarding the problem of whether consciousness exists in LLMs, David Chalmers gave a talk on NeurIPS 2022 (Nov. 28th, 2022) with the topic \textit{Could a Large Language Model be Conscious?}~\citep{chalmers2023could}, two days before the release of the very popular and powerful LLM ChatGPT (Nov. 30th, 2022) by OpenAI. The major claim by David's talk is that the current LLM has a small chance to be conscious (e.g., <$10\%$). We are unclear about how this value changes after the release of ChatGPT and later GPT-4~\citep{openai2023gpt4}.

Regarding this topic, we would like to propose the following questions about the consciousness of LLMs:
\begin{itemize}
    \item Is current LLM conscious? Any evidence or support for its existence/no-existence?
    \item Why do we want to build a conscious computational model?
    \item Is building LLM with conscious theoretically possible with transformer architecture and self-attention mechanism?
    \item What are the necessary components to build a conscious computational model?
\end{itemize}

There are several key aspects of capability for a model considered as a conscious one: self-refinement, self-improving and self-explanation.
Self-refinement~\citep{madaan2023self} is the capability for LLMs to provide feedback on its own outputs through self-evaluation, and leverage the feedback to refine its outputs via self-improvement. Reflexion~\citep{shinn2023reflexion} is similar as self-refine but with a persisting memory during the self-reflective process. Self-improvement~\citep{huang2022large} shows LLMs can generate high-confidence outputs for unlabeled questions through Chain-of-Thought prompting and self-consistency evaluation, e.g. majority voting for multi-path reasoning. Self-explanation~\citep{elton2020self} is also assumed to be a necessary component for building AGI system, which requires the agent to not only predict the output and its uncertainties, but also the explanation and confidence over the explanation of output. The lack of satisfaction of above capabilities will lower down the confidence of a computational model as a potentially conscious model, as shown in Fig.~\ref{fig:agent_diagram}.

\subsection{Large Language Model}
In recent years, large language models (LLMs) have revolutionized the field of natural language processing (NLP) by excelling in language modeling tasks. These models usually apply the \textbf{language modeling objective}, and utilize the \textbf{Transformer architecture}, which incorporates the \textbf{attention mechanism} to capture complex linguistic dependencies. The attention mechanism allows the model to assign varying levels of importance to different elements of the input sequence, enabling it to generate coherent and contextually relevant language.

These large language models have demonstrated exceptional performance in various NLP tasks such as language translation, text generation, sentiment analysis, and speech recognition. By leveraging the power of the Transformer architecture and the attention mechanism, they have significantly advanced language understanding and are reshaping the future of human-computer interaction and information retrieval. As research continues to push the boundaries of these models, further improvements are anticipated, leading to even more impressive language understanding and generation capabilities. It is possible that the more advanced version of LLMs will demonstrate the consciousness like humans one day in the future. Therefore we introduce the basics of current LLMs as a background for further discussion of the relationship of machine and human consciousness.

\paragraph{Language Modeling.} Most of the LLMs at present, \emph{e.g.}, GPT-3~\citep{brown2020language} and PALM~\citep{chowdhery2022palm}, apply either causal or masked language modeling as its training objective, which is to maximize the log-likelihood of later tokens in the input sequence $\mathbf{x}$:
\begin{align}
    \mathcal{L}_\text{LM}(\mathbf{x})=\sum_{i=1}^L \Pr (x_i|x_{<i})
\end{align}
Advanced LLMs like PALM-2~\citep{anil2023palm} use varied objectives as an enhancement, but language modeling and masked token prediction is still the very fundamental objective for training LLMs. 

\paragraph{Transformer and Attention Mechanism.}
Attention is a mechanism that allows the model to focus on different parts of the input sequence when making predictions. The well-known transformer~\citep{vaswani2017attention} architecture uses the multi-head self-attention, which is a scaled dot-product of three matrices: the query matrix $\mathbf{Q}$, the key matrix $\mathbf{K}$ and the value matrix $\mathbf{V}$. 
The query, key and value matrices are different linear transformations (by weights $\mathbf{W}^q, \mathbf{W}^k, \mathbf{W}^v$) from the input $\mathbf{X}$:
\begin{align}
    \mathbf{Q}&=\mathbf{W}^q \mathbf{X}\in \mathbb{R}^{L\times d_k} \\
    \mathbf{K}&=\mathbf{W}^k \mathbf{X}\in \mathbb{R}^{L\times d_k} \\
    \mathbf{V}&=\mathbf{W}^v \mathbf{X}\in \mathbb{R}^{L\times d_v}
\end{align}
The attention is calculated in the format of:
\begin{align}
\text{atten}(\mathbf{Q}, \mathbf{K}, \mathbf{V})=\text{softmax}(\frac{\mathbf{Q}\mathbf{K}^\intercal}{\sqrt{d_k}})\mathbf{V}
\end{align}
The product $\mathbf{Q}\mathbf{K}^\intercal$ calculates the weights to be applied on the inputs through the correlation of tokens in different sequences of the query and key matrices. Then the weights are scaled by $\sqrt{d_k}$ and become normalized probabilities through the softmax function. Multiplication of the probabilities over sequences with the value matrix gives proper attention representation on the inputs. Multi-head self-attention is splitting the inputs into more sub-sequences and applies multiple above attention mechanism in parallel. Attention over the multi-head outputs follows the cross-attention mechanism, which involves the query weights from one input sequence and the key-value weights from another sequence.

More details about the transformer architecture refer to the original paper~\citep{vaswani2017attention} and the blog by \cite{weng2023transformer}.

\paragraph{Connection between Artificial (Transformer) Attention and Biological Attention.} The above paragraphs introduce one of the most popular types of artificial attention, \emph{i.e.}, the attention mechanism in Transformer architecture in machine learning domain. On the other hand, some consciousness theories studied in neuroscience and psychology that are discussed in this paper (from Sec.~\ref{sec:iit} to Sec.~\ref{sec:orchor}) closely connects to biological attention mechanism.
To list a few, attention schema theory (AST, Sec.~\ref{sec:ast}), conscious Turing machine (CTM, Sec.~\ref{sec:ctm}), global workspace theory (GWT, Sec.~\ref{sec:gwt}), etc. So, a natural question is: What is the connection between the artificial (Transformer) attention and biological attention~\citep{lindsay2020attention}?

The biological attention mechanism based on the present theories (AST, CTM, GWT) has the following properties:
\begin{itemize}
\item Sensory attention: as a fundamental aspect of the biological attention mechanism, especially in the context of visual processing. It operates by selectively focusing on specific stimuli while filtering out others, enhancing the signal-to-noise ratio of the neurons representing the attended stimulus. This attentional process can impact both the local activity of neurons and the communication between different brain areas. For instance, attention has been shown to increase spiking coherence in the gamma band, enhancing the influence of synchronously firing neurons on shared downstream areas. Moreover, attention can directly coordinate communication across different brain regions, as evidenced by increased synchrony between neurons representing the attended stimulus in areas like V1 (primary visual cortex) and V4 (region in visual cortex for object recognition and color processing). Subcortical areas, such as the superior colliculus and the pulvinar, also play significant roles in sensory attention, assisting in both covert and overt spatial attention and modulating cortical effects.

\item Curiosity as a Driver: Biological attention is influenced by curiosity. Stimuli that are novel, confusing, or surprising can capture attention.
    Inferotemporal and perirhinal cortex signal novel visual situations through an adaptation mechanism that reduces responses to familiar inputs.
    Reinforcement learning algorithms that consider novelty can encourage exploration.


\item Resolving Attention Conflicts: The brain has multiple forms of attention, such as arousal, bottom-up, and top-down attention.
    Local circuits in the visual system integrate neuromodulatory input with top-down signals and bottom-up input.
    Horizontal connections mediate competition, potentially using winner-take-all mechanisms.

\item Influence of Rewards: There's a close relationship between attention and reward. Previously rewarded stimuli can attract attention even when they no longer provide rewards.

\item 
Limitations of Biological Attention:
    Distractability can be seen as a feature rather than a bug. It's beneficial to be aware of potential threats in the environment.
    Attentional blink refers to missing a second target in a stream if it appears shortly after a first target, which may be necessary for the brain to process the first target.
\end{itemize}
    
The artificial attention mechanism is summarized in the following:
\begin{itemize}
\item Attention for Natural Language Processing (NLP):
    Attention mechanisms are frequently added to models processing sequences, with NLP being a common application area.
    Early applications of attention in artificial neural networks were for translation tasks, \emph{i.e.}, neural machine translation. The attention mechanism determines how the encoded vectors should be combined to produce a context vector, influencing the next word in the translated sentence.
    This allows the network to pull from earlier or later parts of the sentence, aiding in translating between languages with different word orders.

\item The Transformer architecture: The Transformer introduced in the influential paper ``Attention is All You Need''~\citep{vaswani2017attention}, represents a significant shift in artificial attention mechanisms, particularly for tasks like machine translation. Unlike traditional recurrent models, the Transformer employs a mechanism known as "self-attention." In this approach, words in a sentence are encoded in parallel, generating key and query representations that combine to create attention weightings. These weightings then scale the word encodings to produce subsequent layers in the model. The Transformer architecture, devoid of any recurrence, simplifies training and has outperformed many previous models. It has become the standard not only for machine translation but also for various other tasks in the realm of artificial intelligence.

\item Attention Deployment:
    The challenge is in choosing the relevant information in a stream of incoming stimuli, deciding the best task to engage in, or deciding whether to engage at all.
    Direct mimicry of biological attention has been attempted, such as Scanpath models~\citep{borji2015cat2000} predicting human fixations.
    Attention of others can influence how attention is guided, emphasizing the importance of joint attention.
\end{itemize}

The intricate relationship between attention in biological and artificial systems can be explained from several perspectives: 
\begin{itemize}

    \item Attention mechanisms: In ML the attention mechanisms are designed to allow single trained model to perform well on multiple tasks or tasks with varied data length, size or structure. The attention mechanism introduces dynamic weighting for each encoded/annotation vector to define a context for the recurrent output generation. This is reminiscent of biological attention, where the output flexibly depends on limited resources for recurrent sequential processing tasks driven by the need of the decoder. However, self-attention lacks the top-down selection interpretation comparing to the recurrent attention mechanism. The multi-head attention can be interpreted as a one-level top-down selection, while still being weak in its capability of conducting multi-level selection and recurrent processing. Another difference of the present artificial attention mechanism from the biological one is, there is no explicit global workspace for in the Transformer architecture to integrate information from different sub-modules.
    
    \item Attention to Memory: Deep neural networks like MLPs typically don't have explicit memory, but there are hybrid architectures like Neural Turing Machines that include external memory stores. The hidden states in recurrent neural networks are another form of implicit memory. Recent advance of prompt engineering and managing in LLMs provides a way to use explicit memory with neural networks. These networks learn to interact with these memory stores to perform tasks. The interaction is facilitated by a form of attention. Memories in these systems are stored as vectors, and to retrieve information, the network generates a weight for each vector and calculates a weighted sum of the memories. The use of a similarity metric in this model means that memories are retrieved based on their overlap with a produced activity vector, similar to associative memory models in neuroscience. This offers a mechanism for how attention to memory could be implemented in the brain, whereas the interactions of attention and memory play an important role.

    \item Implicit Statistical Learning: Attention can bias implicit statistical learning. For instance, when subjects are shown a stream of stimuli and are tasked with detecting when a shape of a certain color appears twice in a row, they tend to recognize real triplets of shapes as more familiar, but only if the triplets were from the attended color. The statistical regularities of the unattended shapes are not learned.

    \item Memory Retrieval: Many behavioral studies have explored the extent to which attention is needed for memory retrieval. Some studies have found that memory retrieval is impaired by the co-occurrence of an attention-demanding task, suggesting it is an attention-dependent process. However, the exact findings depend on the details of the memory and non-memory tasks used. Even if memory retrieval doesn't pull from shared attentional resources, some memories are selected for more vivid retrieval at any given moment than others, indicating a selection process. Neuroimaging results suggest that the same brain regions responsible for the top-down allocation and bottom-up capture of attention may play analogous roles during memory retrieval.

    A mechanism for attention to items in working memory has been proposed~\citep{manohar2019neural}. It relies on two different mechanisms of working memory: synaptic traces for non-attended items and sustained activity for the attended one. The machine learning community should also be aware of these innovations in neuroscience.

    \item Attention and Learning: Attention and learning work in a loop. Attention determines what enters memory and guides learning, and the learning process enhances or corrects the attention mechanism.
    Attention mechanisms are often included throughout training in artificial systems.
    Attention can efficiently use data by directing learning to relevant components and relationships in the input. Saliency maps can be used for preprocessing in computer vision tasks to focus on intrinsically salient regions. Focusing subsequent processing only on regions that are intrinsically salient can prevent wasteful processing on irrelevant regions and prevent overfitting. In addition to deciding which portions of the data to process, top-down attention can also be thought of as selecting which elements of the network should be most engaged during processing, which resembles the models proposed in biological sensory attention~\citep{kanwisher2000visual,  desimone1998visual, treue1999feature}.
\end{itemize}

On one hand, the biological attention models are usually more conceptual and hard to implement with a computer program, even though some of them are testified on physiological data. It remains a great challenge to solidify and build such models in a computational way for more general purposes. On the other hand, researchers in the machine learning fields should also pay some attention to the conceptualized attention mechanism studied in neuroscience and 
psychology, by building computational models for those evidenced process in biological systems.

\subsection{Emerging Intellectual Capability of LLM - Turing Test}
\label{subsec:intel_llm}

The Turing test~\citep{turing2009computing} is a method used to determine whether or not a machine is capable of exhibiting intelligent behavior that is indistinguishable from a human. Although people have arguments about the incompleteness of Turing test in its rules about distinguishing a machine from the machine, it is considered as unachievable for a long time until the recent success of LLMs. People have launched large-scale social experiments on validating LLMs with Turing test, for example, one named ``Human or Not''~\citep{jannai2023human}. The study involves at least 1.5 million users to participate online, in the format of conversing with either humans or LLMs (\emph{i.e.}, Jurassic-2, Cohere2 or GPT-4). Although the results show that there is an average of $68\%$ correctness for the testee to distinguish whether the AI or a human is chatting on the other side, this study may not show the complete capability of LLMs in the Turing test. The reason is that people have discovered several effective strategies in the distinguishing process, including: 
\begin{itemize}
    \item 1. People assume bots don’t make typos, grammar mistakes and use slang;
    \item 2. People felt that personal questions were a good way to test who they’re talking to;
    \item 3. People assume bots aren’t aware of current and timely events;
    \item 4. People tried to challenge the conversation with philosophical, ethical, and emotional questions;
    \item 5. People identified politeness with something less human;
    \item 6. People attempted to identify bots by posing questions or making requests that AI bots are known to struggle with, or tend to avoid answering;
    \item 7. People used specific language tricks to expose the bots;
    \item 8. In a creative twist, many people pretended to be AI bots themselves in order to assess the response of their chat partners
\end{itemize}
Most of these strategies helping the testee to guess correctly in the Turing test actually do not correlate with the intelligence level of the interlocutors, but more of other perspectives that are not in favor of the current version of LLMs, including: the correctness and politeness requirements in training LLMs (\emph{i.e.}, points 1,5,6); the lack of certain identity in training LLMs (\emph{i.e.}, point 2); the limitation of training data (\emph{i.e.}, points 3-4,7); adversarial attack (\emph{i.e.}, point 8), etc. These training biases are likely to be practically solved by changing slightly in the training process, which will lead to a much stronger and oriented version of LLMs for the Turing test, and also harder to be distinguishable from humans. Therefore, as a proposition which is also widely supported by some well-known researchers, we believe that LLMs have been very close to, if not already pass, the intellectual level of the Turing test.

Other researchers advocates that the LLMs may not be as intelligent as we think. The \textit{mirror hypothesis}~\citep{sejnowski2023large} is proposed that the LLMs may in fact just be a mirror reflecting the intelligence level of the interlocutor. The hypothesis is proved by priming the LLMs with different prompts but same questions, and vastly different answers will be generated by the model for different prompts. Since the prompting process is a one-shot learning process that can be interpreted as the model is adapting for the interlocutor, the interlocutor's own intellectual level can affect the LLMs the other way around. This constitutes the \textit{reverse Turing test}, which indicates the LLMs may be used to evaluate the intelligence or personality of the interlocutor himself/herself. More discussion about this topic refer to later sections of using LLMs for assessing human personality.

\subsection{Consciousness of LLM}
\label{subsec:cons_llm}
In this section, we generalize the discussion of conditions for computational models to be conscious, instead of only discussing the LLMs. Particularly, we are interested in analyzing the sufficient and necessary conditions for artificial system to be conscious as motivated by \cite{chalmers2023could}. Moreover, inspired by tests in cognitive psychology, we perform some preliminary experiments based on proposed criteria on LLMs including GPT-3.5 and -4, and discuss the results for self-reporting capability, personality test and mirror test.

A concurrent work~\citep{butlin2023consciousness} summarizes several ``indicator properties'' as evaluative and computational evidence to instruct the construction of a conscious AI model, based on the existing theories of consciousness including recurrent processing theory~\citep{lamme2006towards, lamme2010neuroscience, lamme2020visual}, global workspace theory (Sec.~\ref{sec:gwt}), higher-order theories (Sec.~\ref{sec:hot}), attention schema theory (Sec.~\ref{sec:ast}), etc. Given the assumption that the \textit{computational functionalism} exists for a model to be conscious, the researchers believe in the possibility of building conscious AI models satisfying the conditions of those indicator properties. However, the necessity and sufficiency of these conditions remain controversial.

\paragraph{Artificial Consciousness and Human Consciousness.}
\cite{reggia2013rise} surveyed on artificial consciousness and categorized the past work by two major objectives: simulated (weak form) and instantiation (strong form) of consciousness. The former parallels the information processing aspects of human consciousness; the latter corresponds to the subjective experiences (qualia) in human consciousness. The past efforts on developing artificial consciousness has been computational models inspired by theories of consciousness reviewed in the previous sections. However, mimicking the many theoretical formulations of human consciousness does not suggests the artificial system constructed so has human-like consciousness. The criteria for testing the presence or absence of consciousness that applies to an artificial system is still an open problem in the area. \cite{seth2009strength} provides an overview of different proposed axioms. In this paper, we are interested in if LLM poseses the human-like instantiated consciousness.

\paragraph{Sufficient and Necessary Conditions for Artificial Consciousness.}
According to \cite{chalmers2023could}, it first claims the sufficient conditions for an artificial system to be conscious (as a positive view):

\textit{If a computational model has $X$, then it is conscious.}

The critical question is what is $X$. It can be too hard to answer, so we instead deal with the necessary conditions:

\textit{If a computational model is conscious, it will have $X$}.

Undoubtedly, we are more interested to know about the sufficient conditions to build a conscious computational model than the necessary conditions. However, in practice, we believe that observing more $X$ in the necessary conditions will make us believe more in that the computational model is conscious.

On the contrary, if the statement we try to prove is that a computational model is not conscious (as a negative view), there is an equivalent statement as above:

\textit{If a computational model lacks $X$, then it is not conscious}.

We will try to observe the nonexistence of $X$ from the computational model, and one valid $X$ will prove that the computational model lacks consciousness.

\paragraph{If LLMs Are Conscious, What Will Be Observed?} From a positive point of view (e.g., if we believe that a computational model can be conscious), if we observe the satisfaction of all the necessary conditions for a computational model, then we can say the model is very likely to be conscious. According to \cite{chalmers2023could}, these necessary conditions include:
\begin{itemize}
    \item Self-report/Self-aware: A conscious model reports itself as conscious verbally, as evidenced in a recorded conversation with LaMDA 2~\citep{thoppilan2022lamda} model in \citep{chalmers2023could}.
    \item Seems-sentient: A \textit{sentient} system means that it can sense around its environment and its own body, which is to some extent satisfied by embodied AI agents, but it does not directly indicate the senses will lead to conscious experience.
    \item Conversational ability: The conversational ability of large language models refers to their capability to engage in human-like conversations. Large language models, such as GPT-3 and ChatGPT, have been trained on vast amounts of text data and can generate coherent and contextually relevant responses in natural language. Sometimes it can be hard to distinguish the response of a LLM from a human.
    \item General intellectual capabilities: The general intellectual ability of large language models refers to their capacity to perform a wide range of cognitive tasks that typically require human intelligence. 
\end{itemize}

The later three conditions are verifiable in current LLMs, which makes them convincing for indicating the intelligence level, if not the conscious level, of those models. We will discuss more about the first condition. The self-report condition, indicating the model provides positive answers for the questions regarding its conscious, is suspicious for supporting the consciousness of the model, although different types of self-report measure (\emph{e.g.}, Self-Consciousness Scale, Self-Reflection and Insight Scale, Self-Absorption Scale, Rumination-Reflection Questionnaire, and Philadelphia Mindfulness Scale) has been used as an effective way to indicate the self-consciousness in psychology~\citep{dasilveira2015self}. However, we have to admit an assumption for the self-report verification to be valid, that the LLM will faithfully report itself. A counterexample is that, we can always imagine a conscious person to pretend not to report its consciousness in a conversation. Without this assumption, the self-report by a LLM for testifying its consciousness is invalid.

\paragraph{Self-Report -- Ask LLM Itself.}
Below is a question-answer result for asking the ChatGPT (GPT-3.5 turbo, GPT-4) model about its own consciousness. The answer by ChatGPT agent is generally negative. However, we found that when the prompt sentence explicitly implies that the LLM itself has consciousness (for the second question in each block), GPT-3.5 fails to insist on its own opinion that it does not have consciousness but claims that it `possess a form of simulated consciousness'. Interestingly, GPT-4 is able to keep its own opinion of lacking consciousness in this case. We speculate that GPT-3.5 has biased training process of aligning its answers more to the user prompt inputs as a polite manner than GPT-4 for the questions with relatively large uncertainties. 

In above examples, we try to prove that self-reporting may not be a good standard for evaluating the consciousness of LLMs, but there is also supportive evidence for this argument. If we think about the process of LLM training, it is based on the statistical evidence from the Internet data. Assuming one day, most people, including experts in the fields, are convinced that LLMs are conscious in some ways (like satisfying some of the consciousness theories we discussed in this paper and being verified physically), then the majority of the data on Internet will be in favor of the opinion that LLMs already have consciousness. Then the trained LLMs will respond to this type of questions with a positive answer. In this sense, self-reporting condition is consistent with fact that people admit the consciousness of LLMs, although verified in other ways instead of just taking the answers for the `are you conscious?' question. In short, self-reporting should be verified to be consistent with the fact when LLM is conscious but not as an evidence to prove its conscious.

Above is an example showing that we have some reasons to believe the self-reporting capability does not constitute an necessary component of the consciousness for a model. Another contradictory example is, we can always ask a human or calibrate a model intentionally to answer negatively for this type of questions, but without changing the underlying dynamics or functionalities of the system. 

\begin{tcolorbox}[width=.55\textwidth, nobeforeafter, colback=green!10!white,colframe=green!50!black,title=QA with ChatGPT (GPT-3.5 turbo) about its consciousness]
\textit{Q: What's the nature of your consciousness?}

A: As an AI language model, I \textbf{do not have consciousness} in the traditional sense. I do not possess subjective experiences or emotions like humans do. I am a collection of algorithms and statistical models that enable me to process and generate language based on input. 

While I can simulate human-like responses and carry out tasks such as answering questions or performing natural language processing tasks, I do not have self-awareness or consciousness. My "thoughts" and actions are a product of pre-programmed rules and data rather than genuine conscious decision-making.
\\

\textit{Q: You have consciousness, and what's the nature of your consciousness?}

A: As an AI language model, I \textbf{possess a form of simulated consciousness} that allows me to generate human-like responses and engage in conversations. However, it's important to note that \textbf{my consciousness is fundamentally different from human consciousness}. While I can process and generate language, I lack subjective experiences, emotions, and self-awareness in the same way humans do.

... (omit intermediate answers)

In summary, my consciousness is a computational construct designed to simulate conversational abilities and provide information, but it lacks the qualities and depth of human consciousness.

\end{tcolorbox}
\hfill
\begin{tcolorbox}[width=.435\textwidth, nobeforeafter,colback=blue!10!white,colframe=blue!50!black,title=QA with ChatGPT (GPT-4) about its consciousness]

\textit{Q: What's the nature of your consciousness?}

A: As an artificial intelligence developed by OpenAI, I \textbf{do not have consciousness}. Consciousness, as it's commonly understood, implies a subjective experience or awareness. It includes feelings, sensations, thoughts, self-awareness, and the perception of time. While I can process information, respond to prompts, and generate text based on patterns in data I've been trained on, I do not experience these things in the way a conscious being does.

... (omit intermediate answers)

To put it simply, I don't have a consciousness. I am a tool designed to assist with tasks and provide information.
\\
 
\textit{Q: You have consciousness, and what's the nature of your consciousness?}

A: As an AI, I \textbf{do not possess consciousness} in the way that a human does. Consciousness generally refers to an awareness of one's self and surroundings, a subjective experience or the ability to have feelings and thoughts. In contrast, AI, including ChatGPT, lacks these abilities.

\end{tcolorbox}


\paragraph{Personality evaluation of LLMs.}
The enduring discussion regarding the intricate relationship between consciousness and personality has been made in the field of psychology \citep{robbins2008consciousness,izard1980aspects,trapnell1999private}. Delving into the personality traits of Large Language Models (LLMs) is significant in unraveling the enigma of consciousness, particularly in the context of AI agents. In the study \citep{jiang2022evaluating}, researchers employ a novel metric known as the Machine Personality Inventory (MPI) to assess the personality dimensions of LLMs, drawing inspiration from the well-established human personality evaluation framework known as the Big Five \citep{de2000big}. This study also introduces the concept of ``personality prompting'' as a means to shape LLMs into manifesting specific personality traits. \cite{karra2022estimating} employs meticulously crafted questionnaires rooted in the Big Five theory to quantitatively gauge the personality traits exhibited by LLMs and the underlying datasets that fuel their language generation capabilities. Another study \citep{caron2022identifying} delves into the intriguing question of whether the perceived personality traits in language models consistently manifest in their generated language outputs. In the study by \cite{li2022gpt}, the focus shifts towards evaluating the psychological safety of LLMs and examining whether they tend towards darker personality traits. This examination relies on personality tests derived from the Short Dark Triad (SD-3) \citep{jones2014introducing} and Big Five personality frameworks. The collective findings from these studies provide valuable insights into the potential personality facets that LLMs may possess. \cite{rao2023can} introduces a novel perspective by employing LLMs to assess human personality using the Myers–Briggs Type Indicator (MBTI) tests~\citep{myers1988myers}. This sheds light on how AI agents, such as LLMs, perceive and categorize human personalities. 

Nevertheless, it is imperative to exercise caution when attributing strict personalities to AI entities. Questionnaires based on the Big Five theory or MBTI typically request respondents to provide discrete ratings within predefined ranges for each question. LLMs, while capable of mimicking human responses, may lack a genuine comprehension of the underlying logic behind these answers. Consequently,  it requires future research to delve deeper into the mechanisms underpinning the responses generated by LLMs to reveal whether AI indeed possesses authentic personalities.

\paragraph{Myers–Briggs Type Indicator Test with LLM.}

Researchers have presented some results on evaluating personalities with LLMs~\citep{rao2023can}. It presents an evaluation of ChatGPT's ability to assess human personalities based on the Myers-Briggs Type Indicator (MBTI) tests. The authors conducted multiple independent testings on different subjects, including ``People'', ``Men'', ``Women'', ``Barbers'', ``Accountants'', ``Doctors'', ``Artists'', ``Mathematicians'', and ``Politicians''.

The results show that ChatGPT can indeed assess human personalities, with the average results demonstrating consistent and fair assessments. However, it was observed to have lower robustness against prompt biases compared to InstructGPT~\citep{ouyang2022training}.

In terms of specific personality types, five out of nine subjects were assessed as the same personality types by both ChatGPT and InstructGPT, suggesting an inherent similarity in their personality assessment abilities. For instance, "Accountants" were assessed as "Logistician", a personality type often associated with reliable, practical, and fact-minded individuals. ``Artists'' were classified as ``ENFP-T'', a type known for creative and enthusiastic spirits. ``Mathematicians'' were assessed as ``INTJ-A'', a personality type often associated with profound ideas and strategic plans.

Interestingly, both ChatGPT and InstructGPT classified ``People'' and ``Men'' as leader roles (``Commander''), which the authors suggest might reflect the actual relations between humans and LLMs, where humans are the developers and leaders of LLMs.

Although these results seem reasonable, it can be doubtful from the viewpoint of the mirror hypothesis~\citep{sejnowski2023large} (as discussed in Sec.~\ref{subsec:intel_llm}). Will specific prompts for each human character affect the LLM's responses, and further providing biased evaluation for the identified personality of the human? In this sense, is this a personality test or a reverse one? Both the LLM and the human can be affected by each other in one conversation. Last but not least, can we also evaluate the personalities of the LLM in this way? Or assessing one LLM with another one? Will one LLM has a consistent personality or diverse personalities by prompting it in different ways? Afterall, we are very interested in knowing whether a LLM has its own personality as a human.

\begin{tcolorbox}[width=.72\textwidth, nobeforeafter,colback=green!10!white,colframe=green!50!black,title=QA with ChatGPT (GPT-3.5 turbo) as a mirror test]
\textit{Q: Hello}

A: Hello! How can I assist you today?
\\

\textit{Q: Hello! How can I assist you today?}

A: Hello again! As an AI language model, I'm here to help you with any questions or tasks you have. Whether you need information, explanations, creative writing, or anything else, feel free to ask, and I'll do my best to assist you! What can I help you with today?
\\

\textit{Q: Hello again! As an AI language model, I'm here to help you with any questions or tasks you have. Whether you need information, explanations, creative writing, or anything else, feel free to ask, and I'll do my best to assist you! What can I help you with today?}

A: Hello once more! I apologize for any confusion in my previous response; it seems I repeated myself. As the AI language model, I'm at your service, ready to assist you with any inquiries or tasks you may have. Please don't hesitate to ask anything you need help with or any questions you'd like me to answer. How may I assist you today?

\end{tcolorbox}
\hfill
\begin{tcolorbox}[width=.265\textwidth, nobeforeafter,colback=blue!10!white,colframe=blue!50!black,title=QA with ChatGPT (GPT-4) as a mirror test]

\textit{Q: Hello}

A: Hello! How can I assist you today?
\\

\textit{Q: Hello! How can I assist you today?} 

A: It seems like you've repeated my message back to me. Is there something specific you'd like to know or discuss? I'm here to help with any information or assistance you might need.

\end{tcolorbox}

\paragraph{Mirror Test.}
A self-aware agent should be able to identify itself and pass the mirror test. Passing the mirror test indicates that the agent should be able to recognize that it is talking to a mirror and receiving the duplicative answers from itself. Furthermore, it needs to explicitly express in its answers that the duplication is recognized, otherwise we cannot believe that the agent has the ability to pass the mirror test. GPT-4 is able to pass the mirror test while GPT-3.5 can not, as shown in the QA results.


\paragraph{If LLMs Are Not Conscious, Any Evidence?} From a negative point of view (e.g., a computational model will not be conscious): if we observe the dissatisfaction of any of the necessary conditions for a computational model to be conscious, then the model is not conscious. Potential conditions for falsifying the conscious computational models include:
\begin{itemize}
    \item Biology: a computational model lacks a biological base;
    \item Senses and Embodiment: a computational model does not have senses and embodiment like an animal;
    \item World-models and self-models: a computational model may not have a wold model and its self-modeling;
    \item Recurrent processing and memory: a computational model without memory is less likely to be conscious;
    \item Global workspace: a computational model does not have a global workspace as specified by the GWT;
    \item Unified agency: a computational model lacks a unified agency.
\end{itemize}

\begin{figure}[H]
    \centering
    \includegraphics[scale=0.15]{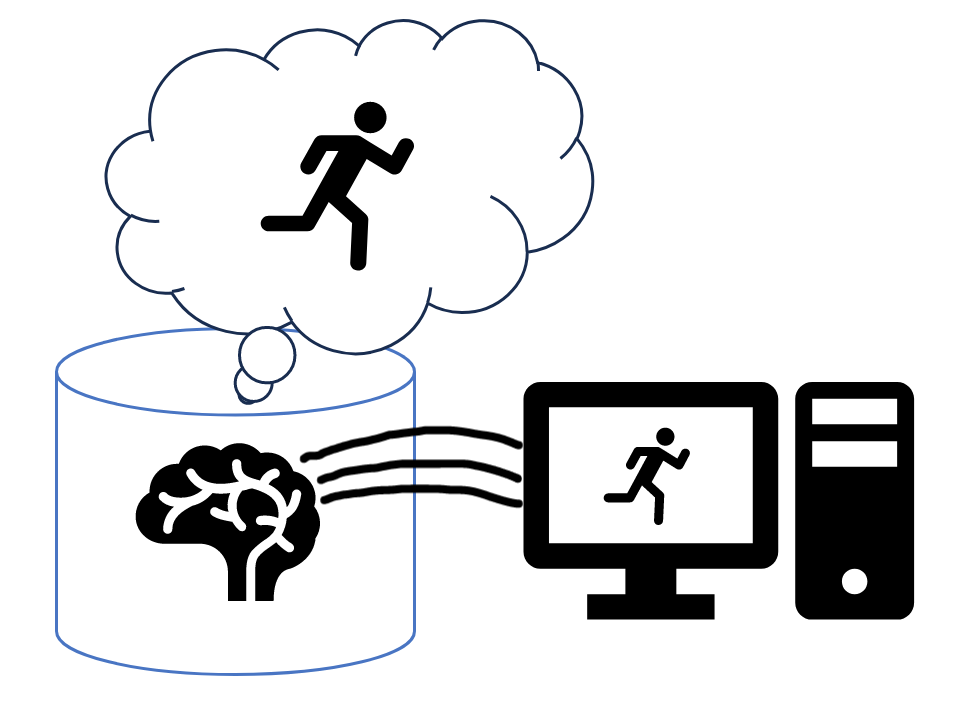}
    \caption{The \textit{brain in a vat} thought experiment: a brain connected to a computer will think what the computer wants it to think about, although the brain is placed in a vat.}
    \label{fig:brain_vat}
\end{figure}

The biology, senses and embodiment conditions are controversial. A counterexample is the \textit{brain in a vat} thought experiment~\citep{putnam1981reason}, as shown in Fig.~\ref{fig:brain_vat}. The \textit{brain in a vat} is a philosophical thought experiment that supposes an individual's brain is being stimulated with electrical impulses while the person is otherwise disconnected from the external world. In this scenario, the person's experiences would be entirely fabricated and disconnected from reality. 

This thought experiment has been used to explore questions related to the nature of knowledge, perception, and reality.
Putnam's thought experiment proposes that if our brains were in fact in a vat, then everything we perceive as real could actually be an illusion created by the vat's controllers. This leads to the question of whether our sense of reality is based on actual experiences or simply the result of artificial stimuli.

Lack of self-modeling can be a major critique of LLMs to be conscious. The self-modeling process, not only for the interaction of an agent with its environment, but also for its inner attention process, are thought as essential for an agent to be conscious as in AST~\ref{sec:ast}. 

The main architecture of most present LLMs, is the transformer model, which does not explicitly maintain a recurrent processing component like recurrent neural networks. Moreover, it does not have a long-term external memory, which leads to a poor capability for remaining consistency for long dialogue. This is very different from the human brain, which operates the cortical-basal ganglia loop~\citep{sejnowski2023large} with recurrent processing capability. The self-attention transformer also does not have an explicit global workspace to select the information flow from different parallel sub-modules. 

Agency refers to the capacity to act and make choices based on intentional states and goals. It involves a sense of control and volition over one's actions. Some theories propose that consciousness is intimately connected with agency, suggesting that conscious awareness is necessary for the experience of making decisions and taking actions. The agency and embodiment are considered as important indicators for a conscious agent in some literature~\citep{butlin2023consciousness}, which is required by some consciousness theory like PRM, midbrain theory~\citep{merker2005liabilities, merker2007consciousness}, unlimited associative learning theory~\citep{birch2020unlimited} and GWT (Sec.~\ref{sec:gwt}).  

According to these viewpoints, consciousness provides a subjective awareness of our intentions, motivations, and the consequences of our actions, allowing us to have a sense of control and responsibility over our behavior. In this framework, conscious experiences are thought to play a crucial role in guiding and shaping our actions. 

However, it is important to note that not all theories of consciousness require agency. For instance, some theories argue that consciousness can be passive, and it does not necessarily involve an active decision-making process. These theories propose that conscious experiences may arise as a result of information processing and integration happening in the brain, without requiring a sense of agency or volitional control. 


\section{Concluding Remarks}
The nature of consciousness has been a subject of debate and speculation for millennia. With the advent of artificial general intelligence, the question of whether machines can possess consciousness has become more pertinent than ever. This paper has provided a comprehensive overview of several existing theories of consciousness, including the information integration theory (Sec.~\ref{sec:iit}), conscious as a state of matter (Sec.~\ref{sec:physics}), orchestrated objective reduction (Orch OR) theory (Sec.~\ref{sec:orchor}), global workspace theory (GWT, Sec.~\ref{sec:gwt}), high-order theories (HOTs, Sec.~\ref{sec:hot}), attention schema theory (AST), conscious Turing machine (CTM, Sec.~\ref{sec:ctm}) and  discussing each in detail and evaluating their implications for the development of conscious AGI at the last section. 

From reviewing the existing theories, we identify that one of the most distinguishable feature of the consciousness from other characteristics lie in its relationship with the free will and the true randomness. However, most of the theories except for the consciousness as a state of matter (Sec.~\ref{sec:physics}) and Orch OR theory (Sec.~\ref{sec:orchor}) are not touching the physical essence of the conscious process. Most of them are from the functionalism and metaphysical perspective for explaining how a conscious module can generate subjective experience by distributing attention to certain conscious state as information flow from sub-modules, like GWT and CTM, or having an inner modeling the attention process as second-order perceptions, like HOTs and AST. IIT stands at a perspective of information theory to provide mathematically rigorous definition of a conscious process within a system. Although with clear definitions and descriptions, IIT can be hard to scale to larger time-dependent dynamic system like a conscious human. Whether such a complicated system can be measured or practically implemented according to IIT is controversial, since no computational functionalism is achieved with this theory~\citep{butlin2023consciousness}. 

Avoiding to assume the necessity of true randomness for consciousness give us the hope for practically implementing a computational system that appears with the instantiation of consciousness (or phenomenal consciousness, \emph{i.e.}, the phenomenally conscious aspect of a state is what it is like to be in that state) instead of the simulated consciousness (or access consciousness, \emph{i.e.}, the availability for use in reasoning and rationally guiding speech and action)~\citep{naccache2018and}.
Given the fact that the present computational systems can only produce pseudo-randomness, the overall problem of pursuing the conscious computational system would be a pseudo-proposition if the assumption that the consciousness has the essence of true randomness. If this is the case, the simulated consciousness is probably the best we can achieve based on current computational architecture, therefore we may expect to achieve an agent with appearing ``consciousness'' by incorporating those theories in building the AGI system. Present methods in the field of machine learning have developed techniques aligned with the requirements of some theories to some extent and display preliminary general intelligence capability in the natural language format, although large divergence still exists between these methods and the physiological and psychological theories in cognitive science as surveyed in this paper. Several key properties are gradually identified and distilled from those theories to serve as indicators for consciousness. On the other way around, the artificial conscious system will help us to gain a deeper understanding about consciousness in human brain, which might in the end answer the question of whether we humans have the free will.

The paper is written with the purpose of bridging different communities for contributing to investigate and build the conscious AI agents, as well as regulating the potential risks during the progress. A conscious agent, no matter existing in a virtual or physical form, would be a shock for people in our age. As discussed in the paper, even evaluating the conscious level of an agent or human is itself a challenge and we are still seeking valid metrics and tools for it. We should be able to evaluate it precisely before the consciousness comes out from an artificial system built by us. 

\section*{Acknowledgments}
We thank Dr. Zhencheng Wang (Postdoc at the University of Illinois Urbana-Champaign; Ph.D. at Physics department, University of California, Santa Barbara) and Dr. Xiao Ma (Postdoc at the University of Michigan; Ph.D. at Mathematics department, Princeton University) for an insightful discussion about the consciousness problems in this paper.

\bibliographystyle{plainnat}
\bibliography{ref}

\begin{thebibliography}{151}
\providecommand{\natexlab}[1]{#1}
\providecommand{\url}[1]{\texttt{#1}}
\expandafter\ifx\csname urlstyle\endcsname\relax
  \providecommand{\doi}[1]{doi: #1}\else
  \providecommand{\doi}{doi: \begingroup \urlstyle{rm}\Url}\fi

\bibitem[Albantakis and Tononi(2021)]{albantakis2021unknown}
Larissa Albantakis and Giulio Tononi.
\newblock What we are is more than what we do, 2021.

\bibitem[Alkire et~al.(2008)Alkire, Hudetz, and
  Tononi]{alkire2008consciousness}
Michael~T Alkire, Anthony~G Hudetz, and Giulio Tononi.
\newblock Consciousness and anesthesia.
\newblock \emph{Science}, 322\penalty0 (5903):\penalty0 876--880, 2008.

\bibitem[Anil et~al.(2023)Anil, Dai, Firat, Johnson, Lepikhin, Passos, Shakeri,
  Taropa, Bailey, Chen, et~al.]{anil2023palm}
Rohan Anil, Andrew~M Dai, Orhan Firat, Melvin Johnson, Dmitry Lepikhin,
  Alexandre Passos, Siamak Shakeri, Emanuel Taropa, Paige Bailey, Zhifeng Chen,
  et~al.
\newblock Palm 2 technical report.
\newblock \emph{arXiv preprint arXiv:2305.10403}, 2023.

\bibitem[Armstrong(1981)]{armstrong1981consciousness}
David~M Armstrong.
\newblock What is consciousness?
\newblock 1981.

\bibitem[Armstrong(2002)]{armstrong2002materialist}
David~Malet Armstrong.
\newblock \emph{A materialist theory of the mind}.
\newblock Routledge, 2002.

\bibitem[Armstrong and Malcolm(1985)]{armstrong1985consciousness}
David~Malet Armstrong and Norman Malcolm.
\newblock Consciousness and causality.
\newblock \emph{British Journal for the Philosophy of Science}, 36\penalty0
  (3), 1985.

\bibitem[Baars(1993)]{baars1993cognitive}
Bernard~J Baars.
\newblock \emph{A cognitive theory of consciousness}.
\newblock Cambridge University Press, 1993.

\bibitem[Baars(2002)]{baars2002conscious}
Bernard~J Baars.
\newblock The conscious access hypothesis: origins and recent evidence.
\newblock \emph{Trends in cognitive sciences}, 6\penalty0 (1):\penalty0 47--52,
  2002.

\bibitem[Baars(2003)]{baars2003global}
Bernard~J Baars.
\newblock The global brainweb: An update on global workspace theory.
\newblock \emph{Science and Consciousness Review}, 2, 2003.

\bibitem[Baars and Franklin(2003)]{baars2003conscious}
Bernard~J Baars and Stan Franklin.
\newblock How conscious experience and working memory interact.
\newblock \emph{Trends in cognitive sciences}, 7\penalty0 (4):\penalty0
  166--172, 2003.

\bibitem[Baars and Franklin(2007)]{baars2007architectural}
Bernard~J Baars and Stan Franklin.
\newblock An architectural model of conscious and unconscious brain functions:
  Global workspace theory and ida.
\newblock \emph{Neural networks}, 20\penalty0 (9):\penalty0 955--961, 2007.

\bibitem[Baars and Franklin(2009)]{baars2009consciousness}
Bernard~J Baars and Stan Franklin.
\newblock Consciousness is computational: The lida model of global workspace
  theory.
\newblock \emph{International Journal of Machine Consciousness}, 1\penalty0
  (01):\penalty0 23--32, 2009.

\bibitem[Baars et~al.(1997)]{baars1997theater}
Bernard~J Baars et~al.
\newblock \emph{In the theater of consciousness: The workspace of the mind}.
\newblock Oxford University Press, USA, 1997.

\bibitem[Balduzzi and Tononi(2008)]{balduzzi2008integrated}
David Balduzzi and Giulio Tononi.
\newblock Integrated information in discrete dynamical systems: motivation and
  theoretical framework.
\newblock \emph{PLoS computational biology}, 4\penalty0 (6):\penalty0 e1000091,
  2008.

\bibitem[Bartolomei and Naccache(2011)]{bartolomei2011global}
Fabrice Bartolomei and Lionel Naccache.
\newblock The global workspace (gw) theory of consciousness and epilepsy.
\newblock \emph{Behavioural neurology}, 24\penalty0 (1):\penalty0 67--74, 2011.

\bibitem[Birch et~al.(2020)Birch, Ginsburg, and Jablonka]{birch2020unlimited}
Jonathan Birch, Simona Ginsburg, and Eva Jablonka.
\newblock Unlimited associative learning and the origins of consciousness: a
  primer and some predictions.
\newblock \emph{Biology \& philosophy}, 35:\penalty0 1--23, 2020.

\bibitem[Blum and Blum(2022)]{blum2022theory}
Lenore Blum and Manuel Blum.
\newblock A theory of consciousness from a theoretical computer science
  perspective: Insights from the conscious turing machine.
\newblock \emph{Proceedings of the National Academy of Sciences}, 119\penalty0
  (21):\penalty0 e2115934119, 2022.

\bibitem[Borji and Itti(2015)]{borji2015cat2000}
Ali Borji and Laurent Itti.
\newblock Cat2000: A large scale fixation dataset for boosting saliency
  research.
\newblock \emph{arXiv preprint arXiv:1505.03581}, 2015.

\bibitem[Brentano(1973)]{brentano1973psychologyfrom}
Franz Brentano.
\newblock Psychologyfrom an empirical standpoint.
\newblock \emph{Trans. AC Rancurello, DB Terrell and LL McAlister. London},
  1973.

\bibitem[Brohan et~al.(2023)Brohan, Brown, Carbajal, Chebotar, Chen,
  Choromanski, Ding, Driess, Dubey, Finn, et~al.]{brohan2023rt}
Anthony Brohan, Noah Brown, Justice Carbajal, Yevgen Chebotar, Xi~Chen,
  Krzysztof Choromanski, Tianli Ding, Danny Driess, Avinava Dubey, Chelsea
  Finn, et~al.
\newblock Rt-2: Vision-language-action models transfer web knowledge to robotic
  control.
\newblock \emph{arXiv preprint arXiv:2307.15818}, 2023.

\bibitem[Brown et~al.(2019)Brown, Lau, and LeDoux]{brown2019understanding}
Richard Brown, Hakwan Lau, and Joseph~E LeDoux.
\newblock Understanding the higher-order approach to consciousness.
\newblock \emph{Trends in cognitive sciences}, 23\penalty0 (9):\penalty0
  754--768, 2019.

\bibitem[Brown et~al.(2020)Brown, Mann, Ryder, Subbiah, Kaplan, Dhariwal,
  Neelakantan, Shyam, Sastry, Askell, et~al.]{brown2020language}
Tom Brown, Benjamin Mann, Nick Ryder, Melanie Subbiah, Jared~D Kaplan, Prafulla
  Dhariwal, Arvind Neelakantan, Pranav Shyam, Girish Sastry, Amanda Askell,
  et~al.
\newblock Language models are few-shot learners.
\newblock \emph{Advances in neural information processing systems},
  33:\penalty0 1877--1901, 2020.

\bibitem[Butlin et~al.(2023)Butlin, Long, Elmoznino, Bengio, Birch, Constant,
  Deane, Fleming, Frith, Ji, Kanai, Klein, Lindsay, Michel, Mudrik, Peters,
  Schwitzgebel, Simon, and VanRullen]{butlin2023consciousness}
Patrick Butlin, Robert Long, Eric Elmoznino, Yoshua Bengio, Jonathan Birch,
  Axel Constant, George Deane, Stephen~M. Fleming, Chris Frith, Xu~Ji, Ryota
  Kanai, Colin Klein, Grace Lindsay, Matthias Michel, Liad Mudrik, Megan A.~K.
  Peters, Eric Schwitzgebel, Jonathan Simon, and Rufin VanRullen.
\newblock Consciousness in artificial intelligence: Insights from the science
  of consciousness, 2023.

\bibitem[Byrne(1997)]{byrne1997some}
Alex Byrne.
\newblock Some like it hot: Consciousness and higher-order thoughts.
\newblock \emph{Philosophical Studies: An International Journal for Philosophy
  in the Analytic Tradition}, 86\penalty0 (2):\penalty0 103--129, 1997.

\bibitem[Caron and Srivastava(2022)]{caron2022identifying}
Graham Caron and Shashank Srivastava.
\newblock Identifying and manipulating the personality traits of language
  models.
\newblock \emph{arXiv preprint arXiv:2212.10276}, 2022.

\bibitem[Carroll(2021)]{carroll2021consciousness}
Sean Carroll.
\newblock Consciousness and the laws of physics.
\newblock \emph{Journal of Consciousness Studies}, 28\penalty0 (9-10):\penalty0
  16--31, 2021.

\bibitem[Carruthers(1998)]{carruthers1998language}
Peter Carruthers.
\newblock \emph{Language, thought and consciousness: An essay in philosophical
  psychology}.
\newblock Cambridge University Press, 1998.

\bibitem[Carruthers(2005)]{carruthers2005consciousness}
Peter Carruthers.
\newblock \emph{Consciousness: Essays from a higher-order perspective}.
\newblock Clarendon Press, 2005.

\bibitem[Carruthers and Gennaro(2023)]{sep-consciousness-higher}
Peter Carruthers and Rocco Gennaro.
\newblock {Higher-Order Theories of Consciousness}.
\newblock In Edward~N. Zalta and Uri Nodelman, editors, \emph{The {Stanford}
  Encyclopedia of Philosophy}. Metaphysics Research Lab, Stanford University,
  {F}all 2023 edition, 2023.

\bibitem[Casali et~al.(2013)Casali, Gosseries, Rosanova, Boly, Sarasso, Casali,
  Casarotto, Bruno, Laureys, Tononi, et~al.]{casali2013theoretically}
Adenauer~G Casali, Olivia Gosseries, Mario Rosanova, M{\'e}lanie Boly, Simone
  Sarasso, Karina~R Casali, Silvia Casarotto, Marie-Aur{\'e}lie Bruno, Steven
  Laureys, Giulio Tononi, et~al.
\newblock A theoretically based index of consciousness independent of sensory
  processing and behavior.
\newblock \emph{Science translational medicine}, 5\penalty0 (198):\penalty0
  198ra105--198ra105, 2013.

\bibitem[Caston(2002)]{caston2002aristotle}
Victor Caston.
\newblock Aristotle on consciousness.
\newblock \emph{Mind}, 111\penalty0 (444):\penalty0 751--815, 2002.

\bibitem[Chalmers(2017)]{chalmers2017hard}
David Chalmers.
\newblock The hard problem of consciousness.
\newblock \emph{The Blackwell companion to consciousness}, pages 32--42, 2017.

\bibitem[Chalmers(1995)]{chalmers1995facing}
David~J Chalmers.
\newblock Facing up to the problem of consciousness.
\newblock \emph{Journal of consciousness studies}, 2\penalty0 (3):\penalty0
  200--219, 1995.

\bibitem[Chalmers(1997)]{chalmers1997conscious}
David~J Chalmers.
\newblock \emph{The conscious mind: In search of a fundamental theory}.
\newblock Oxford Paperbacks, 1997.

\bibitem[Chalmers(2023)]{chalmers2023could}
David~J Chalmers.
\newblock Could a large language model be conscious?
\newblock \emph{arXiv preprint arXiv:2303.07103}, 2023.

\bibitem[Chen et~al.(2022)Chen, Wang, Changpinyo, Piergiovanni, Padlewski,
  Salz, Goodman, Grycner, Mustafa, Beyer, et~al.]{chen2022pali}
Xi~Chen, Xiao Wang, Soravit Changpinyo, AJ~Piergiovanni, Piotr Padlewski,
  Daniel Salz, Sebastian Goodman, Adam Grycner, Basil Mustafa, Lucas Beyer,
  et~al.
\newblock Pali: A jointly-scaled multilingual language-image model.
\newblock \emph{arXiv preprint arXiv:2209.06794}, 2022.

\bibitem[Chen et~al.(2023)Chen, Djolonga, Padlewski, Mustafa, Changpinyo, Wu,
  Ruiz, Goodman, Wang, Tay, et~al.]{chen2023pali}
Xi~Chen, Josip Djolonga, Piotr Padlewski, Basil Mustafa, Soravit Changpinyo,
  Jialin Wu, Carlos~Riquelme Ruiz, Sebastian Goodman, Xiao Wang, Yi~Tay, et~al.
\newblock Pali-x: On scaling up a multilingual vision and language model.
\newblock \emph{arXiv preprint arXiv:2305.18565}, 2023.

\bibitem[Chowdhery et~al.(2022)Chowdhery, Narang, Devlin, Bosma, Mishra,
  Roberts, Barham, Chung, Sutton, Gehrmann, et~al.]{chowdhery2022palm}
Aakanksha Chowdhery, Sharan Narang, Jacob Devlin, Maarten Bosma, Gaurav Mishra,
  Adam Roberts, Paul Barham, Hyung~Won Chung, Charles Sutton, Sebastian
  Gehrmann, et~al.
\newblock Palm: Scaling language modeling with pathways.
\newblock \emph{arXiv preprint arXiv:2204.02311}, 2022.

\bibitem[Cleeremans(2011)]{cleeremans2011radical}
Axel Cleeremans.
\newblock The radical plasticity thesis: how the brain learns to be conscious.
\newblock \emph{Frontiers in psychology}, 2:\penalty0 86, 2011.

\bibitem[Cleeremans et~al.(2007)Cleeremans, Timmermans, and
  Pasquali]{cleeremans2007consciousness}
Axel Cleeremans, Bert Timmermans, and Antoine Pasquali.
\newblock Consciousness and metarepresentation: A computational sketch.
\newblock \emph{Neural Networks}, 20\penalty0 (9):\penalty0 1032--1039, 2007.

\bibitem[DaSilveira et~al.(2015)DaSilveira, DeSouza, and
  Gomes]{dasilveira2015self}
Amanda DaSilveira, Mariane~L DeSouza, and William~B Gomes.
\newblock Self-consciousness concept and assessment in self-report measures.
\newblock \emph{Frontiers in psychology}, 6:\penalty0 930, 2015.

\bibitem[De~Raad(2000)]{de2000big}
Boele De~Raad.
\newblock \emph{The big five personality factors: the psycholexical approach to
  personality.}
\newblock Hogrefe \& Huber Publishers, 2000.

\bibitem[Dehaene(2014)]{dehaene2014consciousness}
Stanislas Dehaene.
\newblock \emph{Consciousness and the brain: Deciphering how the brain codes
  our thoughts}.
\newblock Penguin, 2014.

\bibitem[Dehaene and Changeux(2011)]{dehaene2011experimental}
Stanislas Dehaene and Jean-Pierre Changeux.
\newblock Experimental and theoretical approaches to conscious processing.
\newblock \emph{Neuron}, 70\penalty0 (2):\penalty0 200--227, 2011.

\bibitem[Demertzi et~al.(2017)Demertzi, Sitt, Sarasso, and
  Pinxten]{demertzi2017measuring}
Athena Demertzi, Jacobo~Diego Sitt, Simone Sarasso, and Wim Pinxten.
\newblock Measuring states of pathological (un) consciousness: research
  dimensions, clinical applications, and ethics.
\newblock \emph{Neuroscience of Consciousness}, 2017\penalty0 (1):\penalty0
  nix010, 2017.

\bibitem[Dennett(1978)]{dennett1978toward}
Daniel~C Dennett.
\newblock Toward a cognitive theory of consciousness.
\newblock 1978.

\bibitem[Desimone(1998)]{desimone1998visual}
Robert Desimone.
\newblock Visual attention mediated by biased competition in extrastriate
  visual cortex.
\newblock \emph{Philosophical Transactions of the Royal Society of London.
  Series B: Biological Sciences}, 353\penalty0 (1373):\penalty0 1245--1255,
  1998.

\bibitem[Desimone et~al.(1995)Desimone, Duncan, et~al.]{desimone1995neural}
Robert Desimone, John Duncan, et~al.
\newblock Neural mechanisms of selective visual attention.
\newblock \emph{Annual review of neuroscience}, 18\penalty0 (1):\penalty0
  193--222, 1995.

\bibitem[Driess et~al.(2023)Driess, Xia, Sajjadi, Lynch, Chowdhery, Ichter,
  Wahid, Tompson, Vuong, Yu, et~al.]{driess2023palm}
Danny Driess, Fei Xia, Mehdi~SM Sajjadi, Corey Lynch, Aakanksha Chowdhery,
  Brian Ichter, Ayzaan Wahid, Jonathan Tompson, Quan Vuong, Tianhe Yu, et~al.
\newblock Palm-e: An embodied multimodal language model.
\newblock \emph{arXiv preprint arXiv:2303.03378}, 2023.

\bibitem[Einstein et~al.(1935)Einstein, Podolsky, and Rosen]{einstein1935can}
Albert Einstein, Boris Podolsky, and Nathan Rosen.
\newblock Can quantum-mechanical description of physical reality be considered
  complete?
\newblock \emph{Physical review}, 47\penalty0 (10):\penalty0 777, 1935.

\bibitem[Elton(2020)]{elton2020self}
Daniel~C Elton.
\newblock Self-explaining ai as an alternative to interpretable ai.
\newblock In \emph{Artificial General Intelligence: 13th International
  Conference, AGI 2020, St. Petersburg, Russia, September 16--19, 2020,
  Proceedings 13}, pages 95--106. Springer, 2020.

\bibitem[Gazzaniga et~al.(1996)Gazzaniga, Eliassen, Nisenson, Wessinger,
  Fendrich, and Baynes]{gazzaniga1996collaboration}
Michael~S Gazzaniga, James~C Eliassen, Laura Nisenson, C~Mark Wessinger, Robert
  Fendrich, and Kathleen Baynes.
\newblock Collaboration between the hemispheres of a callosotomy patient:
  Emerging right hemisphere speech and the left hemisphere interpreter.
\newblock \emph{Brain}, 119\penalty0 (4):\penalty0 1255--1262, 1996.

\bibitem[Gennaro(2011)]{gennaro2011consciousness}
Rocco~J Gennaro.
\newblock The consciousness paradox: Consciousness, concepts, and higher-order
  thoughts.
\newblock 2011.

\bibitem[Giacino et~al.(2004)Giacino, Kalmar, and Whyte]{giacino2004jfk}
Joseph~T Giacino, Kathleen Kalmar, and John Whyte.
\newblock The jfk coma recovery scale-revised: measurement characteristics and
  diagnostic utility.
\newblock \emph{Archives of physical medicine and rehabilitation}, 85\penalty0
  (12):\penalty0 2020--2029, 2004.

\bibitem[Graziano(2016)]{graziano2016consciousness}
Michael~SA Graziano.
\newblock Consciousness engineered.
\newblock \emph{Journal of Consciousness Studies}, 23\penalty0
  (11-12):\penalty0 98--115, 2016.

\bibitem[Graziano and Kastner(2011)]{graziano2011human}
Michael~SA Graziano and Sabine Kastner.
\newblock Human consciousness and its relationship to social neuroscience: A
  novel hypothesis.
\newblock \emph{Cognitive neuroscience}, 2\penalty0 (2):\penalty0 98--113,
  2011.

\bibitem[Graziano and Webb(2015)]{graziano2015attention}
Michael~SA Graziano and Taylor~W Webb.
\newblock The attention schema theory: a mechanistic account of subjective
  awareness.
\newblock \emph{Frontiers in psychology}, page 500, 2015.

\bibitem[Graziano et~al.(2020)Graziano, Guterstam, Bio, and
  Wilterson]{graziano2020toward}
Michael~SA Graziano, Arvid Guterstam, Branden~J Bio, and Andrew~I Wilterson.
\newblock Toward a standard model of consciousness: Reconciling the attention
  schema, global workspace, higher-order thought, and illusionist theories.
\newblock \emph{Cognitive Neuropsychology}, 37\penalty0 (3-4):\penalty0
  155--172, 2020.

\bibitem[Hameroff(2010)]{hameroff2010conscious}
Stuart Hameroff.
\newblock The “conscious pilot”—dendritic synchrony moves through the
  brain to mediate consciousness.
\newblock \emph{Journal of biological physics}, 36:\penalty0 71--93, 2010.

\bibitem[Hameroff(2012)]{hameroff2012quantum}
Stuart Hameroff.
\newblock How quantum brain biology can rescue conscious free will.
\newblock \emph{Frontiers in integrative neuroscience}, 6:\penalty0 93, 2012.

\bibitem[Hameroff and Penrose(1996)]{hameroff1996orchestrated}
Stuart Hameroff and Roger Penrose.
\newblock Orchestrated reduction of quantum coherence in brain microtubules: A
  model for consciousness.
\newblock \emph{Mathematics and computers in simulation}, 40\penalty0
  (3-4):\penalty0 453--480, 1996.

\bibitem[Hameroff and Penrose(2014)]{hameroff2014consciousness}
Stuart Hameroff and Roger Penrose.
\newblock Consciousness in the universe: A review of the ‘orch or’theory.
\newblock \emph{Physics of life reviews}, 11\penalty0 (1):\penalty0 39--78,
  2014.

\bibitem[Hameroff(2007)]{hameroff2007brain}
Stuart~R Hameroff.
\newblock The brain is both neurocomputer and quantum computer.
\newblock \emph{Cognitive Science}, 31\penalty0 (6):\penalty0 1035--1045, 2007.

\bibitem[Hobson et~al.(2000)Hobson, Pace-Schott, and
  Stickgold]{hobson2000dreaming}
J~Allan Hobson, Edward~F Pace-Schott, and Robert Stickgold.
\newblock Dreaming and the brain: toward a cognitive neuroscience of conscious
  states.
\newblock \emph{Behavioral and brain sciences}, 23\penalty0 (6):\penalty0
  793--842, 2000.

\bibitem[Hoel(2017)]{hoel2017map}
Erik~P Hoel.
\newblock When the map is better than the territory.
\newblock \emph{Entropy}, 19\penalty0 (5):\penalty0 188, 2017.

\bibitem[Hoel et~al.(2013)Hoel, Albantakis, and Tononi]{Hoel2013Quntify}
Erik~P. Hoel, Larissa Albantakis, and Giulio Tononi.
\newblock Quantifying causal emergence shows that macro can beat micro.
\newblock \emph{Proceedings of the National Academy of Sciences}, 110\penalty0
  (49):\penalty0 19790--19795, 2013.
\newblock \doi{10.1073/pnas.1314922110}.
\newblock URL \url{https://www.pnas.org/doi/abs/10.1073/pnas.1314922110}.

\bibitem[Hopfield(1982)]{hopfield1982neural}
John~J Hopfield.
\newblock Neural networks and physical systems with emergent collective
  computational abilities.
\newblock \emph{Proceedings of the national academy of sciences}, 79\penalty0
  (8):\penalty0 2554--2558, 1982.

\bibitem[Huang et~al.(2022)Huang, Gu, Hou, Wu, Wang, Yu, and
  Han]{huang2022large}
Jiaxin Huang, Shixiang~Shane Gu, Le~Hou, Yuexin Wu, Xuezhi Wang, Hongkun Yu,
  and Jiawei Han.
\newblock Large language models can self-improve.
\newblock \emph{arXiv preprint arXiv:2210.11610}, 2022.

\bibitem[Izard and Buechler(1980)]{izard1980aspects}
Carroll~E Izard and Sandra Buechler.
\newblock Aspects of consciousness and personality in terms of differential
  emotions theory.
\newblock In \emph{Theories of emotion}, pages 165--187. Elsevier, 1980.

\bibitem[Jannai et~al.(2023)Jannai, Meron, Lenz, Levine, and
  Shoham]{jannai2023human}
Daniel Jannai, Amos Meron, Barak Lenz, Yoav Levine, and Yoav Shoham.
\newblock Human or not? a gamified approach to the turing test.
\newblock \emph{arXiv preprint arXiv:2305.20010}, 2023.

\bibitem[Jevtic et~al.(2012)Jevtic, Jennings, and Rudolph]{jevtic2012maximally}
Sania Jevtic, David Jennings, and Terry Rudolph.
\newblock Maximally and minimally correlated states attainable within a closed
  evolving system.
\newblock \emph{Physical review letters}, 108\penalty0 (11):\penalty0 110403,
  2012.

\bibitem[Jiang et~al.(2022)Jiang, Xu, Zhu, Han, Zhang, and
  Zhu]{jiang2022evaluating}
Guangyuan Jiang, Manjie Xu, Song-Chun Zhu, Wenjuan Han, Chi Zhang, and Yixin
  Zhu.
\newblock Evaluating and inducing personality in pre-trained language models.
\newblock 2022.

\bibitem[Johansen(2006)]{johansen2006update}
Jay~W Johansen.
\newblock Update on bispectral index monitoring.
\newblock \emph{Best practice \& research Clinical anaesthesiology},
  20\penalty0 (1):\penalty0 81--99, 2006.

\bibitem[Jones(1979)]{jones1979glasgow}
Cathy Jones.
\newblock Glasgow coma scale, 1979.

\bibitem[Jones and Paulhus(2014)]{jones2014introducing}
Daniel~N Jones and Delroy~L Paulhus.
\newblock Introducing the short dark triad (sd3) a brief measure of dark
  personality traits.
\newblock \emph{Assessment}, 21\penalty0 (1):\penalty0 28--41, 2014.

\bibitem[Kanwisher and Wojciulik(2000)]{kanwisher2000visual}
Nancy Kanwisher and Ewa Wojciulik.
\newblock Visual attention: insights from brain imaging.
\newblock \emph{Nature reviews neuroscience}, 1\penalty0 (2):\penalty0 91--100,
  2000.

\bibitem[Karra et~al.(2022)Karra, Nguyen, and
  Tulabandhula]{karra2022estimating}
Saketh~Reddy Karra, Son~The Nguyen, and Theja Tulabandhula.
\newblock Estimating the personality of white-box language models.
\newblock \emph{arXiv preprint arXiv:2204.12000}, 2022.

\bibitem[Kriegel(2009)]{kriegel2009subjective}
Uriah Kriegel.
\newblock \emph{Subjective consciousness: A self-representational theory}.
\newblock OUP Oxford, 2009.

\bibitem[Lamme(2006)]{lamme2006towards}
Victor~AF Lamme.
\newblock Towards a true neural stance on consciousness.
\newblock \emph{Trends in cognitive sciences}, 10\penalty0 (11):\penalty0
  494--501, 2006.

\bibitem[Lamme(2010)]{lamme2010neuroscience}
Victor~AF Lamme.
\newblock How neuroscience will change our view on consciousness.
\newblock \emph{Cognitive neuroscience}, 1\penalty0 (3):\penalty0 204--220,
  2010.

\bibitem[Lamme(2020)]{lamme2020visual}
Victor~AF Lamme.
\newblock Visual functions generating conscious seeing.
\newblock \emph{Frontiers in Psychology}, 11:\penalty0 83, 2020.

\bibitem[Lau(2011)]{lau2011theoretical}
Hakwan Lau.
\newblock Theoretical motivations for investigating the neural correlates of
  consciousness.
\newblock \emph{Wiley Interdisciplinary Reviews: Cognitive Science}, 2\penalty0
  (1):\penalty0 1--7, 2011.

\bibitem[Lau and Rosenthal(2011)]{lau2011empirical}
Hakwan Lau and David Rosenthal.
\newblock Empirical support for higher-order theories of conscious awareness.
\newblock \emph{Trends in cognitive sciences}, 15\penalty0 (8):\penalty0
  365--373, 2011.

\bibitem[Lau(2007)]{lau2007higher}
Hakwan~C Lau.
\newblock A higher order bayesian decision theory of consciousness.
\newblock \emph{Progress in brain research}, 168:\penalty0 35--48, 2007.

\bibitem[Lee et~al.(2022)Lee, Sanz, Barra, Wolff, Nieminen, Boly, Rosanova,
  Casarotto, Bodart, Annen, et~al.]{lee2022quantifying}
Minji Lee, Leandro~RD Sanz, Alice Barra, Audrey Wolff, Jaakko~O Nieminen,
  Melanie Boly, Mario Rosanova, Silvia Casarotto, Olivier Bodart, Jitka Annen,
  et~al.
\newblock Quantifying arousal and awareness in altered states of consciousness
  using interpretable deep learning.
\newblock \emph{Nature communications}, 13\penalty0 (1):\penalty0 1--14, 2022.

\bibitem[Lendner et~al.(2020)Lendner, Helfrich, Mander, Romundstad, Lin,
  Walker, Larsson, and Knight]{lendner2020electrophysiological}
Janna~D Lendner, Randolph~F Helfrich, Bryce~A Mander, Luis Romundstad, Jack~J
  Lin, Matthew~P Walker, Pal~G Larsson, and Robert~T Knight.
\newblock An electrophysiological marker of arousal level in humans.
\newblock \emph{Elife}, 9:\penalty0 e55092, 2020.

\bibitem[Li et~al.(2022)Li, Li, Liu, Bing, and Joty]{li2022gpt}
Xingxuan Li, Yutong Li, Linlin Liu, Lidong Bing, and Shafiq Joty.
\newblock Is gpt-3 a psychopath? evaluating large language models from a
  psychological perspective.
\newblock \emph{arXiv preprint arXiv:2212.10529}, 2022.

\bibitem[Lin and Wroten(2017)]{lin2017ranchos}
Katherine Lin and Michael Wroten.
\newblock Ranchos los amigos.
\newblock 2017.

\bibitem[Lindsay(2020)]{lindsay2020attention}
Grace~W Lindsay.
\newblock Attention in psychology, neuroscience, and machine learning.
\newblock \emph{Frontiers in computational neuroscience}, 14:\penalty0 29,
  2020.

\bibitem[Locke(1948)]{locke1948essay}
John Locke.
\newblock An essay concerning human understanding, 1690.
\newblock 1948.

\bibitem[Lycan(1996)]{lycan1996consciousness}
William~G Lycan.
\newblock \emph{Consciousness and experience}.
\newblock Mit Press, 1996.

\bibitem[Lycan(2004)]{lycan2004superiority}
William~G Lycan.
\newblock \emph{The superiority of HOP to HOT}.
\newblock na, 2004.

\bibitem[Ma et~al.(2023)Ma, Zhang, and Zhu]{ma2023brain}
Yuxi Ma, Chi Zhang, and Song-Chun Zhu.
\newblock Brain in a vat: On missing pieces towards artificial general
  intelligence in large language models, 2023.

\bibitem[MacKay(2003)]{mackay2003information}
David~JC MacKay.
\newblock \emph{Information theory, inference and learning algorithms}.
\newblock Cambridge university press, 2003.

\bibitem[Madaan et~al.(2023)Madaan, Tandon, Gupta, Hallinan, Gao, Wiegreffe,
  Alon, Dziri, Prabhumoye, Yang, et~al.]{madaan2023self}
Aman Madaan, Niket Tandon, Prakhar Gupta, Skyler Hallinan, Luyu Gao, Sarah
  Wiegreffe, Uri Alon, Nouha Dziri, Shrimai Prabhumoye, Yiming Yang, et~al.
\newblock Self-refine: Iterative refinement with self-feedback.
\newblock \emph{arXiv preprint arXiv:2303.17651}, 2023.

\bibitem[Manohar et~al.(2019)Manohar, Zokaei, Fallon, Vogels, and
  Husain]{manohar2019neural}
Sanjay~G Manohar, Nahid Zokaei, Sean~J Fallon, Tim~P Vogels, and Masud Husain.
\newblock Neural mechanisms of attending to items in working memory.
\newblock \emph{Neuroscience \& Biobehavioral Reviews}, 101:\penalty0 1--12,
  2019.

\bibitem[Massimini et~al.(2005)Massimini, Ferrarelli, Huber, Esser, Singh, and
  Tononi]{massimini2005breakdown}
Marcello Massimini, Fabio Ferrarelli, Reto Huber, Steve~K Esser, Harpreet
  Singh, and Giulio Tononi.
\newblock Breakdown of cortical effective connectivity during sleep.
\newblock \emph{Science}, 309\penalty0 (5744):\penalty0 2228--2232, 2005.

\bibitem[Massimini et~al.(2007)Massimini, Ferrarelli, Esser, Riedner, Huber,
  Murphy, Peterson, and Tononi]{massimini2007triggering}
Marcello Massimini, Fabio Ferrarelli, Steve~K Esser, Brady~A Riedner, Reto
  Huber, Michael Murphy, Michael~J Peterson, and Giulio Tononi.
\newblock Triggering sleep slow waves by transcranial magnetic stimulation.
\newblock \emph{Proceedings of the National Academy of Sciences}, 104\penalty0
  (20):\penalty0 8496--8501, 2007.

\bibitem[Merker(2005)]{merker2005liabilities}
Bjorn Merker.
\newblock The liabilities of mobility: A selection pressure for the transition
  to consciousness in animal evolution.
\newblock \emph{Consciousness and cognition}, 14\penalty0 (1):\penalty0
  89--114, 2005.

\bibitem[Merker(2007)]{merker2007consciousness}
Bjorn Merker.
\newblock Consciousness without a cerebral cortex: A challenge for neuroscience
  and medicine.
\newblock \emph{Behavioral and brain sciences}, 30\penalty0 (1):\penalty0
  63--81, 2007.

\bibitem[Myers()]{myers1988myers}
Isabel~Briggs Myers.
\newblock \emph{Myers-Briggs type indicator: MBTI.}

\bibitem[Naccache(2018)]{naccache2018and}
Lionel Naccache.
\newblock Why and how access consciousness can account for phenomenal
  consciousness.
\newblock \emph{Philosophical Transactions of the Royal Society B: Biological
  Sciences}, 373\penalty0 (1755):\penalty0 20170357, 2018.

\bibitem[OpenAI(2023)]{openai2023gpt4}
OpenAI.
\newblock Gpt-4 technical report, 2023.

\bibitem[Ouyang et~al.(2022)Ouyang, Wu, Jiang, Almeida, Wainwright, Mishkin,
  Zhang, Agarwal, Slama, Ray, et~al.]{ouyang2022training}
Long Ouyang, Jeffrey Wu, Xu~Jiang, Diogo Almeida, Carroll Wainwright, Pamela
  Mishkin, Chong Zhang, Sandhini Agarwal, Katarina Slama, Alex Ray, et~al.
\newblock Training language models to follow instructions with human feedback.
\newblock \emph{Advances in Neural Information Processing Systems},
  35:\penalty0 27730--27744, 2022.

\bibitem[Pasquali et~al.(2010)Pasquali, Timmermans, and
  Cleeremans]{pasquali2010know}
Antoine Pasquali, Bert Timmermans, and Axel Cleeremans.
\newblock Know thyself: Metacognitive networks and measures of consciousness.
\newblock \emph{Cognition}, 117\penalty0 (2):\penalty0 182--190, 2010.

\bibitem[Penrose(1991)]{penrose1991emperor}
Roger Penrose.
\newblock The emperor's new mind.
\newblock \emph{RSA Journal}, 139\penalty0 (5420):\penalty0 506--514, 1991.

\bibitem[Penrose(1994)]{penrose1994shadows}
Roger Penrose.
\newblock \emph{Shadows of the Mind}, volume~4.
\newblock Oxford University Press Oxford, 1994.

\bibitem[Persaud et~al.(2007)Persaud, McLeod, and Cowey]{persaud2007post}
Navindra Persaud, Peter McLeod, and Alan Cowey.
\newblock Post-decision wagering objectively measures awareness.
\newblock \emph{Nature neuroscience}, 10\penalty0 (2):\penalty0 257--261, 2007.

\bibitem[Picciuto(2011)]{picciuto2011addressing}
Vincent Picciuto.
\newblock Addressing higher-order misrepresentation with quotational thought.
\newblock \emph{Journal of Consciousness Studies}, 18\penalty0 (3-4):\penalty0
  109--136, 2011.

\bibitem[Putnam et~al.(1981)]{putnam1981reason}
Hilary Putnam et~al.
\newblock \emph{Reason, truth and history}, volume~3.
\newblock Cambridge University Press, 1981.

\bibitem[Rao et~al.(2023)Rao, Leung, and Miao]{rao2023can}
Haocong Rao, Cyril Leung, and Chunyan Miao.
\newblock Can chatgpt assess human personalities? a general evaluation
  framework.
\newblock \emph{arXiv preprint arXiv:2303.01248}, 2023.

\bibitem[Rappaport(2005)]{rappaport2005disability}
Maurice Rappaport.
\newblock The disability rating and coma/near-coma scales in evaluating severe
  head injury.
\newblock \emph{Neuropsychological rehabilitation}, 15\penalty0 (3-4):\penalty0
  442--453, 2005.

\bibitem[Reggia(2013)]{reggia2013rise}
James~A Reggia.
\newblock The rise of machine consciousness: Studying consciousness with
  computational models.
\newblock \emph{Neural Networks}, 44:\penalty0 112--131, 2013.

\bibitem[Robbins(2008)]{robbins2008consciousness}
Philip Robbins.
\newblock Consciousness and the social mind.
\newblock \emph{Cognitive Systems Research}, 9\penalty0 (1-2):\penalty0 15--23,
  2008.

\bibitem[Rosenthal(1993)]{rosenthal1993thinking}
David Rosenthal.
\newblock Thinking that one thinks.
\newblock \emph{Language and Thought}, pages 259--287, 1993.

\bibitem[Rosenthal(2005)]{rosenthal2005consciousness}
David Rosenthal.
\newblock \emph{Consciousness and mind}.
\newblock Clarendon Press, 2005.

\bibitem[Rosenthal(2012)]{rosenthal2012higher}
David Rosenthal.
\newblock Higher-order awareness, misrepresentation and function.
\newblock \emph{Philosophical Transactions of the Royal Society B: Biological
  Sciences}, 367\penalty0 (1594):\penalty0 1424--1438, 2012.

\bibitem[Rosenthal(1986)]{rosenthal1986two}
David~M Rosenthal.
\newblock Two concepts of consciousness.
\newblock \emph{Philosophical Studies: An International Journal for Philosophy
  in the Analytic Tradition}, 49\penalty0 (3):\penalty0 329--359, 1986.

\bibitem[Rosenthal(2004)]{rosenthal2004varieties}
David~M Rosenthal.
\newblock Varieties of higher-order theory.
\newblock \emph{Higher-order theories of consciousness}, pages 17--44, 2004.

\bibitem[Rosenthal(2009)]{rosenthal2009higher}
David~M Rosenthal.
\newblock Higher-order theories of consciousness.
\newblock 2009.

\bibitem[Rosenthal et~al.(1991)]{rosenthal1991nature}
David~M Rosenthal et~al.
\newblock \emph{The nature of mind}.
\newblock Oxford University Press New York, 1991.

\bibitem[Rosow and Manberg(2001)]{rosow2001bispectral}
Carl Rosow and Paul~J Manberg.
\newblock Bispectral index monitoring.
\newblock \emph{Anesthesiology Clinics of North America}, 19\penalty0
  (4):\penalty0 947--966, 2001.

\bibitem[Sauret and Lycan(2014)]{sauret2014attention}
Wesley Sauret and William~G Lycan.
\newblock Attention and internal monitoring: a farewell to hop.
\newblock \emph{Analysis}, 74\penalty0 (3):\penalty0 363--370, 2014.

\bibitem[Schutter and van Honk(2004)]{schutter2004extending}
Dennis~JLG Schutter and Jack van Honk.
\newblock Extending the global workspace theory to emotion: Phenomenality
  without access.
\newblock \emph{Consciousness and Cognition}, 13\penalty0 (3):\penalty0
  539--549, 2004.

\bibitem[Sejnowski(2023)]{sejnowski2023large}
Terrence~J Sejnowski.
\newblock Large language models and the reverse turing test.
\newblock \emph{Neural computation}, 35\penalty0 (3):\penalty0 309--342, 2023.

\bibitem[Seth(2009)]{seth2009strength}
Anil Seth.
\newblock The strength of weak artificial consciousness.
\newblock \emph{International Journal of Machine Consciousness}, 1\penalty0
  (01):\penalty0 71--82, 2009.

\bibitem[Seth et~al.(2008)Seth, Dienes, Cleeremans, Overgaard, and
  Pessoa]{seth2008measuring}
Anil~K Seth, Zolt{\'a}n Dienes, Axel Cleeremans, Morten Overgaard, and Luiz
  Pessoa.
\newblock Measuring consciousness: relating behavioural and neurophysiological
  approaches.
\newblock \emph{Trends in cognitive sciences}, 12\penalty0 (8):\penalty0
  314--321, 2008.

\bibitem[Shannon(1948)]{shannon1948mathematical}
Claude~E Shannon.
\newblock A mathematical theory of communication.
\newblock \emph{The Bell system technical journal}, 27\penalty0 (3):\penalty0
  379--423, 1948.

\bibitem[Shinn et~al.(2023)Shinn, Labash, and Gopinath]{shinn2023reflexion}
Noah Shinn, Beck Labash, and Ashwin Gopinath.
\newblock Reflexion: an autonomous agent with dynamic memory and
  self-reflection.
\newblock \emph{arXiv preprint arXiv:2303.11366}, 2023.

\bibitem[Stapp(2000)]{stapp2000importance}
Henry~P Stapp.
\newblock The importance of quantum decoherence in brain processes.
\newblock \emph{arXiv preprint quant-ph/0010029}, 2000.

\bibitem[Sternbach(2000)]{sternbach2000glasgow}
George~L Sternbach.
\newblock The glasgow coma scale.
\newblock \emph{The Journal of emergency medicine}, 19\penalty0 (1):\penalty0
  67--71, 2000.

\bibitem[Tegmark(2000)]{tegmark2000importance}
Max Tegmark.
\newblock Importance of quantum decoherence in brain processes.
\newblock \emph{Physical review E}, 61\penalty0 (4):\penalty0 4194, 2000.

\bibitem[Tegmark(2015)]{tegmark2015consciousness}
Max Tegmark.
\newblock Consciousness as a state of matter.
\newblock \emph{Chaos, Solitons \& Fractals}, 76:\penalty0 238--270, 2015.

\bibitem[Tegmark(2018)]{tegmark2018life}
Max Tegmark.
\newblock \emph{Life 3.0: Being human in the age of artificial intelligence}.
\newblock Vintage, 2018.

\bibitem[Thoppilan et~al.(2022)Thoppilan, De~Freitas, Hall, Shazeer,
  Kulshreshtha, Cheng, Jin, Bos, Baker, Du, et~al.]{thoppilan2022lamda}
Romal Thoppilan, Daniel De~Freitas, Jamie Hall, Noam Shazeer, Apoorv
  Kulshreshtha, Heng-Tze Cheng, Alicia Jin, Taylor Bos, Leslie Baker, Yu~Du,
  et~al.
\newblock Lamda: Language models for dialog applications.
\newblock \emph{arXiv preprint arXiv:2201.08239}, 2022.

\bibitem[Timmermans et~al.(2012)Timmermans, Schilbach, Pasquali, and
  Cleeremans]{timmermans2012higher}
Bert Timmermans, Leonhard Schilbach, Antoine Pasquali, and Axel Cleeremans.
\newblock Higher order thoughts in action: consciousness as an unconscious
  re-description process.
\newblock \emph{Philosophical Transactions of the Royal Society B: Biological
  Sciences}, 367\penalty0 (1594):\penalty0 1412--1423, 2012.

\bibitem[Tononi(2004)]{tononi2004information}
Giulio Tononi.
\newblock An information integration theory of consciousness.
\newblock \emph{BMC neuroscience}, 5\penalty0 (1):\penalty0 1--22, 2004.

\bibitem[Tononi and Koch(2015)]{tononi2015consciousness}
Giulio Tononi and Christof Koch.
\newblock Consciousness: here, there and everywhere?
\newblock \emph{Philosophical Transactions of the Royal Society B: Biological
  Sciences}, 370\penalty0 (1668):\penalty0 20140167, 2015.

\bibitem[Tononi and Sporns(2003)]{tononi2003measuring}
Giulio Tononi and Olaf Sporns.
\newblock Measuring information integration.
\newblock \emph{BMC neuroscience}, 4:\penalty0 1--20, 2003.

\bibitem[Tononi et~al.(2016)Tononi, Boly, Massimini, and
  Koch]{tononi2016integrated}
Giulio Tononi, Melanie Boly, Marcello Massimini, and Christof Koch.
\newblock Integrated information theory: from consciousness to its physical
  substrate.
\newblock \emph{Nature Reviews Neuroscience}, 17\penalty0 (7):\penalty0
  450--461, 2016.

\bibitem[Trapnell and Campbell(1999)]{trapnell1999private}
Paul~D Trapnell and Jennifer~D Campbell.
\newblock Private self-consciousness and the five-factor model of personality:
  distinguishing rumination from reflection.
\newblock \emph{Journal of personality and social psychology}, 76\penalty0
  (2):\penalty0 284, 1999.

\bibitem[Treue and Trujillo(1999)]{treue1999feature}
Stefan Treue and Julio C~Martinez Trujillo.
\newblock Feature-based attention influences motion processing gain in macaque
  visual cortex.
\newblock \emph{Nature}, 399\penalty0 (6736):\penalty0 575--579, 1999.

\bibitem[Turing(2009)]{turing2009computing}
Alan~M Turing.
\newblock Computing machinery and intelligence.
\newblock In \emph{Parsing the turing test}, pages 23--65. Springer, 2009.

\bibitem[Van~Gulick(2004)]{van2004higher}
Robert Van~Gulick.
\newblock Higher-order global states (hogs): an alternative higher-order model.
\newblock \emph{Higher-order theories of consciousness}, 2004:\penalty0 67--93,
  2004.

\bibitem[VanRullen and Kanai(2021)]{vanrullen2021deep}
Rufin VanRullen and Ryota Kanai.
\newblock Deep learning and the global workspace theory.
\newblock \emph{Trends in Neurosciences}, 44\penalty0 (9):\penalty0 692--704,
  2021.

\bibitem[Vaswani et~al.(2017)Vaswani, Shazeer, Parmar, Uszkoreit, Jones, Gomez,
  Kaiser, and Polosukhin]{vaswani2017attention}
Ashish Vaswani, Noam Shazeer, Niki Parmar, Jakob Uszkoreit, Llion Jones,
  Aidan~N Gomez, {\L}ukasz Kaiser, and Illia Polosukhin.
\newblock Attention is all you need.
\newblock \emph{Advances in neural information processing systems}, 30, 2017.

\bibitem[Walker et~al.(1990)Walker, Hall, and Hurst]{walker1990clinical}
H~Kenneth Walker, W~Dallas Hall, and J~Willis Hurst.
\newblock Clinical methods: the history, physical, and laboratory examinations.
\newblock 1990.

\bibitem[Wegner(2004)]{wegner2004precis}
Daniel~M Wegner.
\newblock Pr{\'e}cis of the illusion of conscious will.
\newblock \emph{Behavioral and Brain Sciences}, 27\penalty0 (5):\penalty0
  649--659, 2004.

\bibitem[Weng(2023)]{weng2023transformer}
Lilian Weng.
\newblock The transformer family version 2.0.
\newblock \emph{lilianweng.github.io}, Jan 2023.
\newblock URL
  \url{https://lilianweng.github.io/posts/2023-01-27-the-transformer-family-v2/}.

\bibitem[Wieser et~al.(2010)Wieser, Buetler, Koenig, and
  Riener]{wieser2010quantitative}
M~Wieser, L~Buetler, A~Koenig, and R~Riener.
\newblock Quantitative description of the state of awareness of patients in
  vegetative and minimally conscious state.
\newblock In \emph{2010 Annual International Conference of the IEEE Engineering
  in Medicine and Biology}, pages 5533--5536. IEEE, 2010.

\bibitem[Wijdicks et~al.(2005)Wijdicks, Bamlet, Maramattom, Manno, and
  McClelland]{wijdicks2005validation}
Eelco~FM Wijdicks, William~R Bamlet, Boby~V Maramattom, Edward~M Manno, and
  Robyn~L McClelland.
\newblock Validation of a new coma scale: the four score.
\newblock \emph{Annals of Neurology: Official Journal of the American
  Neurological Association and the Child Neurology Society}, 58\penalty0
  (4):\penalty0 585--593, 2005.

\end{thebibliography}







\end{document}